***Journal of Physics: Photonics*** **Roadmap**

# Roadmap on UV-C photodetectors: materials, applications and industry perspectives

**Fabien Massabuau[1,29,30], Drew Riley[2], Paul Meredith[2], Tilman Weiss[3], Damanpreet Kaur[4], Yuichi Oshima[5], Robert W Martin[1], Eva Monroy[6], Le Chen[7], Hongwei Liang[8], Hong Yin[7], Keyun Gu[5], Meiyong Liao[5], Yaonan Hou[2,9], Fa Cao[10,11], Xiaosheng Fang[10], Ruiheng Li[12], Guoqiang Peng[12], Zhiwen Jin[12], Lijie Li[9], Nasim Zarrabi[13], Sebastian Wood[13], Jesper Skottfelt[14], Susan E S Spesyvtseva[1,15], Jonathan McKendry[1,15], Christopher G Leburn[1,15], Daniel K L Oi[1], Ryan Pereira[16], Graeme Moore[17], Tom Lendrem[18], Christoph Wagner[19], David Maestre[20], Emilio Nogales[20], David J Rogers[21], Eric Sandana[21], Michel Chamberlin[21], Ferechteh H Teherani[21], Bianchi Méndez[20], Yana Suchikova[22], Marina Konuhova[23], Anatoli I Popov[23], Luís F. da Silva[24], Eduard Llobet[25,26,27], Sangjin Yoon[28], Dohyung Kim[28], Sangwoo Hong[28] and Seung Hwan Ko[28]**

[1] Department of Physics, SUPA, University of Strathclyde, Glasgow G4 0NG, United Kingdom
[2] Centre for Integrative Semiconductor Materials and Department of Physics, Swansea University Bay Campus, Crymlyn Burrows, Swansea SA2 8EN, United Kingdom
[3] sglux GmbH, Berlin, Germany
[4] Department of Physics and Materials Science and Engineering, Jaypee Institute of Information Technology Noida, Uttar Pradesh, India
[5] National Institute for Materials Science, Tsukuba 305-0044, Japan
[6] Université Grenoble Alpes, CEA, Grenoble INP, IRIG, PHELIQS, F-38000 Grenoble, France
[7] State Key Laboratory of High Pressure and Superhard Materials, College of Physics, Jilin University, Changchun 130012, China
[8] School of Integrated Circuits, Dalian University of Technology, Dalian 116024, China
[9] Electronic and electrical engineering, Swansea University, Bay Campus, Swansea SA1 8EN, United Kingdom
[10] Department of Materials Science, Institute of Optoelectronics, State Key Laboratory of Molecular Engineering of Polymers, Fudan University, Shanghai 200433, China
[11] State Key Laboratory of Flexible Electronics (LoFE) & Institute of Advanced Materials (IAM), School of Materials Science and Engineering, Nanjing University of Posts and Telecommunication (NJUPT), Nanjing 210023, China
[12] School of Physical Science and Technology, Lanzhou Center for Theoretical Physics, Key Laboratory of Theoretical Physics of Gansu Province, Key Laboratory of Quantum Theory and Applications of MoE, and Gansu Provincial Research Center for Basic Disciplines of Quantum Physics, Lanzhou University, Lanzhou, Gansu 730000, China
[13] Electromagnetic and Electrochemical Technologies, National Physical Laboratory, Hampton Road, Teddington TW11 0LR, United Kingdom
[14] Centre for Electronic Imaging, School of Physical Sciences, The Open University, Walton Hall, Milton Keynes MK7 6AA, United Kingdom
[15] Institute of Photonics, SUPA, University of Strathclyde, Glasgow, United Kingdom
[16] The Lyell Centre, Heriot Watt University, Edinburgh EH14 4AS, United Kingdom
[17] Scottish Water, 6 Buchanan Gate, Glasgow G33 6FB, United Kingdom
[18] Badger Meter UK Ltd, Oldham, United Kingdom
[19] Badger Meter Austria GmbH, Vienna, Austria
[20] Departamento de Física de Materiales, Universidad Complutense de Madrid, 28040 Madrid, Spain
[21] Nanovation, 8 route de Chevreuse, 78117 Châteaufort, France
[22] Scientific Department, Berdyansk State Pedagogical University, 69061 Zaporizhzhia, Ukraine
[23] Institute of Solid State Physics, University of Latvia, Kengaraga 8, Street, LV-1063 Riga, Latvia
[24] LM2N, Federal University of São Carlos, São Carlos, Brazil

[25] Universitat Rovira i Virgili, MINOS, School of Engineering, Tarragona, Spain
[26] IU-RESCAT, Research Institute in Sustainability, Climatic Change and Energy Transition, URV, Vila-seca, Spain
[27] TecnATox - Centre for Environmental, Food and Toxicological Technology, URV, Tarragona, Spain
[28] Department of Mechanical Engineering, Seoul National University, Seoul, Republic of Korea

[29] Guest Editor of the Roadmap.
[30] Author to whom any correspondence should be addressed.

E-mails: f.massabuau@strath.ac.uk

## Abstract

UV-C photodetectors are poised to play an increasingly important role in future photonic technologies, driven by the rapid emergence of UV-C light sources and new wide bandgap semiconductors. These advances are enabling new levels of spectral selectivity, radiation hardness, sensitivity, and device integration, while opening opportunities across a broad range of applications. This roadmap provides a comprehensive overview of the current landscape of UV-C photodetection, spanning established and emerging material platforms ($Ga_2O_3$, AlGaN, BN, diamond, MgZnO, 2-dimensional materials, metal halide perovskites, micro-electromechanical systems), and their applications in metrology, astronomy, communications, environmental monitoring, fire detection, missile warning, gas sensing, and medical diagnostics. By identifying opportunities, bottlenecks, and future directions, this roadmap aims to support both newcomers and established researchers, with the aim of accelerating the translation of UV-C photodetectors into impactful technologies.

## Contents

# 1. Introduction

**Fabien Massabuau**[1]

[1] Department of Physics, SUPA, University of Strathclyde, Glasgow, United Kingdom

E-mail: f.massabuau@strath.ac.uk

Photodetectors have a long history, with early observations of light-to-electricity phenomena dating back to 1839, when A. E. Becquerel reported the photovoltaic effect [1], followed by W. Smith's observation of photoconductivity in selenium (Se) [2]. Since then, advances in materials and devices have enabled the exploitation of much of the electromagnetic spectrum, with the greatest progress occurring in the visible and infrared (IR) regions, particularly for applications in solar energy conversion and telecommunications. In comparison, the detection of ultraviolet (UV) radiation remains a considerably less mature technology.

The UV region of the electromagnetic spectrum spans wavelengths from 400 to 100 nm and is commonly divided into UV-A (400–320 nm), UV-B (320–280 nm), and UV-C (280–100 nm). Of relevance to this roadmap is the UV-C range, which has attracted growing attention in recent years owing to its germicidal properties and the parallel development of UV-C light-emitting diodes (LEDs) [3]. These advances have increased interest in both the generation and detection of UV-C radiation across a range of scientific and technological contexts.

Legacy UV-C sensing technologies rely on silicon (Si), silicon carbide (SiC), photomultiplier tubes (PMTs), microchannel plate detector (MCPs), charge-coupled devices camera (CCD), and complementary metal-oxide semiconductor camera (CMOS). However, the emergence of a broad range of wide-bandgap semiconductors for UV-C detection over the past decade (Figure 1) has the potential to reshape the field. These materials offer several attractive characteristics for photodetection, including intrinsic spectral selectivity and radiation hardness. Leading developing and exploratory candidates include gallium oxide ($Ga_2O_3$), aluminium gallium nitride (AlGaN), boron nitride (BN), diamond, magnesium zinc oxide (MgZnO), several 2-dimensional materials (*e.g.* h-BN, $Ca_2Nb_3O_{10}$, $Sr_2Nb_3O_{10}$), chloride-based metal halide perovskites ($CsPbCl_3$), as well as their integration into micro-electromechanical systems (MEMS). Together, these developments are expanding the range of possible applications, strengthening established uses while enabling new opportunities across metrology, astronomy, communications, environmental monitoring, fire detection, missile warning, gas sensing, and medical diagnostics.

Interest in this field is driven not only by scientific progress but also by growing industrial demand. Although market estimates vary between reports, UV sensing has consistently been identified as a growing sector over several years. In 2020, the global UV sensor market (encompassing UV-A, UV-B and UV-C devices) was estimated at approximately USD 2.69 billion [4]. By 2025, this figure had risen to around USD 8.54 billion, with continued growth projected into the 2030s at a compound annual growth rate of approximately 25% [5-6] – we note that some reports use more conservative values [7]. While such forecasts should be interpreted with appropriate caution, they nevertheless highlight the increasing commercial relevance of UV, and UV-C, sensing technologies.

The aim of this roadmap is to provide a comprehensive overview of the UV-C photodetection landscape that would be useful to both newcomers and established researchers in the field. It seeks to present an interdisciplinary account of progress in UV-C photodetection, beginning with a general introduction to photodetectors and followed by discussions of incumbent and emerging material platforms, and finishing with a presentation of key applications and industry perspectives from experts across the relevant disciplines. In doing so, the roadmap aims to identify current bottlenecks and clarify the requirements needed for these technologies to achieve commercial deployment.

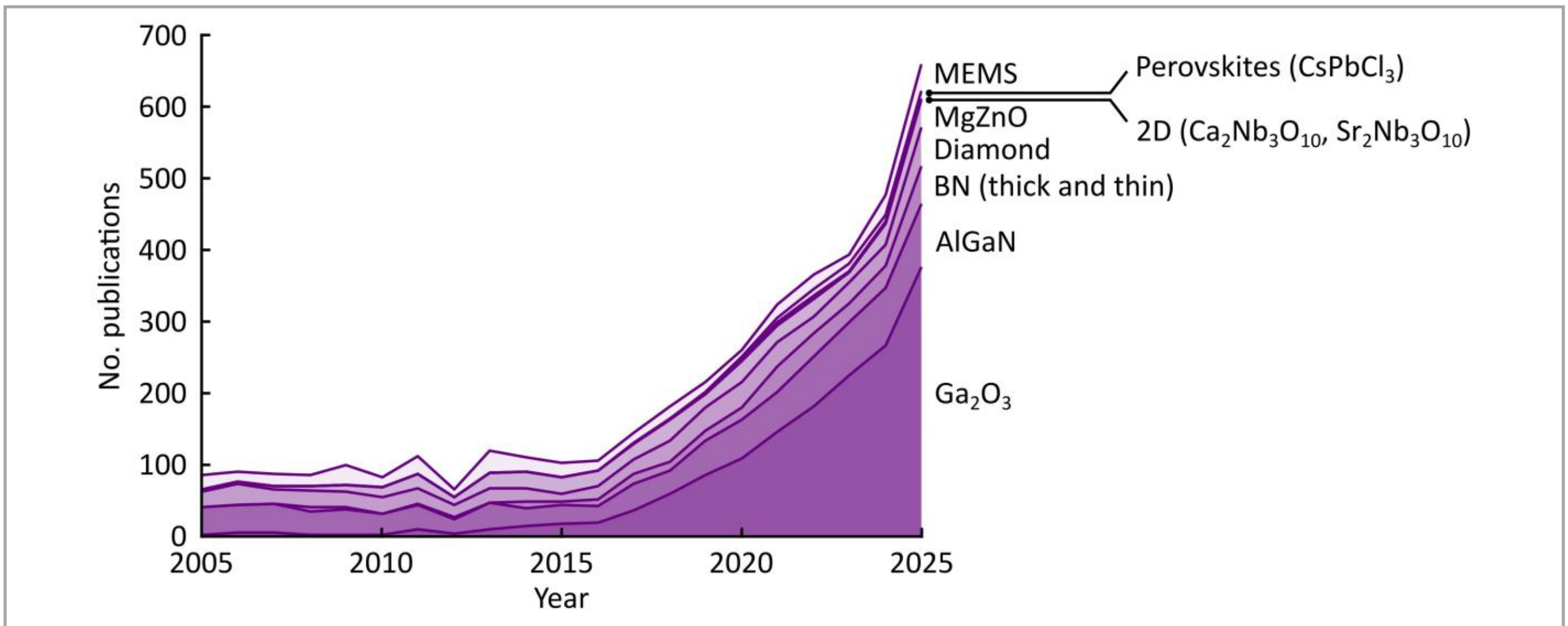

**Figure 1.** Number of publications per year for emerging UV-C detector material technologies. Produced using Scopus using the queries “detectors” (and synonyms) AND “UV-C” (and synonyms) AND “material name” (and synonyms).

## Acknowledgements

The author acknowledges support from the UK Space Agency Enabling Technologies Programme (Grant No. UKSAG23_0043_ETP4-052).

# 2. Introduction to photodetectors: performance figures of merit

**Drew Riley[1] and Paul Meredith[1]**

[1] Centre for Integrative Semiconductor Materials and Department of Physics, Swansea University Bay Campus, Crymlyn Burrows, Swansea, Wales, United Kingdom, SA2 8EN

E-mail: d.b.riley@swansea.ac.uk

The appropriateness and relevance of photodetector Performance Figures of Merit (FOM) will always depend on the prescribed application, be it imaging, sensing, communications, or otherwise. Although this roadmap is limited to UV-C photodetection using wide bandgap semiconductors exhibiting photoconductive or photovoltaic characteristics, we are, in this introductory section, agnostic to considerations of application-specific performance priorities. In general, one must consider a comprehensive suite of FOM covering all aspects of performance. Here we outline that suite before discussing the most common UV-C semiconductor photodetector device architectures.

**Photodetector Figures of Merit.**

The current output by a photodetector ($I$) will depend on many intrinsic factors of the device such as architecture, carrier mobility, carrier lifetime, light absorption strength, and contact extraction efficiency. However, if we take a high-level 'external' view, the primary objective of any photodetector is to transduce an optical signal with incident radiation power ($\Phi$), at wavelength ($\lambda$), and modulation frequency ($f$) into a measurable output signal (current or voltage). With this in mind, $I$ will be dependent on these extrinsic factors as well as the device active area ($A$), temperature ($T$), and bias on the detector ($V_b$) as well as the frequency bandwidth of the detecting circuit ($\Delta f$), thus $I \equiv I(\Phi, \lambda, f, A, T, V_b, \Delta f)$. If this current is above the noise floor of the photodetector, then the transduction signal can be meaningfully reconstructed. In light of this, the photodetector field has adopted a more-or-less agreed upon set of FOM which not only serve as comparator benchmarks for various technologies, but also as application-specific quantifiers. These FOM are as follows:

*Noise Spectral Density (NSD)*. The primary noise sources in photodetection are dependent on the chosen device architecture (see Common Architectures section below) and intrinsic consideration such as the chosen semiconductor(s) and the nature of the contacts (ohmic, Schottky, or otherwise). However, at low frequencies ($f < 500$ Hz) all photodetectors display NSD dominated by 1/f (or 'flicker') noise ($i_{1/f}$), which every electronic circuit is subject to. This flicker noise is a somewhat ethereal quantity whose physical origins are not always clear. By virtue of its frequency dependence, it is also often referred to as 'pink', as opposed to 'white' which indicates a frequency independence. At higher frequencies the noise floor is limited by either the shot noise ($i_{shot}$ – uncorrelated arrival of electrons at the contacts), or Johnson-Nyquist (thermal) noise ($i_{thermal}$ – thermal fluctuations leading to current). These, and perhaps other noise sources such as generation-recombination noise ($i_{g-r}$), add to create the NSD which indicates the lower limit of current that can be considered signal as a function of frequency. NSD is most accurately measured using a super swept-heterodyne spectrum analyser. Care must be taken to ensure that the noise floor of the detection circuit is lower than that of the device and that the bandwidth of the measurement is accurately reported. Figure 1a exemplifies the shape of the device and circuit NSD, and we can represent the total NSD as:

$$\langle i_{noise}^2(f)\rangle = \left[\langle i_{shot}^2\rangle + \langle i_{g-r}^2\rangle + \langle i_{thermal}^2\rangle + \langle i_{1/f}^2(f)\rangle\right] = \left[2q\langle i_d\rangle + \frac{4k_BT}{R_{sh}} + i_{1/f}^2(f)\right]\cdot\Delta f$$

where additionally, $q$ is the charge, $i_d$ the DC dark current, $k_B$ Boltzmann's constant, and $R_{sh}$ the device shunt resistance.

*Spectral Responsivity (R).* Once the NSD is understood the photocurrent $(I_{\text{ph}})$ can be evaluated as the difference between output current generated under illumination and the NSD. The Responsivity $(R)$ is then defined as:

$$R(f,\lambda,A,T,V_b,\Delta f) = \frac{I_{\text{ph}}(f,\lambda,A,T,V_b,\Delta f)}{\Phi(f,\lambda)}$$

which is closely related to the external quantum efficiency ($\eta = R \times hc/\lambda$) though the unit of $R$ [A/W] is more convenient for photodetectors compared to a fractional efficiency. Figure 1b shows an illustration of a potential responsivity of a UV photodetector with a band edge near 4 eV.

*Gain.* The multiplicative effect on $R$ due to extrinsic or intrinsic features, as examples; bias, carrier regeneration, or carrier recycling from traps. Figure 1b exemplifies a linear gain on $R$. It is important to note that all FOMs must be measured under the same conditions that created gain in $I_{\text{ph}}$. It is not uncommon in the literature to see, for example, $R$ reported under gain conditions and NSD not, which can lead to dramatic overestimations of Specific Detectivity (see below).

*Linear Dynamic Range (LDR).* At high light intensities $R$ can be limited by bimolecular recombination or series resistance [1,2], while at low light intensities carrier trapping and de-trapping, shunt resistance or capacitive effects can limit the photocurrent [1,2,3]. The linear dynamic range is defined, in dB, as the range over which the Spectral Responsivity is constant (where the slope of $I_{\text{ph}}$ vs. $\Phi$ is unity).

*Noise Equivalent Power (NEP) and Detectivity.* NEP is defined as the radiation power that will result in a current equal to the NSD for a given $\lambda$, $f$, and $\Delta f$. The NEP can be found from a direct measurement of a $I_{\text{ph}}$ vs $\Phi$, or from extrapolating this to the measured noise floor over the LDR. This is exemplified in Figure 1c where the NEP is shown for both cases. The Detectivity is defined as the inverse of the NEP [4].

*Specific Detectivity* ($D^*$). The most crucial FOM for photodetectors. From a measurement of the NEP $D^*$ can be determined as

$$D^*(f,\lambda) = \frac{\sqrt{A\Delta f}}{\text{NEP}(f,\lambda)}$$

expressed in the unit of Jones [cmHz$^{1/2}$/W]. Note that $R$ is determined by both intrinsic and extrinsic factors, 6 in total, whereas $D^*$ is determined strictly by extrinsic factors $(f,\lambda)$. Hence, $D^*$ is an intrinsic comparison of photodetector platforms under certain measurement conditions and is explicitly a measure of the minimum signal that can be transduced at a given speed [5]. Often $R$ will be related to $D^*$ through $D^* = \sqrt{A\Delta f} \times R(\lambda)/\left(\langle i_{noise}^2(f)\rangle\right)^{1/2}$. However, this relies on the assumption that the LDR extends down to the NEP and that $R(\lambda)$ and $i_{noise}(f)$ are measured under the same electrical conditions, c.f. our earlier comment that $i_{noise}$ must be measured under the same conditions that produce the gain used to measure $R$. These assumptions can be particularly problematic when assessing the validity of literature claims of $D^* > 10^{12}$ Jones and should not be used to calculate $D^*$ in novel material systems or device architectures where a full understanding of $R$ and the NSD is limited or incomplete [6]. Potential differences are exemplified in Figure 1d, where $D^*(\lambda)$ is shown calculated from the NEP in Figure 1c and from $R(\lambda)$ in Figure 1b.

*Rise time/fall time* ($\tau_r$). The time for a photodetector response to increase/decrease between 10% and 90% of the saturation value when under a square-wave modulated radiation field, as exemplified in Figure 1f.

*3dB bandwidth* ($f_{3dB}$). $f_{3dB}$ is the frequency at which the Responsivity is reduced by 3dB from the low frequency Responsivity as measured from a Bode Plot, exemplified in Figure 1a.

*Roll-off.* The derivative of the Bode Plot where $f \gg f_{3dB}$, expressed in dB/decade.

*Frequency Response.* Often the frequency response of a photodetector is considered a low-pass filter (LPF) described by one time constant $\tau$, with a frequency response given by [7].

$$R(f) \propto \frac{\sqrt{\tau}}{[1+(2\pi f\tau)^2]^{1/2}}$$

If one accepts this premise and that the response time of the photodetector is *RC* limited then $\tau_r \cong 0.35/f_{3dB}$, and the roll-off is -10 dB/decade, as exemplified by the green curves in Figure 1a. However, many detectors cannot be defined by one time constant [8], while filters such as the Butterworth Filter (BWF) may better explain the high frequency characteristics [9]. The photodetector frequency response modelled by an $n^{\text{th}}$ order BWF would be

$$R(f) \propto \frac{\sqrt{\tau}}{[1+(2\pi f\tau)^{2n}]^{1/2}}$$

Resulting in variable roll-off and $f_{3dB}$ for a given $\tau$, as exemplified by the dashed lines in Figure 1a, where $f_{\text{3dB}}$ and the roll-off decrease with increasing *n*. Importantly $\tau_r$, $f_{3dB}$, and roll-off should not be considered related via a LPF especially when evaluating novel materials, rather each should be measured independently using a Bode Plot and time series analysis of $\tau_r$.

Once the frequency response is understood $D^*(f)$ can be measured at the optimum wavelength $(\lambda_{\text{optimum}})$. Care must be taken as accurate measurements of NEP require high gain amplifiers which often have limited bandwidth, while utilizing $R(f)$ requires precise understanding of the NSD. Figure 1f highlights potential differences in $D^*(f)$ when utilizing various NSD and high frequency responses.

In summary to fully characterise a photodetector it is necessary to report the NSD (Figure 1a), $R(\lambda)$ (Figure 1b), $D^*(\lambda)$ (Figure 1d), a Bode Plot of $R(f,\lambda_{\text{optimum}})$ (Figure 1a) from which $f_{3dB}$ and roll-off is evaluated, a photodetector response curve with a square wave modulated radiation field to evaluate $\tau_r$ (Figure 1e), and $D^*(f,\lambda_{\text{optimum}})$ (Figure 1f). Importantly, all these must be measured under the conditions that create gain, if gain is an important feature of the device [5]. This holistic approach to understanding photodetector performance is absolutely crucial when investigating new materials systems or architectural approaches. The equations presented above contain rich physics which must be understood not only to report accurate comparative FOM, but also to engineer performance for a particular application. This is (and will continue to be) a central task in bringing forward viable UV-C photodetectors using wide and ultra-wide gap semiconductors such as the $Ga_2O_3$ family. In this specific case (for example) one could anticipate a complex trap-and-defect-dominated electronic landscape which could be ideal for promoting gain but likewise impose high noise floors which drastically limits Specific Detectivity and Frequency Response.

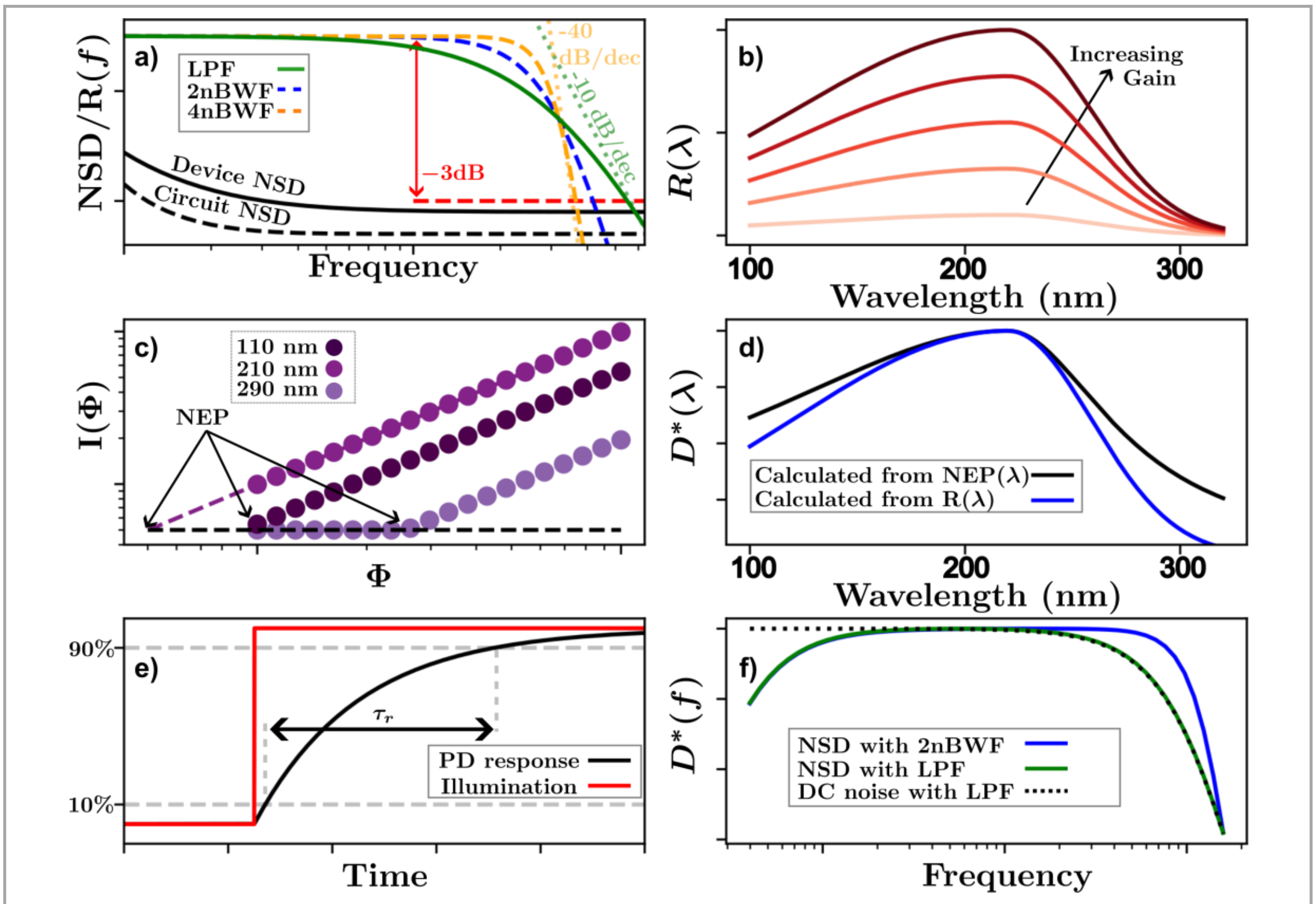

**Figure 1.** Essential Performance FOM for photodetectors irrespective of application and use scenario. a) Bode Plot. Expected NSD shapes for a device (black-solid line) and the associated detection circuit (black-dashed line). Responsivity at the optimum wavelength for a photodetector modelled after a Low-Pass Filter (LPF) (solid-green line) and a Butterworth Filter (BWF) (dashed lines). Marked is the -3dB line (red-dashed) and the roll-off for the LPF at -10dB/decade (green-dotted line) and the 4$^{th}$ order BWF -40 dB/decade (orange-dotted line). b) Exemplified Responsivity as a function of wavelength for various linear gains. c) Photodetector current response as a function of light power for various wavelengths. Indicated is the Noise Equivalent Power (NEP) for the case where the NEP is directly measured (290 nm), the case where NEP is equal to the lowest light intensity (110 nm), and where the LDR is interpolated down to the NEP (110 nm). d) Specific Detectivity ($D^*$) as a function of wavelength. Calculated from the NEP as in panel c (black-solid line), and from the responsivity as in panel b (blue-solid line). e) Modulated square-wave time-series analysis of photodetector response. Indicated is the photodetector rise time $\tau_r$. f) $D^*$ as a function of frequency. Calculated from the NSD and a 2nd order BWF (blue-solid line), the NSD and LPF shown in panel a (green-solid line), and a DC noise level and a LPF (black-dotted line).

**Common Architectures**

Having established the application, material and architecture agnostic Performance FOM, we now turn to the most common semiconductor-based photodetector architectures utilising photoconductive or photovoltaic mechanisms for light-to-electrical transduction.

*Photoconductors* utilise variations in conductivity due to photocarrier concentration in a lone semiconducting layer with dual ohmic contacts. Gain is defined by the ratio of the carrier lifetime to transit time. Bias simultaneously probes the conductivity and increases gain via the transit time. There are two basic photoconductor concepts, notably:

a. *Intrinsic photoconductors* utilise a single intrinsic semiconductor where photocarriers are generated via band-to-band transitions.
b. *Extrinsic photoconductors* exploit doping levels in an insulating semiconductor to affect photocarrier concentration, resulting in majority carrier photoconduction.

*Schottky Barrier Photodiodes* utilise a pairing of one ohmic and one Schottky contact resulting in a majority carrier depletion region near the metal-semiconductor interface. The Fermi-level becomes pinned to that of the semiconductor, resulting in a relatively small and uncontrollable majority carrier extraction barrier and, consequentially, high NSD. Two routes to generating photocarriers are possible:

a. *Band-to-band* transitions in the doped semiconductor create electron-hole pairs which can overcome the Schottky barrier under bias.
b. *Internal photoemission* occurs when optical transitions in the metal create majority carriers with enough energy to overcome the Schottky barrier and inject into the semiconductor.

*Metal-semiconductor-metal photodiodes* utilise two Schottky contacts. This has the effect of reducing the junction capacitance creating high-speed (>GHz) photodetectors at the expense of Responsivity and hence Specific Detectivity.

*Phototransistors* utilise a transistor architecture in which the switching is determined by photocarriers.

a. *The bi-polar junction transistor* geometry operates using a photoconductor as the base-collector junction modulating the on-off state of the transistor.
b. *The field effect transistor* geometry modulates the on-off state of the source-drain channel through a photovoltage applied at the gate terminal.

*p-n junction photodiodes* utilise band-to-band transitions in both *p*-type and *n*-type layers where all photocarriers can potentially be separated and extracted. Bias is often applied to increase the external quantum efficiency in a 'gain-like' mechanism.

*p-i-n photodiodes* consist of an intrinsic region between heavily dope *n*- and *p*-type regions. Under reverse bias the depletion region can extend throughout the intrinsic region. The photocarriers that contribute to photocurrent are created in the intrinsic layer and extracted through ohmic contacts. Defect generated carriers in the doped regions are the primary source of the NSD.

*Avalanche photodiodes (APDs)* are heavily doped *p*-*n* junctions designed to detect very weak optical signals. Large bias results in high electric fields and energetic electrons which, under the right conditions, can ionize the lattice resulting in a multiplicative effect across the so called 'avalanche region'. Additional unique FOMs can exist for APDs such as the ionization rates for electrons and holes.

The choice of architecture is driven primarily by the priority performance FOM for any particular application. But, in the context of emerging photodetector technologies, it is often also driven by material-level constraints such as (for example) the lack of availability of a *p*-type or *n*-type route to doping (c.f. p-doping in $\beta$-$Ga_2O_3$), non-ohmic contacts or high defect densities. This is the situation that currently exists in the field of UV-C photodetectors and to lesser or greater extents in other spectral windows such as the near IR.

**Acknowledgments**

P.M. acknowledges funding from the UK Research and Innovation project MANTISS 'Molecular Absorbers for Novel Transducing Imaging and Sensing Semiconductor Photodetectors' (UKRI4483).

# 3. Incumbent UV-C photodetection technology

**Tilman Weiss**[1]

[1] sglux GmbH, Berlin, Germany

E-mail: weiss@sglux.de

### Status

Current approaches to measure UV-C radiation employ photodiodes and vacuum discharge tubes. The photodiodes may consist of SiC [1], AlGaN [2], UV enhanced Si [3], or Si detector chips with conversion filters [4]. Other photodetector concepts exist employing various wide band-gap semiconductor materials such as $Ga_2O_3$ [5] [6], diamond [7] or titanium dioxide ($TiO_2$) [8]. These new wide-gap semiconductors will be discussed in the next chapters of this review, however, they have not been introduced to the market yet or did not work out ($TiO_2$). While analytical or academic niche applications also use PMTs, MCPs, and CCD/CMOS imaging systems for UV-C detection, Si and SiC-based photodetectors are the dominant technologies in terms of market size.

The use of the above presented detectors depends on the application. For fire detection at the presence of sunlight the discharge tubes show off their advantages due to their excellent rejection of sun's UV radiation (UV-A and UV-B). Thus, they are able to just detect the fire's UV-C radiation. All currently available photodiodes, with or without filters cannot be applied for fire detection due to their no-zero sensitivity in the UV-A and UV-B wavelength range. A disadvantage of the discharge tubes is their limited lifetime and fragility. Flame detection, e.g. combustion control (while sunlight is absent) is possible with all semiconductor-based detector concepts. Another important application is the control of UV-C sources used for water purification (germ inactivation by UV-C radiation) where the radiation to be measured is high ($1mW/cm^2$ or more). Accordingly, the best possible radiation hardness is required. SiC photodiodes are regarded as the most radiation hard commercially available solution [9] [10]. The excellent radiation hardness make SiC photodiodes an excellent choice as radiation detectors in medical applications or high-energy physics [11]. In addition they had been space qualified and used in measurement devices on Mars [12]. SiC photodiodes are also available with VUV (vacuum UV) sensitivity enabling the measurement of Xe-Excimer lamps and synchrotron radiation between 120 and 400 nm (Figure 1). For low signal applications SiC APDs are available.

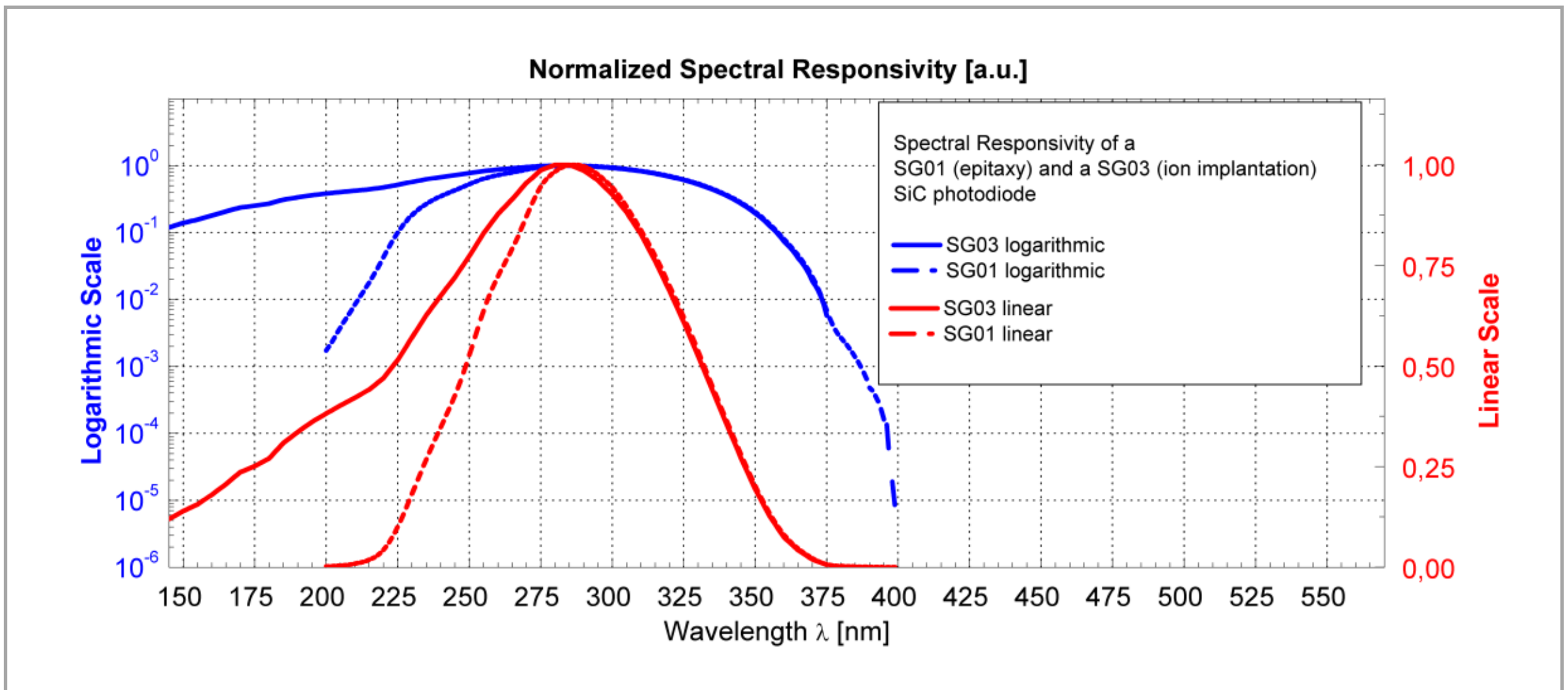

**Figure 1.** Spectral responsivity of the sglux standard SiC photodiodes (epitaxy) and the sglux VUV photodiodes (ion implantation).

### Current and future challenges

A challenge is detecting a flame or an arc in the presence of sunlight. Discharge tubes can be used here with certain restrictions, but they have the disadvantages mentioned above. A suitable photodiode with a sufficient solar rejection rate (eight orders of magnitude) would enable new applications in the area of fire detection. Such a photodiode is currently not available. Although SiC photodiodes have a rejection rate for visible light (above 400 nm) of more than ten orders of magnitude [9], the needed filters to reject the sun's UV-A and UV-B radiation do not offer the needed optical density. It is currently not possible producing optical filter that transmit UV-C while rejecting UV-A and UV-B with an optical density of more than eight orders of magnitude. The bandgap of AlGaN photodiodes can be modified through the adjustment of the Al-to-Ga ratio, thereby resulting in photodiodes that exhibit a specific spectral response within the UV-C wavelength range. However, the UV-C to UV-A, UV-B, VIS rejection rate of currently commercially available AlGaN photodiodes is 4-5 orders of magnitude only. SiC-based spectrometers would offer new possibilities in analytical chemistry. The development of a 512 pixel SiC line array was successful. The assembly of a hybrid SiC photodiode line array and Si-based readout integrated circuits could be demonstrated. The integration into a commercially available spectrometer resulted in a spectral resolution of 0.5 nm with a wavelength range of 230 and 400nm. However, a fully SiC-based integrated solution is still an open research topic. The challenge is the development of a SiC-based CMOS technology as known for Si-based image sensors.

In the field of VUV, SiC photodiodes need sapphire or $MgF_2$ windows to avoid degradation of the short wavelength sensitivity. To become a dependable replacement for UV-enhanced Si photodiodes as reference device for synchrotron radiation sources or in UV astronomy SiC-photodiodes have to be further improved.

### Advances in science and technology to meet challenges

First prototypes of a SiC-based 64-pixel UV-image sensor with integrated readout have been manufactured in the Netherlands at the TU Delft in collaboration with the Fraunhofer IISB Erlangen, Germany [13]. The building blocks of the 64-pixel image sensor contain three transistor active pixels which be addressed and read. Furthermore, a comparator and unity-gain buffer were implemented. The image-sensor has various integrated signal outputs: An analog and single digital output as well as a 2-bit analog-digital converter was integrated to allow the demonstration of a real digital SiC-based CMOS image sensor. The development of standard integrated building blocks for a SiC-based CMOS technology opens the way for SiC-based transimpedance amplifier circuit. Thus, manufacturing of complete on-wafer sensors allows much smaller UV detector design. In combination with an analog-digital converter a I2C bus design seems possible in the future.

## Concluding remarks

When it comes to measuring UV-C radiation, the use of SiC has become the standard over other approaches. The reason is its high radiation hardness, a property that photodiodes based on other materials do not possess. A true innovation would be the development of a UV photodiode capable of detecting only the UV-C component of a fire, insensitive to sunlight. This would enable entirely new products in the field of fire detection. However, there are currently no promising concepts.

# 4. Emerging materials for UV-C photodetection

## 4.1. Gallium oxide ($Ga_2O_3$)

**Damanpreet Kaur[1], Yuichi Oshima[2] and Fabien Massabuau[3*]**

[1] Department of Physics and Materials Science and Engineering, Jaypee Institute of Information Technology Noida, Uttar Pradesh, India
[2] National Institute for Materials Science, Tsukuba, Japan
[3] Department of Physics, SUPA, University of Strathclyde, Glasgow, United Kingdom

*E-mail: f.massabuau@strath.ac.uk

### Status

Although $Ga_2O_3$ has been investigated since the mid-20$^{th}$ century, its development truly accelerated around 2010 with the emergence of its potential in power electronic and UV-C photodetection. This highly polymorphic compound exists in five known phases, labelled α (rhombohedral, $R\bar{3}c$), β (monoclinic, $C2/m$), κ (orthorhombic, $Pna2_1$), γ (cubic, $Fd\bar{3}m$), and δ (cubic, $Ia\bar{3}$) [1]. Among these, β-$Ga_2O_3$ is the thermodynamically stable phase and can be produced via bulk or through homo- or heteroepitaxy [2]. The α-, κ-, and γ- phases are metastable but can be stabilised using heteroepitaxy, most commonly on sapphire [3]. Meanwhile the δ-phase is unstable.

$Ga_2O_3$ is inherently suited for UV-C detection due to its bandgap energy of approximately 4.3-5.3 eV, depending on the polymorph (including amorphous $Ga_2O_3$). This wide bandgap enables solar-blind spectral response with natural rejection of wavelengths above 280 nm, eliminating the need for additional filtering. This has encouraged significant research effort on UV photodetector devices, particularly using photoconductor, Schottky barrier and metal-semiconductor-metal photodiode architectures, which offer fabrication simplicity and promising responsivity [4].

One of the most striking feature of $Ga_2O_3$-based photodetectors is their exceptionally high responsivity. Values exceeding 1 A/W at 250 nm have long been reported – already higher than commercial UV-enhanced Si photodiodes – with some studies claiming responsivities reaching as high as to $10^5$-$10^7$ A/W [4]. Paradoxically, such high responsivities have been reported for high-quality bulk or low-quality heteroepitaxial films alike. Polyakov *et al.* analysed this paradox in their review linking these high gains to hole trapping, potentially from hole polaron or $V_{Ga}$-related deep acceptor mechanisms [5]. However, these same mechanisms also give rise to slow temporal responses, typically on the order of several seconds. [4].

$Ga_2O_3$ also distinguishes itself from other wide bandgap semiconductors through its polymorphism, which offers both opportunities and challenges for device engineering. While β-$Ga_2O_3$ remains the most studied phase, the metastable phases also exhibit attractive properties for photodetector applications. In particular, α-$Ga_2O_3$ offers enhanced scope for bandgap engineering to due to its isomorphism with other $M_2O_3$ (M: metal) [6], and κ-$Ga_2O_3$ exhibits spontaneous polarisation [7]. Proof-of-concept photodetectors have been demonstrated for all polymorphs, though their figures of merit are generally lower than those of β-$Ga_2O_3$ – largely due to the lower maturity of the material and less extensive research effort. Polymorphism has still to be fully exploited in a device context.

### Current and future challenges

Despite these attractive properties, the development of $Ga_2O_3$ for UV-C photodetector applications remains constrained by several material-level and device-level challenges (Figure 1).

*Defects understanding*. $Ga_2O_3$ is a relatively young material, and the mechanisms limiting device performance are not yet fully understood. Point defects can introduce deep trap states causing sub-bandgap absorption, degrade spectral selectivity and prolong response time. However, accurate attribution of these defects remains difficult. Heteroepitaxy additionally introduces high densities of extended defects – dislocations, stacking faults and grain boundaries – that remain poorly studied. Understanding the properties and impact of these extended defects will be essential to the development of future devices, particularly in metastable phases. It will be also critical to establish these properties for each polymorph, as studies are suggesting strong phase-dependence of material properties, including polaronic [8], optical [9], thermal [10] and radiation tolerance [11].

*p-doping*. Realising p-type doping in $Ga_2O_3$ constitutes a major material challenge due to the material's relatively flat valence band and the absence of suitable shallow acceptors [12]. Various doping (*e.g.* H, N, Mg, P, Zn) and co-doping strategies have been explored, but with limited success and significant complications from intrinsic donor compensation. The lack of effective p-type doping drastically restricts the range of possible device architectures, preventing the realisation of p-n, p-i-n or APDs.

*Cost*. The cost of $Ga_2O_3$ substrates and epitaxial wafers remains high due to the early developmental stage of the material. Prices vary with orientation, quality and wafer size, but 2-inch substrates and epiwafers typically exceed $1,400 and $2,000, respectively. Reducing material cost is essential for broader technological adoption. Heteroepitaxy on foreign substrates presents an attractive avenue for cost reduction, but the limited understanding of the defects generated in these processes currently hinders that route.

*Suitability for practical applications*. Despite the attractive high responsivities of $Ga_2O_3$-based photodetectors, their response time – typically several seconds [4] – are too slow for many practical applications. Faster detectors usually show reduced responsivity, revealing a difficult performance trade-off challenge for enhancing both parameters simultaneously. Additionally, other important performance metrics such as specific detectivity and noise equivalent power have received comparatively little attention but remain essential for real-world deployment.

*Low technology readiness level (TRL)*. Although numerous proof-of-concept devices have been reported, $Ga_2O_3$-based UV photodetectors remain generally at low TRL, with limited application demonstration or system-level integration. Early examples include line-of-sight and non-line-of-sight communications [13, 14], water quality monitoring [15], fire detection [16], gas sensing [17], or imaging arrays [18].

### Advances in science and technology to meet challenges

Progress across several areas is helping address the key challenges outlined above – schematised in Figure 1.

*Epitaxy*. Recent advances in epitaxy address several challenges simultaneously. Growth on large-area substrates reduces costs, while improved crystal quality enables advanced characterisation to resolve defect and understand their fundamental properties. Furthermore, progress in producing high-purity, low-defect-density films provides new opportunities to optimise device performance for target application, assuming the underlying mechanisms behind performance losses can be clearly identified.

*Defect engineering*. Advances in understanding and controlling defects have enabled targeted engineering of material properties. For example, a $V_O$ concentration gradient method has been shown to simultaneously improve photodetector responsivity and response time [19]. Defect control strategies also include tuning formation energies to reduce compensation [20] or explore co-doping approaches [21] aimed at enabling p-type conductivity.

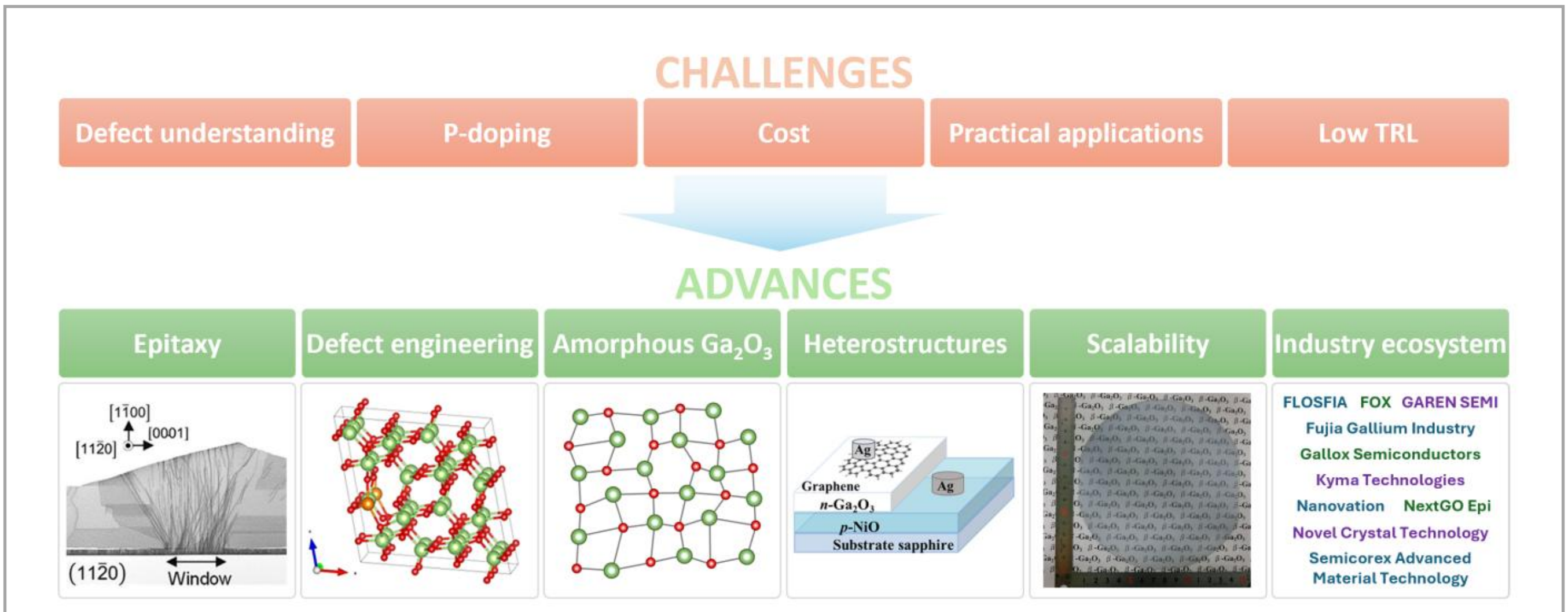

**Figure 1.** Schematic diagram of challenges and advances of $Ga_2O_3$-based UV-C photodetectors. (Epitaxy) Reproduced with permission from [26] copyright © 2025 National Institute for Materials Science, reprinted by permission of Informa UK Limited, trading as Taylor & Francis Group, www.tandfonline.com on behalf of National Institute for Materials Science. 2025 The Author(s). Published by National Institute for Materials Science in partnership with Taylor & Francis Group; (Defect engineering) Reprinted from [20]; (Heterostructures) Adapted with permission from [23]. Copyright 2020 American Chemical Society; (Scalability) Reprinted from [27].

*Amorphous $Ga_2O_3$*. In contrast to defect reduction strategies, several studies have shown that amorphous $Ga_2O_3$ can deliver attractive photodetector performance [22]. The absence of grain boundaries reduces electron scattering, and $V_O$-induced photoconductive gain can be high – albeit at the expense of slow response. Moreover, the low temperature processing of amorphous $Ga_2O_3$ and its compatibility with large area fabrication offer significant cost advantages.

*Heterostructures*. Improved material growth and understanding have enabled the development of more complex device architectures. Heterostructures are primarily used to circumvent the lack of p-type $Ga_2O_3$ by combining it with p-doped wide bandgap semiconductors (*e.g.* NiO, GaN, $Cr_2O_3$) [23], alloying with p-type compounds ($\alpha$-$Ir_2O_3$, $\alpha$-$Rh_2O_3$) [24], or engineering band alignment using other n-type material or $Ga_2O_3$ polymorphs [25]. These approaches enable faster response times and allow photovoltaic device operation.

*Scalability*. $Ga_2O_3$ wafers are commercially available, with most suppliers offering 2- or 4-inch products. Moving forward, Novel Crystal Technology began shipping of 6-inch bulk (001) $\beta$-$Ga_2O_3$ wafers in 2026 and aims to supply 8-inch wafers by 2035. A collaboration with Kyma Technologies has also been announced to commercialise 6-inch $\beta$-$Ga_2O_3$ epiwafers by 2027. For the metastable phases, FLOSFIA introduced 4-inch $\alpha$-$Ga_2O_3$ epiwafers in 2025. The move toward larger wafers and new industrial entrants will improve accessibility and accelerate technology adoption.

*Industry ecosystem*. We are now witnessing the emergence of a $Ga_2O_3$ industry ecosystem, with established companies (*e.g.* Novel Crystal Technology, Kyma Technologies, Semicorex Advanced Material Technology, FLOSFIA, Nanovation) alongside recent spin-outs (*e.g.* NextGO Epi, FOX, GAREN SEMI, Fujia Gallium Industry) now supplying bulk and epitaxial materials, as well as entrepreneurial initiatives in the device sector (e.g. FLOSFIA, Nanovation, Gallox Semiconductors). Together, these provide an ecosystem supporting future commercial development of $Ga_2O_3$-based photodetectors.

### Concluding remarks

$Ga_2O_3$ holds significant promise for UV-C photodetectors, yet progress is hampered by key material and device challenges, including limited understanding of killer defects, absence of effective p-type doping, high material cost, and slow response times. Recent advances in large-area wafer availability, improved epitaxy, defect engineering strategies, and device concepts based on heterostructure and amorphous $Ga_2O_3$ are helping mitigate these constraints and advance the technology.

At present, $Ga_2O_3$ UV-C photodetectors are best suited to applications where temporal response is not a primary requirement, for example in exoplanet spectroscopy, environmental monitoring, or gas sensing. While performance requirements ultimately depend on the target application, achieving faster operation without compromising responsivity would considerably expand the accessible markets. Continued improvement across materials, epitaxy, and device architecture will be essential to unlock the full technological potential of $Ga_2O_3$.

## Acknowledgements

The authors acknowledge support from the UK Space Agency Enabling Technologies Programme (Grant No. UKSAG23_0043_ETP4-052).

# 4.2. Aluminium gallium nitride (AlGaN)

**Robert W Martin[1*] and Eva Monroy[2]**

[1] Department of Physics, SUPA, University of Strathclyde, Glasgow G4 0NG, United Kingdom
[2] Université Grenoble Alpes, CEA, Grenoble INP, IRIG, PHELIQS, F-38000 Grenoble, France

*E-mail: r.w.martin@strath.ac.uk

### Status

The development of AlGaN UV-C photodetectors began in the late 1990s, driven by the need for solar-blind UV detection in aerospace, environmental monitoring, and industrial applications [1]. Interest in AlGaN for UV-C photodetection lies in its tunable direct bandgap ranging from 3.4 eV for GaN to over 6 eV for AlN, allowing excellent rejection of visible and near-UV light without the need for filters. AlGaN promises very low minority carrier densities and, by extension, dark currents (with $<10^{-7}$ A/cm$^2$ demonstrated up to 50 V [2]). Additionally, AlGaN can be grown on various substrates and integrated into diverse device architectures (for example, Schottky, metal-semiconductor-metal, p–i–n, APDs, field-effect phototransistors, focal plane arrays (FPAs)) [3,4]. Commercially available devices include Schottky photodiodes that cover the spectral range 210-280 nm with typical responsivities of 0.05-0.07 A/W [5]. APDs are available with peak gains of 300,000 for 220 nm illumination [2]. Figure 1 shows a typical device structure, with p-side uppermost to mitigate "carry- over" of the Mg dopant and an inclined mesa to offset breakdown.

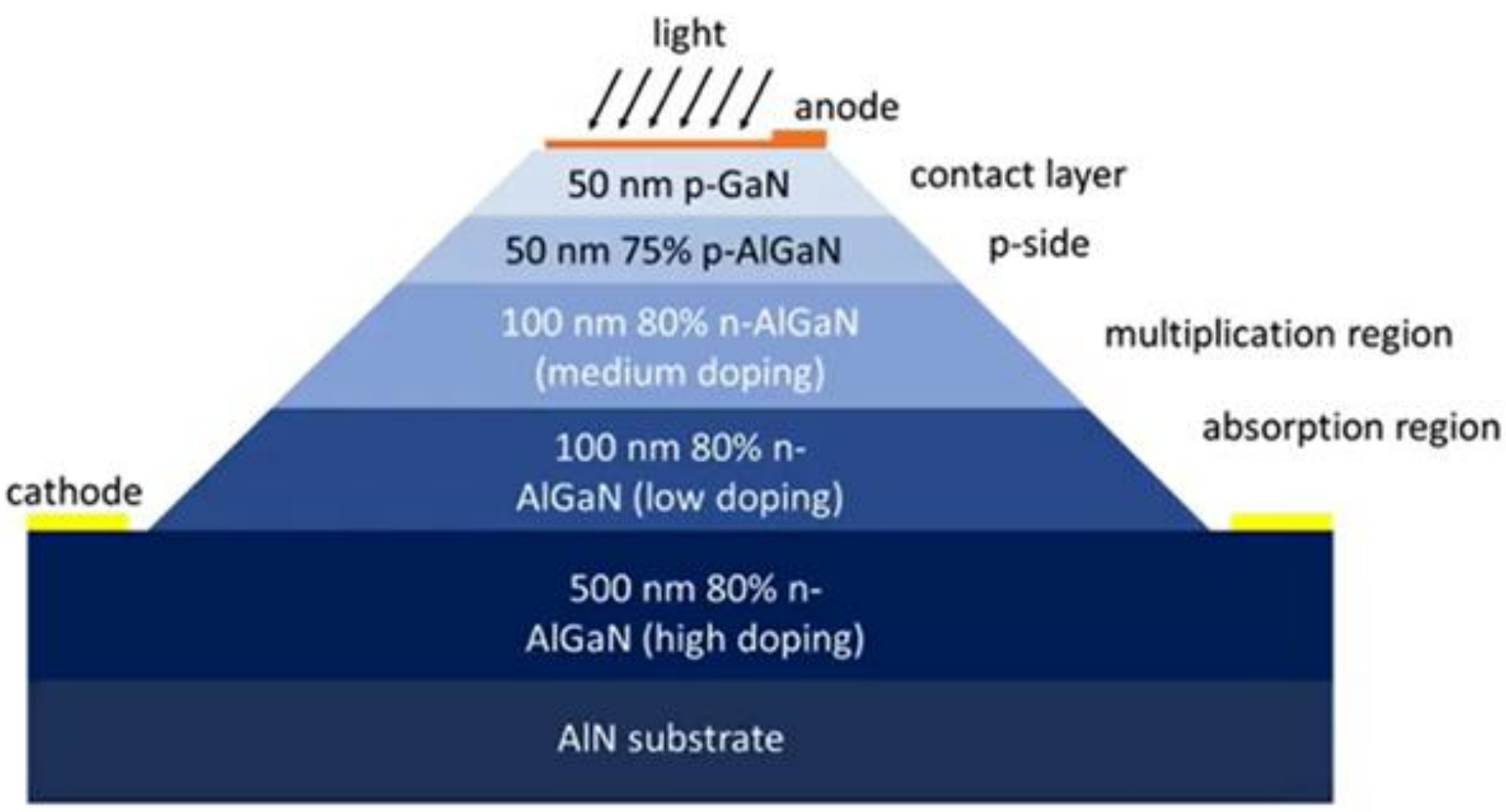

**Figure 1.** Schematic of an Adroit Materials UV-C photodetector structure, produced p-side up on an AlN substrate with sloping mesa sidewalls and designed for front-side illumination. From [6].

Most common substrate options include sapphire (UV transparent) and silicon (111) (low cost, integration capabilities), although the increasing availability of bulk AlN substrates is very attractive in order to reduce defects arising from lattice and thermal mismatch [2,7,8], potentially enabling ultra-high-performance devices.

Progress in the fabrication of AlGaN nanostructures and nanotechnologies has opened pathways to improved performance, notably through single or ensemble nanowire-based detectors [9,10] and nanoplasmonics to enhance absorption [3]. Nanowire geometries offer key advantages, including efficient strain relaxation, which enables the growth of high-AlN content AlGaN with improved crystal quality. In addition, nanowires can exhibit an "antenna effect," whereby the optical absorption cross-section exceeds the electrical collection area, effectively decoupling optical and electrical device dimensions and enhancing light–matter interaction without increasing capacitance. In addition, hybrid device concepts such as p-AlGaN/n-$Ga_2O_3$ heterostructures [11] offer complementary material properties, in this case with enhanced spectral selectivity.

Photodetectors based on AlGaN/GaN transistor architectures—particularly high-electron-mobility transistors (HEMTs)—are a highly promising approach for UV detection [12,13]. These devices exploit the 2-dimensional electron gas (2DEG) formed at the AlGaN/GaN interface, enabling high carrier mobility and efficient carrier transport. The internal gain mechanisms in phototransistors can dramatically enhance responsivity without relying on slow trapping processes, thereby preserving rapid temporal response. As a result, these devices can significantly improve the responsivity–speed product, effectively overcoming the intrinsic trade-off between sensitivity and speed that limits conventional photodetectors.

The robustness of AlGaN-devices is well known, with transistor operation above 1000°C and resistance to a wide range of chemicals, making them well suited for harsh environments. Resistance to radiation, such as protons, neutrons and higher energy electrons, is becoming increasingly relevant for applications in space and high energy density environments. AlGaN devices have a very high level of radiation hardness and this is expected to increase with the use of native AlN substrates due to the reduction in the number of pre-existing defects [6].

### Current and future challenges

The high-AlN contents required for UV-C detection lead to significant lattice mismatch with commonly-used sapphire substrates. This results in high dislocation densities and degraded device performance, including reduced responsivity and increased leakage current, noise, persistent photoconductivity and risk of premature breakdown in APD devices. Improved access to high-quality large-area AlN substrates will be significant in overcoming these challenges [6]. For example, APDs grown on single crystal AlN have shown two orders of magnitude improvement in "gain×efficiency" product [2]. Progress has also been made using epitaxial lateral overgrowth [14] and nanopatterned sapphire [4], bringing advantages of cost and scalability but they have not yet matched the quality achieved on AlN substrates. A very high UV transparency for the bulk AlN is needed to allow back illumination, which is preferred due to the significantly higher hole (than electron) ionization coefficient [15].

The growth and processing of high-quality, high-AlN layers remain costly and technically demanding, and more progress in achieving efficient and controllable p-type doping is needed. These constraints limit widespread commercial adoption, confining AlGaN UV-C photodetectors to niche applications, such as space missions or harsh industrial environments. Developing large-area or flexible AlGaN-based UV-C detectors for emerging applications (*e.g.* wearables or Internet of Things (IoT) sensors) will require advances in nanostructuring, epitaxial layer transfer, and integration onto flexible or unconventional substrates—areas that currently face significant material and processing challenges. Also, scaling AlGaN detectors into FPAs [4] for imaging applications remains difficult due to challenges in achieving uniform epitaxy and consistent doping across large areas. While AlGaN offers high breakdown fields suitable for APD operation, progress is needed to achieve reliable, low-noise UV-C APDs which are currently limited by breakdown at the mesa edge, defect-related issues and unstable gain mechanisms [2,3,8,9].

Controlling defects and dislocations in AlGaN, even when grown on bulk AlN, remains a challenge. Screw dislocations associated with hexagonal hillocks limit performance and device size. Compensating defects, such as oxygen vacancy complexes and carbon substitutions, need to be controlled. Surface defects can lead to leakage currents, low yield and unstable gain. As well as countering these with higher quality substrates there have been other fabrication approaches, including selective ion-implantation and passivation [8,13].

AlGaN photodetectors generally suffer from persistent photoconductivity, leading to long photocurrent decay times although heating and passivation have been shown to help [13,16]. This is less of an issue with devices operating under bias, such as APDs, with response times of 20 ns reported [6].

### Advances in science and technology to meet challenges

As with other UV-C devices, increased availability of large area AlN substrates will be a major advance. The leading method is sublimation growth [7] and further desirable advances include increasing growth rate, reducing degradation of crucible or furnace and greater control of thermal stresses and crystallinity. Other possibilities are hydride vapor phase epitaxy (HVPE) on AlN Seeds and flux growth. HVPE has potential for much higher growth rates but work is needed to manage parasitic reactions, strain and stoichiometry.

There is hope for improved p-doping. For example, it has been shown how the polarization fields in short-period AlGaN superlattices can be employed to greatly enhance this [18], and such structures are now used in AlGaN-based UV-C LEDs. Similar approaches are being used in AlGaN photodetectors [4] but can be extended, learning from the LED experience in order to increase hole density and mobility without the cost of doping-related defects and inhomogeneities.

Further use of advanced multilayers, such as separate absorption and multiplication (SAM) structures [16] will help improve performance. Tailoring the strength and direction of the inbuilt polarization fields and the organization of different compositions of AlGaN should allow enhanced doping, breakdown voltage and gain. Furthermore, the valence band offsets in the multilayer structures can be used to promote hole ionization in APDs.

There is promise in developing photonic integrated circuits incorporating AlGaN UV-C-detectors [19], including possibilities with UV-C LEDs doubling up as photodetectors [20]. Hybrid solutions integrating AlGaN UV-C detectors with other materials, such as growth on Si and integration with CMOS or flexible materials would be significant advances. Likewise, the use of nanoparticles to enhance light trapping and further work on nanowires or other AlGaN nanostructures. Developments with multilayer structures and substrate transparency to facilitate back-illuminated APDs will pave the way for easier integration and packaging layouts (e.g. for arrays) using flip-chip technology.

Single photon detection using APDs in Geiger mode has been shown to be achievable [4, 6, 10] and such devices offer advantages, included reduced size and operating requirements, for a growing range of applications such as UV-C optical communications and quantum technologies. Back illumination and separate absorption and multiplication structures are advantageous, but more work is needed to achieve sufficiently high and stable multiplication gain for devices operating with avalanche breakdown. Again, reducing the relevant dislocations and defects are challenges here, due to their links with carrier capture and scattering as well as leakage.

#### Concluding remarks

The compound semiconductor AlGaN has intrinsic suitability and advantages for UV-C photodetection, for example its tuneable bandgap crossing the full UV-C range, its resistance to radiation and chemical degradation and its successful use in a range of other optoelectronic devices. UV-C photodetectors with a range of types and architectures are already available and there is plenty of scope for improved performance and diversification of applications. Challenges remain in terms of achieving the sufficiently low dislocation and defect densities needed for higher performing detectors, but steady developments in the quality and size of AlN substrates allow for optimism here. Inherent characteristics of AlGaN-based structures, including the intense in-built fields and high breakdown fields, offer routes to enhanced devices or new applications such as improved p-doping or single photon detection. Looking ahead, increasing integration of AlGaN UV-C detectors into hybrid systems, such as photonic integrated circuits, imaging arrays and flexible devices, is expected to drive further innovation and broaden the application landscape.

## Acknowledgements

We are grateful to Dr. R. Kirste of Adroit Materials for input and advice.

# 4.3. Boron nitride (BN)

**Le Chen[1], Hongwei Liang[2] and Hong Yin[1*]**

[1] State Key Laboratory of High Pressure and Superhard Materials, College of Physics, Jilin University, Changchun 130012, China
[2] School of Integrated Circuits, Dalian University of Technology, Dalian 116024, China

*E-mail: hyin@jlu.edu.cn (Hong Yin)

### Status

Boron nitride (BN) possesses several allotropes including cubic (c-BN), hexagonal (h-BN), wurtzitic (w-BN), and rhombohedral (r-BN) phase, all of which exhibit ultra-wide bandgap (typically 6.0–6.4 eV) and other distinctive properties including high band-edge absorption coefficient ($7\times10^5$ cm$^{-1}$ [1], enabling ~99% absorption within just a few tens of nanometres of thickness), exceptional dielectric strength ($\kappa$~3-4 for h-BN [2] and 1-2 for amorphous phase [3]), chemical inertness, thermal stability and mechanical robustness. These attributes have enabled BN as a promising semiconducting material for UV-C photodetection [4], particularly attractive under extreme conditions such as continuous radiation, high temperatures, and corrosive environments.

To achieve device applications, large-area single crystals compatible with advanced microelectronic fabrication are highly desirable for BN. UV-C photodetection based on c-BN has been rarely reported due to the challenges in synthesizing high-quality c-BN films (Figure 1a, 1d) [5-6]. In contrast, significant advances on the synthesis of h-BN epitaxial films and bulk single crystals via various techniques, such as chemical vapor deposition (CVD) [7], metal–organic vapor phase epitaxy (MOVPE) [1, 8], ion beam sputtering deposition (IBSD) [9], submicron-spacing vapor deposition [10] and metal flux methods [11], have led to a bloom of device studies. The film quality has been significantly improved in terms of crystalline quality, crystal size (both lateral and thickness as Figure 1b-c) [9, 11], epitaxial orientation (Figure 1e-f) [7, 8], and material availability. However, high synthesis temperatures (1300–1700°C) [1, 7-8, 11], or catalytic metals [11], are typically required to compensate for the slow growth kinetics of BN, which often involves a complicated subsequent transfer process. Alternatively, direct deposition of h-BN on dielectric substrates often results in the formation of disordered turbostratic BN (t-BN) [8] due to insufficient catalytic activity. These issues would severely impact on the device performance based on h-BN films. Thus far, achieving controllable growth of high-quality BN has remained a key objective in the field.

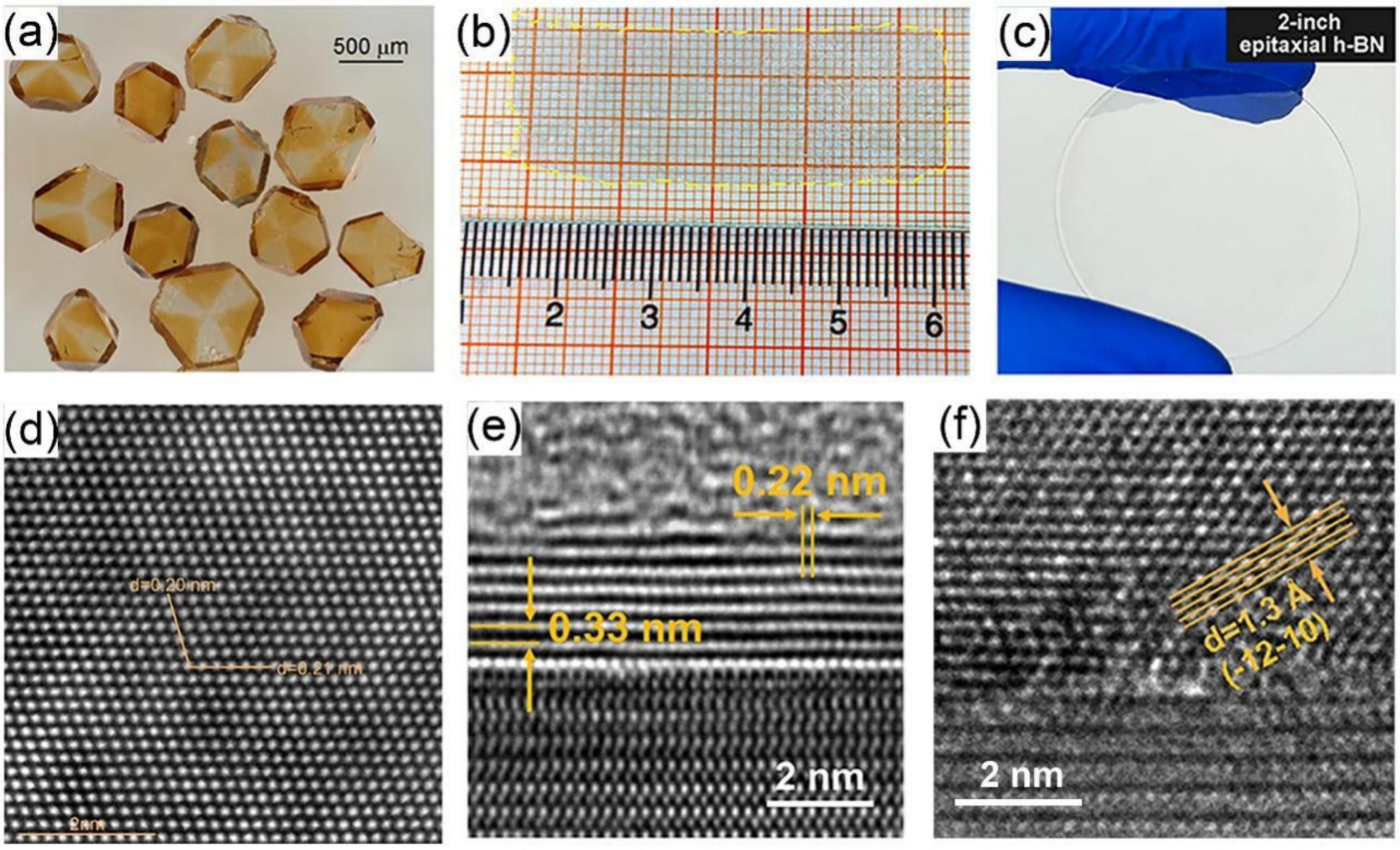

**Figure 1.** Thick BN films/bulk single crystals for UV photodetection. Optical image of the (a) c-BN single crystals (Reproduced with permission from [6], Copyright 2024, Springer Nature), (b) Centimetre-scale h-BN single-crystal (Reproduced with permission from [11], Copyright 2025, Elsevier) and (c) h-BN thick films deposited on sapphire substrates (Reproduced with permission from [9]. Copyright 2025, Wiley-VCH). High-resolution electron microscopy of (d) c-BN crystals (Reproduced with permission from [6], Copyright 2024, Springer Nature), (e) Van der Waals epitaxial h-BN with the c-plane orientation (Reproduced with permission from [10], Copyright 2025, Wiley-VCH) and (f) m-plane vertically epitaxial h-BN film (Reproduced with permission from [9], Copyright 2025, Wiley-VCH).

On the device front, BN-based detectors with various architectures including photoconductive [7-9] and photovoltaic junction devices [6, 11], have been extensively investigated (Figure 2a-b). While each configuration presents certain trade-offs, they collectively demonstrate favourable performance for BN in the short-wave region of the UV-C band, achieving high spectral selectivity, low noise, fast response, and operational stability in harsh environments, proving its significant potential for both routine and complex real-world detection tasks. In addition to their crystalline counterparts, both the composite paper/films made from cost-effective BN nanosheets [12-13] and low-temperature, rapid-growth amorphous BN [14] have also shown promise in detection capability, offering a comparable bandgap and easy availability. Recently, impressive progress on the demonstration of novel functionalities such as polarization-sensitive detection (Figure 2c-d) [9] and imaging (Figure 2e) with detector arrays [15], underscores the expanding utility of BN in this field.

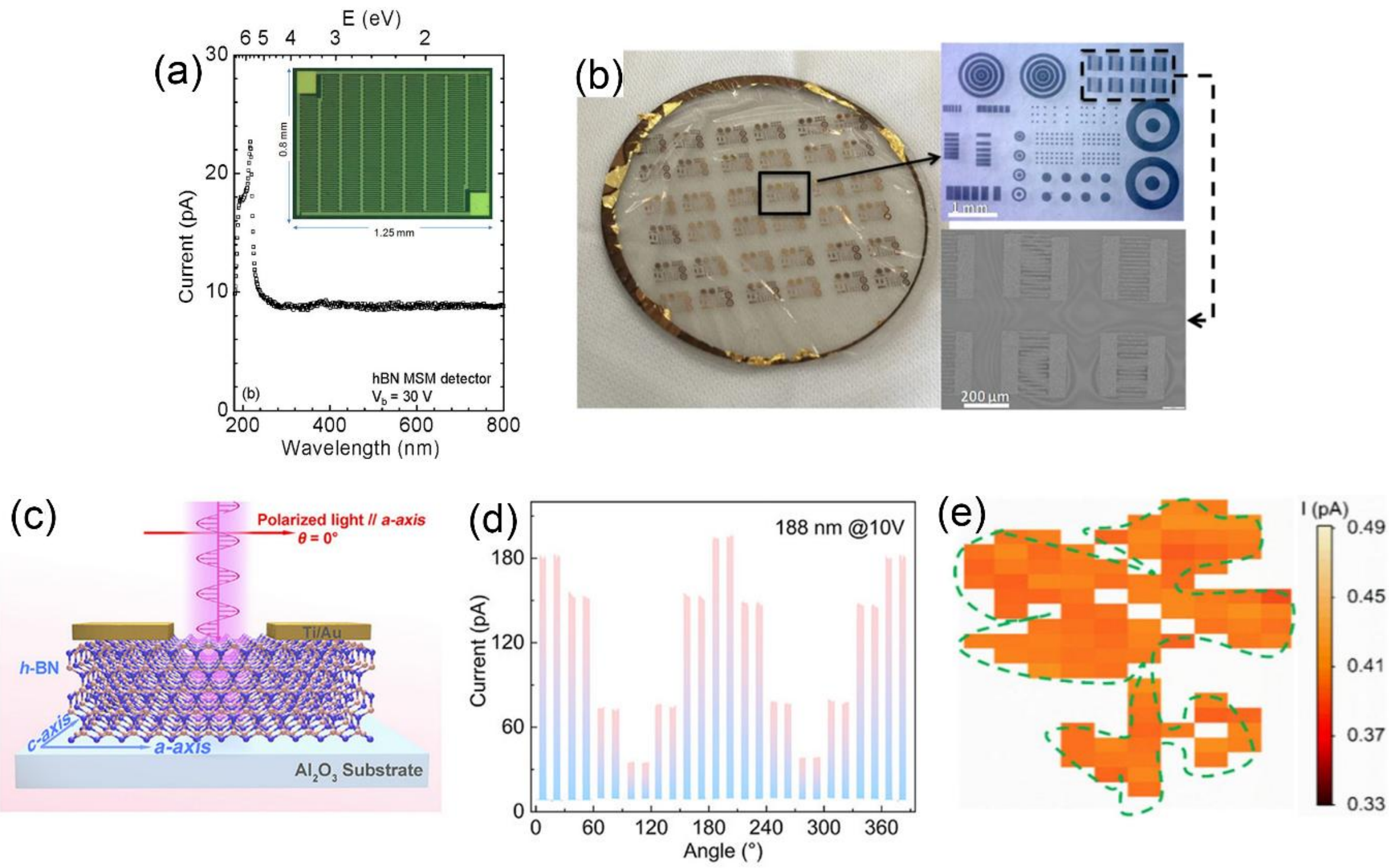

**Figure 2.** Photodetectors based on h-BN. (a) Spectral response of the h-BN planar photodetector. The inset is microscope image of the device. Reprinted from [1], with the permission of AIP Publishing. (b) Optical image of the photodetector fabricated on an h-BN wafer. The insets display microscope images of the devices. Reproduced from [8], CC-BY 4.0. (c) Schematic of the polarized sensitive photodetection device and (d) Polarization-resolved dynamic photoresponse at different polarized angle. Reproduced with permission from [9]. Copyright 2025, Wiley-VCH. (e) Imaging function of the photodetectors arrays for sensing a "rose" flower-pattern with UV-C light illumination. Reproduced with permission from [15]. Copyright 2025, Elsevier.

### Challenges in BN material/device for advanced photodetectors

Although significant progress has been made in BN-based UV-C photodetectors, critical challenges spanning material synthesis, device fabrication, and device performance must be systematically addressed to establish mature and scalable production routes from lab to industry.

*Material synthesis*. In terms of material preparation, the efficient and low-cost production of wafer-scale, uniform, low-defect, and high-crystallized h-BN films with appropriate thickness remains the primary bottleneck. Existing growth or assembly methods often involve trade-offs among material quality and fabrication complexity, which directly limits its availability and further applications.

*Device fabrication*. From the perspective of device fabrication, the compatibility of layered h-BN with conventional semiconductor manufacturing techniques requires further improvement. Etching processes, for instance, should be refined to mitigate the degradation of crystal qualities of h-BN [16]. Ohmic contacts with appropriate metals that match the work function of BN are crucial for efficient carrier injection and collection, which is particularly difficult due to its ultra-wide bandgap.

*Device performance*. In spite of excellent spectral selectivity, the performance of BN-based detectors is still lower than expected in terms of dark current (typically ranging from fA to a few nA), responsivity (several μA $W^{-1}$ to several mA $W^{-1}$), and response speed (advanced devices can achieve tens of microseconds, while most are on the millisecond or even second level) and is challenging their practical applications. For instance, photoconductive detectors usually have slow and noisy responses, and photovoltaic-type detectors simultaneously possess faster response times and a low noise but with a limitation of responsivity. These factors should be strategically balanced to achieve optimal overall performance.

Future research is expected to go beyond wavelength and intensity detection, exploring multidimensional optical information such as polarization state and phase, thereby enabling advanced capabilities including multi-mode UV communication and high-resolution imaging [9]. Addressing these challenges is essential to unlocking the full potential of BN for next-generation UV-C photodetectors with superior performance and extended functionality.

**Advances in science and technology to meet challenges**

The pursuit of stable, efficient, and practical devices is still facing challenges. The practical device applications of BN lie in significant advances on precisely controlled growth of large-area, single crystals that are compatible with established microelectronic fabrications. For instance, the stacking-controlled growth of r-BN thick crystals offers efficient nonlinear optical characteristics and facilitates compact integrated photonics applications [17]. The regulation of epitaxial orientation and growth mode of h-BN thick films by overcoming the most thermodynamically favorable van der Waals growth allows novel functionalities and specific detection applications [9]. Moreover, substrate engineering plays an important role in not only guiding the epitaxial growth [18], but also in serving as an effective barrier against impurity diffusion [19].

For device fabrications, the technical compatibility should be considered. During device integration, the silicon semiconductor production technologies need to be modified to meet the requirements of BN-based detectors. For instance, strategies that have successfully achieved ohmic contact in other wide-bandgap semiconductors can be leveraged. These include band engineering-designed metal stacks, the formation of metal nitride interlayers, and the introduction of heavily doped regions to facilitate tunneling, all aimed at reducing contact resistance.

For performance enhancement, the design of novel device architectures represents a key strategy to overcome trade-offs between key performance metrics. Forming heterostructures with other materials is an important way to regulate the electronic structure and properties of BN, thereby achieving more functionality. Moreover, construction of p–i–n BN junctions offer a promising route to spatially separate the carrier generation (intrinsic layer) and transport (depletion layer) regions, thereby simultaneously achieving high internal gain which corresponds to high responsivity, and fast response. To expand the application prospects, new functions for detecting higher-dimensional information such as polarization states can be implemented. Achieving low-symmetry surfaces through nucleation and growth regulation to overcome the conventional energetically favorable c-plane growth orientation [9], or employing patterning processes to introduce macroscopic low-symmetry crystal array structures [20], have been proven effective in enabling its effective polarization detection capability.

**Concluding remarks**

Owing to its exceptional intrinsic properties, BN has garnered considerable interest over the past decade for applications in UV-C photodetection. The field remains young and rapidly evolving, and as a result, researchers are increasingly exploring its full potential for achieving high sensitivity, new functionalities, highly robustness operations in diversely demanding scenarios. To meet the requirements of next-generation detection systems, which are trending toward higher integration, miniaturization, intelligence, and multifunctionality, synchronized advances are essential in several key areas, including developing cost-effective yet precise synthesis methods, innovating device architectures to extend the performance limits of detector metrics, introducing new functional capabilities, and improving the fabrication processes of both BN-based devices and detection systems. The participation of experts from the related fields, including materials science, solid-state physics, optical and electrical engineering, is expected to collaborate in bringing this technology to its full potential.

**Acknowledgements**

Financial support from the National Natural Science Foundation of China (Grant No. 12335011) is gratefully acknowledged. H.Y. also thanks to the financial support from the Key R&D Project of Scientific Development of Jilin Province of China (20240302099GX).

# 4.4. Diamond

**Keyun Gu and Meiyong Liao***

Research Center for Electronic and Optical Materials, National Institute for Materials Science, Tsukuba, Japan

E-mail: meiyong.liao@nims.go.jp

### Status

Diamond's exceptional intrinsic properties, such as a wide direct bandgap energy (~5.5 eV), outstanding radiation hardness, and remarkable chemical stability, render it a promising material platform for UV-C photodetection [1]. Researchers have long endeavored to realize diamond photodetectors that fulfill the "5S" criteria—high sensitivity (responsivity), rapid response speed, high spectral selectivity, elevated signal-to-noise ratio, and enduring stability. Early efforts had been made in using polycrystalline diamond for photodetectors with either metal-semiconductor-metal structures [2] or Schottky barrier photodiodes [3]. However, their performance was limited by non-carbon phases and grain boundaries. Since the 2000s, the development of single-crystal diamond (SCD) has enabled the development of high performance UV-C photodetectors that meet the 5S requirements [4] [5]. In 2005, the first boron-doped diamond Schottky barrier photodiodes were fabricated, demonstrating high thermal stability and distinct forward/reverse photoresponses [6]. Later, metal-semiconductor-metal photodetector based on SCD epilayer delivered a responsivity of 6 $A \cdot W^{-1}$ at 220 nm under 3 V bias with a distinct gain and a UV-C/visible light rejection ratio of $10^8$ [7]. The high responsivity and the photoconductive mechanism of the photodetector based on SCD were elucidated. It was demonstrated that boron doping concentration in a homoepitaxial layer on a nitrogen-doped substrate can modulate dark current, responsivity, and speed via positive persistent photoconductivity (PPPC) [8]. Recently, 3D architecture photodetectors that enhanced electrical contact conductivity and carrier collection were developed, which showed a responsivity of 9.94 $A \cdot W^{-1}$ at 5 V [9]. By using laser irradiation, all carbon-photodetectors and image sensors based on diamond were demonstrated, which showed a DUV responsivity of 21.8 $A \cdot W^{-1}$ and a specific detectivity (D*) of $1.39 \times 10^{12}$ Jones [10]. Synergistic interactions between H-/O- surface terminations and deep defects enabled ultrahigh gain ($>10^5$) at 5 V low voltage and tunable photoresponse performance [11]. Meanwhile, SCD wafer sizes have scaled from 9 mm in 2009 [12] to 0.5 inches by 2012 [13], 2 inches by 2014 [14], and 3.5 inches by 2017 [15], which would push forward the development of large-area UV-C imaging. In Figure 1, we briefly summarize the development of typical diamond photodetectors. The convergence of scalable high-quality SCD wafers and high-performance low-voltage photodetectors allow the development of integrated SCD UV-C arrays for ultimate applications.

### Current and future challenges

Recent years have witnessed rapid progress in the growth of high-quality SCD and device design of SCD photodetectors. Wafer sizes have expanded to 3.5 inches by 2017, yet high dislocation densities (~$10^7$ $cm^{-2}$) and residual stresses in these larger boules continue to limit electronic integration [15]. The growth of large SCD wafers with high crystal quality, low dislocation density, and controllable impurity doping remains a great challenge. Dopant incorporation varies with substrate temperature, and gas chemistry [16], leading to nonuniform dark currents and inconsistent photoconductive gains across a wafer. Realizing reliable n-type diamond is another major hurdle. Conventional n-type dopants (e.g., P, S) exhibit deep donor levels and low solubility in the diamond lattice, resulting in carrier concentrations that are too low for practical use and activation energies so high that free-electron populations at room temperature are negligible [17]. This prevents the fabrication of standard p–n junctions and forces reliance on metal-semiconductor-metal or Schottky barrier photodiode architectures. Advanced contact schemes, such as asymmetric Schottky/ohmic Ti/WC–WC configurations and laser-induced graphitic carbon electrodes have significantly enhanced device performance. However, their long-term stability under continuous, high-flux UV-C irradiation and repeated thermal cycling has yet to be validated. Interface traps at the diamond–metal interface, arising from dangling bonds, fabrication residues, or lattice mismatch need dedicated investigation [18]. Three-dimensional and vertical device architectures have enhanced carrier collection efficiency. However, their complex geometries introduce significant integration challenges, complicating lithographic patterning and wafer bonding. Ultrahigh gain at low bias has been achieved through surface states and defect engineering. However, these surface states are thermally and chemically unstable—water, oxygen, and ionic adsorption can alter their density and distribution, causing gain variability and degradation [19]. Overcoming these intertwined challenges, in cost-effective, large scale, and high quality SCD growth, precise doping, junction formation, durable contacts, interface state control, stable surface engineering, and scalable architectures, will be critical for translating diamond UV-C photodetectors from laboratory demonstrations into reliable, mass-manufacturable UV-C photodetection technologies.

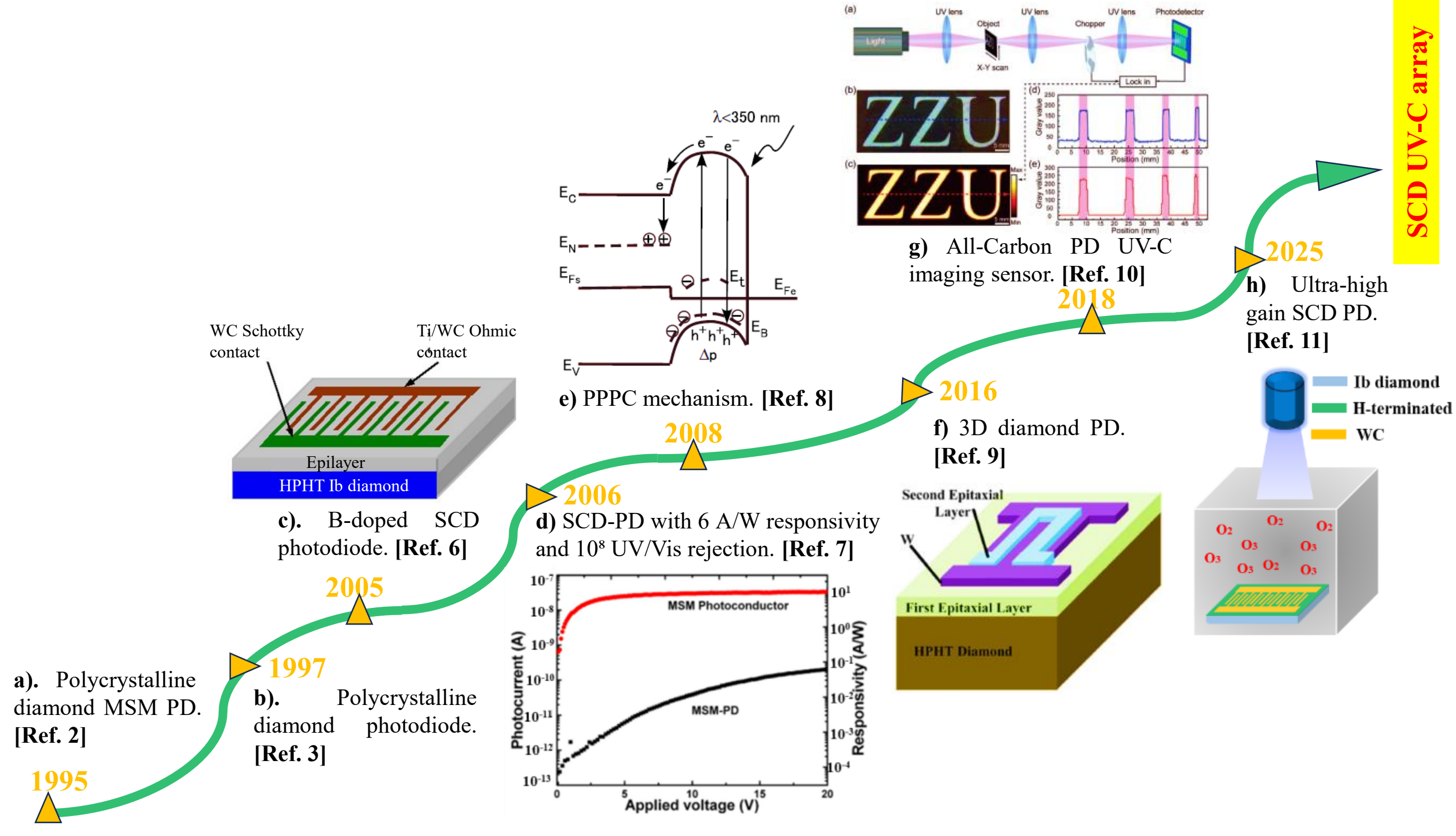

**Figure 1.** Timeline of the development of diamond UV-C photodetectors. (a)–(b) Early polycrystalline diamond-based metal-semiconductor-metal photodetectors and Schottky photodiodes. (c)–(e) Boron-doped SCD photodetectors with improved thermal stability and gain performance (f) Bottom-up fabricated 3D structured SCD photodetector. (g) All-carbon photodetector enabled by in situ laser-induced graphitic electrodes. (h) Type Ib diamond-based photodetector with H-/O-terminated surface engineering. (Reprinted with permission from Refs. 6, 7@AIP publishing, Ref.8@APS publishing, and Refs.9-11@Wiley publishing.)

### Advances in science and technology to meet challenges

The growth of SCD currently involves (i) homoepitaxial growth on SCD substrates, which are limited in size, (ii) the mosaic overgrowth approach, using high-quality SCD seed crystals in a microwave plasma chemical vapor deposition (MPCVD) chamber, and (iii) heteroepitaxy on Ir-buffered $Al_2O_3$ or Si. While homoepitaxial growth yields high crystal quality, increasing efforts have been devoted to developing large-area SCD wafers for scalable electronic applications. Impurity control has been vastly improved by optimizing gas-phase precursors, substrate temperature profiles, and reactor flow dynamics. To tackle the long-standing n-type doping challenge, co-doping strategies and high-pressure, high-temperature treatments are under investigation to increase dopant solubility and activate deep donors. Interface trap densities can be reduced through a multi-step process: initial $O_2$/Ar plasma cleaning removes organic residues and dangling bonds, followed by atomic-layer etching to smooth microscopic roughness. This graded interface not only minimizes trap states but also enhances metal contact stability. Stable surface states, essential for high gain and fast response, can be achieved by optimizing hydrogen plasma and ozone treatments to create the ideal termination, then preserved by an ultrathin passivation layer that shields against moisture, oxygen, and ionic contamination. Finally, selecting contact metals with work functions matched to the diamond's doping type, facilitates the formation of ohmic or Schottky contacts, optimizing barrier height or contact resistance to enhance device performance. Coupling these tailored electrodes with robust surface-passivation schemes not only stabilizes the metal–diamond interface against Fermi-level pinning and chemical degradation but also delivers long-term reliable UV-C photodetectors. Scalable MPCVD growth with high crystal quality, well controlled dopants and impurities, atomic layer deposition (ALD) for interface passivation, and the selection of metal electrodes with optimized work functions, the performance of diamond UV-C photodetectors would be further improved.

### Concluding remarks

In summary, diamond's intrinsic advantages, including its wide bandgap energy, radiation hardness, and chemical stability, make it a promising semiconductor for UV-C photodetection under solar-blind conditions. The progress in crystal growth, defect and impurity control, device design, and surface/interface engineering has facilitated the development of diamond UV-C detectors, with desirable performance demonstrated in laboratories to meet the 5S requirements. Diamond UV-C photodetectors that simultaneously have ultra-high gain over $10^5$, response speed lower than milliseconds, and solar-blind ratio over $10^6$ remains a future challenge. SCD wafers with low dislocation densities, well-controlled dopant levels, and stable surfaces/interfaces are still needed to further optimize device performance. A broader range of applications remains to be explored, particularly in imaging sensors, where large-area SCD wafers are in demand.

### Acknowledgements

We gratefully acknowledge support from JSPS KAKENHI (Grant Nos. 24H00287, 22K18957, and 15H03999) and the Advanced Research Infrastructure for Materials and Nanotechnology in Japan (ARIM) of MEXT (Grant No. JPMXP1224NM5055).

# 4.5. Magnesium zinc oxide (MgZnO)

**Yaonan Hou**[1,2]

[1] Electronic and electrical engineering, Swansea University, Bay Campus, SA1 8EN, Swansea, United Kingdom
[2] Centre for Integrative Semiconductor Materials (CISM), Swansea University, Bay Campus, SA1 8EN, Swansea, United Kingdom
E-mail: yaonan.hou@swansea.ac.uk

### Status

The growing interest in ZnO and its alloys arose because of the direct wide bandgap of 3.37 eV, [1] combined with a large exciton binding energy of 60 meV,[2] which inspired the development of low-threshold solid-state UV lasers working at room temperature.[3] Due to this, the enthusiasm for ZnO has been growing rapidly since the 1990s. Despite the lack of stable p-type dopant, which has become a persistent challenge for the fabrication of high-efficiency light-emitting devices, its alloys, such as aluminium-doped zinc oxide (AZO) and indium gallium zinc oxide (IGZO), have been commercialised for transparent electrodes and thin film transistors (TFTs) for display driving panels. Meanwhile, the UV-C photodetector fabricated from the ternary alloy MgZnO by bandgap engineering is also under fast development towards broad applications, as reviewed in this roadmap.

There are several advantages of MgZnO UV-C photodetectors compared to those fabricated from other materials. For the first, the bandgap of MgZnO can be tuned from 3.37 to 7.8 eV, [4] covering full UV spectral range. Secondly, the elements of the alloy are non-toxic and abundantly available on the Earth, indicating the potential of low-cost and environment-friendly UV-C devices. Thirdly, the energy band positions, including the conduction band minimum and the valence band maximum allows MgZnO UV-C photodetectors grown on Si substrate without stray light response from Si, [5] which enables the integration with CMOS compatible readout circuit.

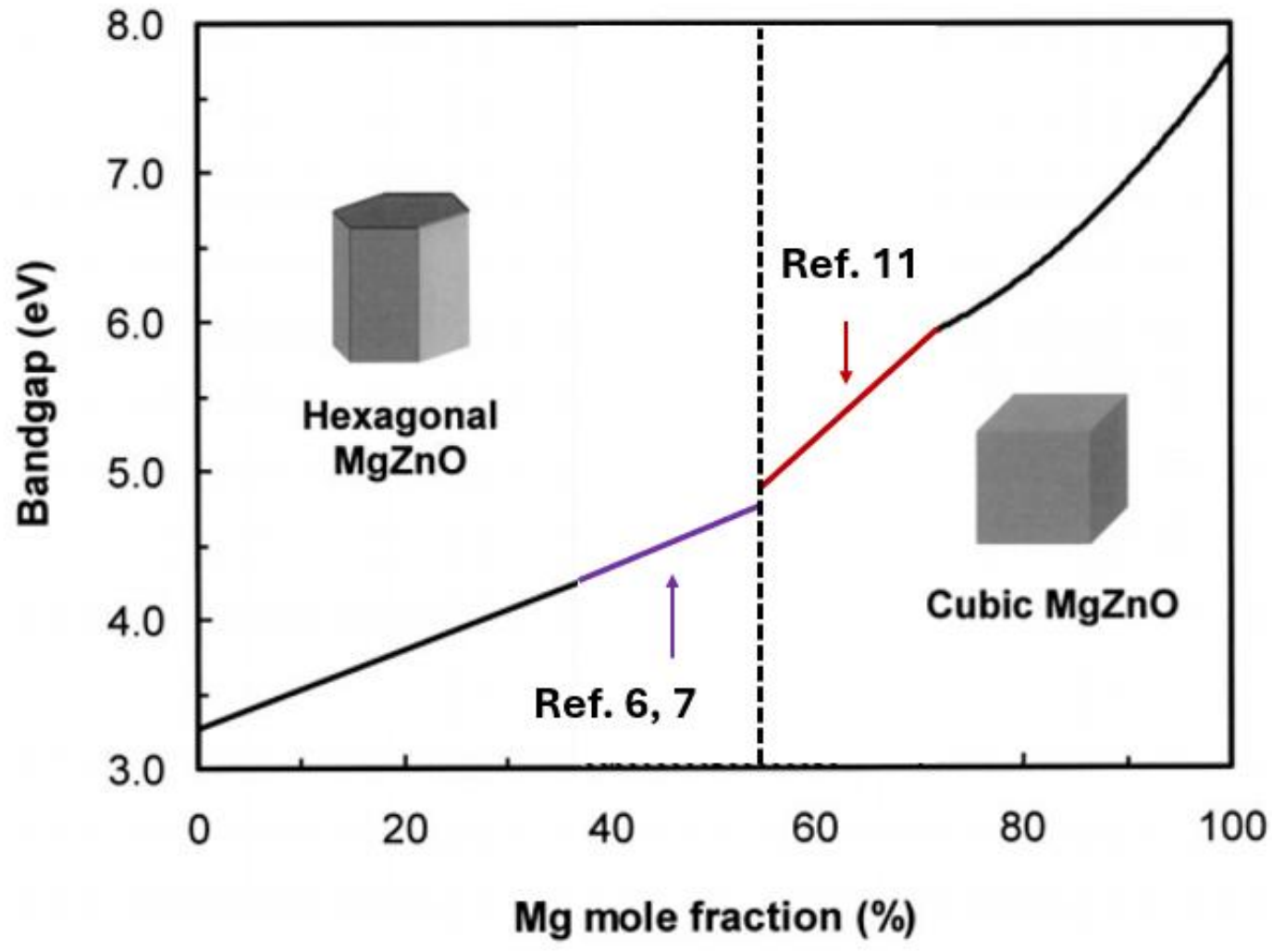

**Figure 1** Bandgap energy coverage by wurtzite and cubic MgZnO. The figure is adapted from [12] with the permission of AIP Publishing, Copyright 2025 AIP Publishing LLC.

One of the challenges lies in monocrystal MgZnO synthesis. Due to the crystal lattice difference between ZnO (hexagonal) and MgO (cubic), there is a limited solid solubility of MgO in ZnO (~4 mol%) in equilibrium conditions. Phase segregation tends to happen during the material growth with high-Mg content, hindering the bandgap from being extended to UV-C range (beyond 4.43 eV, corresponding to 280 nm) and practical applications in UV-C detection. Fortunately, this challenge has been overcome by turning component compositions using nonequilibrium epitaxial growth techniques such as molecular beam epitaxy (MBE), metalorganic chemical vapor deposition (MOCVD), pulsed laser deposition (PLD) and so on.[6-10] The best records demonstrate that the energy bandgap of single-phased wurtzite MgZnO has been achieved up to 4.55 eV whilst the cubic MgZnO achieved down to 4.77 eV with similar approaches (Figure 1b), [6, 7, 11, 12] nearly covering the full UV-C spectrum. Based on the bandgap engineering, many types of MgZnO UV-C detectors have been realised, such as solar-blind photodetectors on Si, dual-band or multi-band UV-C photodetectors and even avalanche UV-C photodetectors. [4,5,7,13, 14]

### Current and future challenges

*Epitaxy*. Although the single-phase wurtzite and cubic MgZnO epitaxial films are available now, the growth often requires extremely fine control of the growth conditions such as the flux ratio, temperature and pressure for obtaining such metastable materials. The stringent growth conditions also require a proper buffer layer due to the lack of large-area ZnO substrates for homoepitaxial growth. The state-of-the-art MgZnO films are grown on the lattice-mismatched c-plane sapphire (lattice mismatch of 18% with ZnO) or (111) Si (lattice mismatch of 16% with ZnO). For example, high-Mg content wurtzite MgZnO grown on Si or sapphire by MBE needs MgO/MgZnO buffer or toxic BeO buffer to avoid the strain-induced surface cracking or phase segregation. [14, 15] Even though, the thickness of most of UV-C MgZnO films is limited to a few hundred nanometers to sustain the strain from the lattice mismatch with the substrate, as well as the lattice difference between ZnO and MgO.

*Defects*. Native point defects ubiquitously exist in oxide semiconductors, including $Ga_2O_3$ and MgZnO. Such defects are commonly present as vacancies (e.g., $V_O$ and $V_{Zn}$), interstitials (e.g., $Zn_i$) or substitutionals (e.g., $Zn_O$ and $O_{Zn}$), and their charged states and pairs. [16] Among these, $V_O$ and its charged state (i.e., $V_o$ and $V_o$*) are the dominant deep-level defects, which are intrinsically n-type with lower formation entropy than others. Thus, they are stable at room temperature and can hardly be compensated with external dopants, which adversely affect the material and the device performances. One major challenge is that the slow photo-generated carrier recombination process through these defects will lead to persistent photocurrent, [4] substantially decreasing the response speed (or reducing the bandwidth). So far, the best results for MgZnO photodetectors are in a few hundred nanoseconds, [7] which is believed that $V_o$ density reduction due to Mg-O bonding in the epitaxial growth. The other challenge is the issue of stable p-type doping due to the presence of such stubborn n-type defects, which limits the design of MgZnO UV-C photodetector to planar metal-semiconductor-metal or heterojunction structures. Therefore, the device characteristics and applications are constrained compared with their counterparts, such as AlGaN.

*Chemical stability.* MgZnO inherits the ionic properties of ZnO (despite the covalency of the $sp^3$ hybridisation of Zn-O bonds) and MgO, indicating a low chemical stability, including the surface chemical status by absorption/desorption of oxygen and moisture from the air. This causes issues in stable performance and lifetime of the UV-C photodetectors (though the advantage is wet-etching compatibility).

### Advances in science and technology to meet challenges

Previous growth methods primarily focused on utilising planar sapphire or Si substrates, often introducing large strains due to the lattice mismatch and the dislocations propagating into the optical active layer. One feasible solution could be the growth on pre-patterned substrates, as was used in growing high-crystalline III-Vs. [17, 18] In such methods, the dislocations formed due to the lattice mismatch will be confined in the patterns instead of propagating up to the epitaxial layers. At the same time, the strain will be partially or fully released by the patterns with a size at the micrometre or nanometre scale. The buffer layer also plays a critical role in wide bandgap MgZnO growth. Most of the epitaxial growth utilised a single buffer layer (e.g., MgO and ZnO) or a double-layer buffer (BeO/MgZnO). [6, 13, 14] The adoption of multiple buffers, such as a MgO/MgZnO super lattice, would be an additional consideration to accommodate the strain and reduce the dislocations. Graded Mg content in the epitaxial layers can also be considered in the growth techniques to improve the MgZnO quality.

In order to reduce the point defects (e.g., $V_0$), growth environments need to be carefully controlled, normally requiring an O-rich condition. Besides, doping with reactive elements, such as Al and Ga, would have a strong bond with O, potentially reducing $V_0$ density. In addition, postgrowth annealing in an oxygen environment can also be considered as an effective approach.

Chemical instability is an inherent property of MgZnO. Surface passivation with dielectric coatings (e.g., $Al_2O_3$ or $SiO_2$) is a straightforward and effective way to enhance the device performance. But it is worth noting that the deposition method (e.g., precursors used in ALD and PECVD) should not disrupt the original properties of MgZnO.

## Concluding remarks

Compared to commercialised ZnO-based alloys such as AZO and IGZO, MgZnO remains under development as a promising candidate for UV-C photodetectors, offering advantages such as a tunable operating wavelength and compatibility with Si technology. Although the material's potential of MgZnO has not been fully unlocked, a positive progress is that the recent research has shifted toward low-cost fabrication techniques, such as PLD and CVD. [19, 20] With ongoing advancements in growth and fabrication methods, high-performance MgZnO UV-C photodetectors are envisaged to become commercially viable in the future.

## Acknowledgements

The author acknowledges the Royal Society for supporting oxide semiconductors under IEC\NSFC\242145. This publication is also partly supported by the Swansea University through SACEME seed corn funding.

# 4.6. 2D materials

**Fa Cao*[1,2] and Xiaosheng Fang*[1]**

[1] College of SmartMaterials and Future Energy, State Key Laboratory of Molecular Engineering of Polymers, Fudan University, Shanghai, PR China.
[2] State Key Laboratory of Flexible Electronics (LoFE) & Institute of Advanced Materials (IAM), School of Materials Science and Engineering, Nanjing University of Posts and Telecommunication (NJUPT), Nanjing 210023, P. R. China

E-mail: iamfcao@njupt.edu.cn (F. Cao) xshfang@fudan.edu.cn (X. S. Fang)

### Status

2D ultrawide bandgap semiconductors have gained increasing interest in UV-C photodetectors due to their unique mechanical (flexibility) and physical properties (high optical transparency, tunable electrical conductivity, high carrier mobility, ultra-high gate dielectrics)[1]. Unlike the extensively studied 2D narrow bandgap semiconductor materials in visible and IR photodetecting areas[2], 2D ultrawide bandgap semiconductors based UV-C photodetectors are less investigated and there is plenty of room to be explored. Currently, some of the most actively researched 2D ultrawide bandgap semiconductors materials for UVC photodetectors include h-BN (Eg = 5.7 eV)[3], β-$Ga_2O_3$ (Eg = 4.8 eV)[4,5], and emerging niobates (Eg > 3.8 eV)[6-8], as shown in Figure 1. For example, h-BN PD with ultrahigh rejection ratio ($R_{220}/R_{290}$=13990) was reported [3]. Quasi-2D β-$Ga_2O_3$ (thickness ~ 75 nm) was fabricated by thermal oxidation of a GaSe nanosheet (few-layer GaSe nanoflakes obtained by mechanical exfoliation and dry transfer onto $SiO_2$/Si substrates were oxidized at 850 $^oC$ for 2 h in an $O_2$/Ar (90%/10%) atmosphere to successfully produce β-$Ga_2O_3$ nanoflakes.), realizing a responsivity of 25 mA/W [4]. Although the above mentioned 2D UV-C photodetectors have achieved significant progress, we note that the photocurrent is typically at the pA level, which requires high electrical signal detection accuracy.

Recently, a 2D $Ca_2Nb_3O_{10}$ UV-C photodetector realized a high responsivity of 14.9 A/W and a detectivity of $8.7 \times 10^{13}$ Jones (at 3 V bias and 280 nm illumination wavelength)[7]. To further improve the photodetecting properties of the materials, the 2D nanosheet was doped with Ta, forming a 2D $Ca_2Nb_{3-x}Ta_xO_{10}$ perovskite which achieved a peak responsivity of 469.5 A/W – 7000 times greater than that of $Ca_2Nb_3O_{10}$ at 1 V bias [8]. Besides, the 2D $Ca_2Nb_3O_{10}$ nanosheet can be easily attached to paper substrate to fabricate flexible UV-C photodetector [9].

While these investigations provide a dramatic starting point for new types of 2D UV-C photodetectors, there is plenty of room for exploration. For instance: (1) basic research focusing on the physical and photoelectric properties of the niobates; (2) understanding the large gap between theoretical and experimental performance; and (3) solving incompatibilities with current integrated technology. UV-C photodetector based on ultrawide bandgap semiconductors is a show strong potential, however it is still unclear which material will excel or whether new ones need developing.

### Current and future challenges

*Low cost, large scale, and uniform synthesis*. The development 2D UV-C photodetectors currently poses a formidable challenge for both materials and application (Figure 2). For materials, first, we may consider low cost, new fabrication methods, since the previous reported methods (e.g. chemical vapor deposition, ion intercalation [10]) are time-consuming or complex. Besides, a large scale and uniform fabrication method is needed to enable integration of UV-C photodetector chips. Commonly used methods, such as CVD, liquid phase stripping, and chemical reduction, find it difficult to meet uniformity requirements of number of layers, size and substrate coverage of 2D materials.

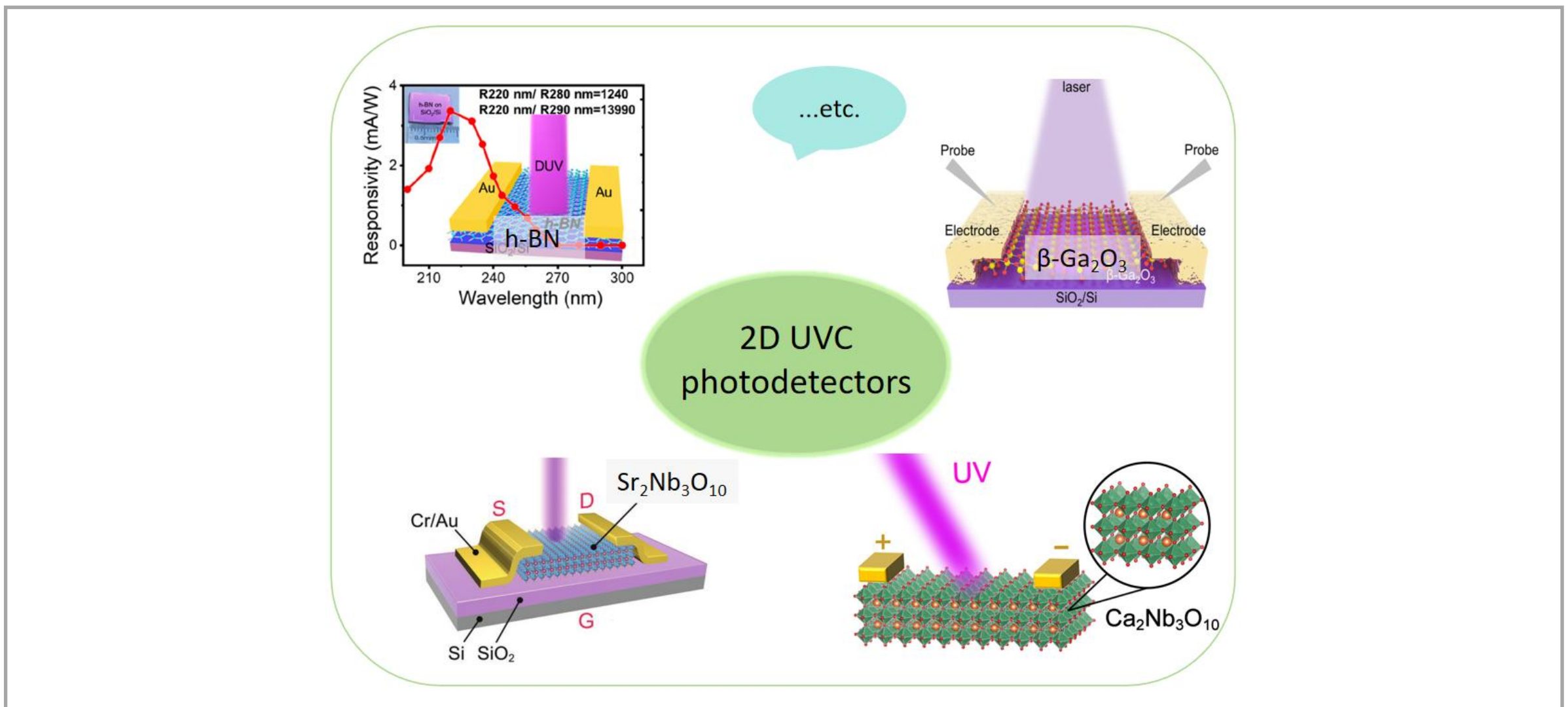

**Figure 1.** UVC photodetectors based on various 2D semiconductors, reprinted with permission from. [3] Copyright 2020 American Chemical Society. [4]Copyright 2025 The Royal Society of Chemistry. [6] Copyright 2020 WILEY-VCH. [7] Copyright 2020 American Chemical Society.

*Defect control at surfaces and interfaces*. When a low-cost and large-scale synthesis method of 2D ultrawide bandgap semiconductors is realized, we should consider surface and interface challenges, including vacancy defects, edge defects, inter-layer defects, strain defects, and oxidation/hydrogenation defects. Defects in 2D ultrawide bandgap semiconductors can a significant impact on the electronic and optical properties of the material–affecting electronic structure, optical absorption, luminescence and other properties of the material. For example, vacancy defects or boundary defects usually cause local changes in the density of electronic states, which may lead to electronic state localization or increase in carrier scattering, thus affecting the conductivity or semiconducting properties. Vacancy defects may lead to change in local electronic states, affecting optical absorption.

*Integration and intelligent applications*. For application, in addition to the highly integrated requirements mentioned above, we should also take into account the future requirements for multifunctional devices and the development of artificial intelligence (intelligent adaptive UV-C photodetector chip which can meet our demands for sensing and memory or even calculation in sensor).

### Advances in science and technology to meet challenges

To address the significant challenges of providing low-cost, high figures of merit, and intelligent UV-C chip, we think the following approaches can be referred to.

*Heterostructure optimization*. By designing appropriate heterostructures, such as constructing 2D heterostructures with graphene, we can optimize the performance of 2D ultrawide bandgap semiconductors UV-C photodetector and improve their photoelectric conversion efficiency and response speed.

*Defect engineering*. Reducing defects through surface treatment (such as quantum dot modification, plasma treatment, surface functional group modification etc.), doping and other methods, should improve the quality of 2D materials, and subsequently the stability and reliability of the photodetector.

*Integrated design*. To meet the needs of practical applications, future UV-C photodetectors need to develop in the direction of integration. The newly proposed charge-assisted oriented assembly film-formation process may play a guiding role in fabrication large scale 2D ultrawide bandgap semiconductors materials for integration [11]. 2D ultrawide bandgap semiconductors materials can be integrated with other functional modules to form a multifunctional or intelligent UV-C sensing system. Furthermore, we can also develop in-sensor memory and computing integrated devices, that is, the preliminary processing of UV-C signal at the sensing end, will be a future trend [12].

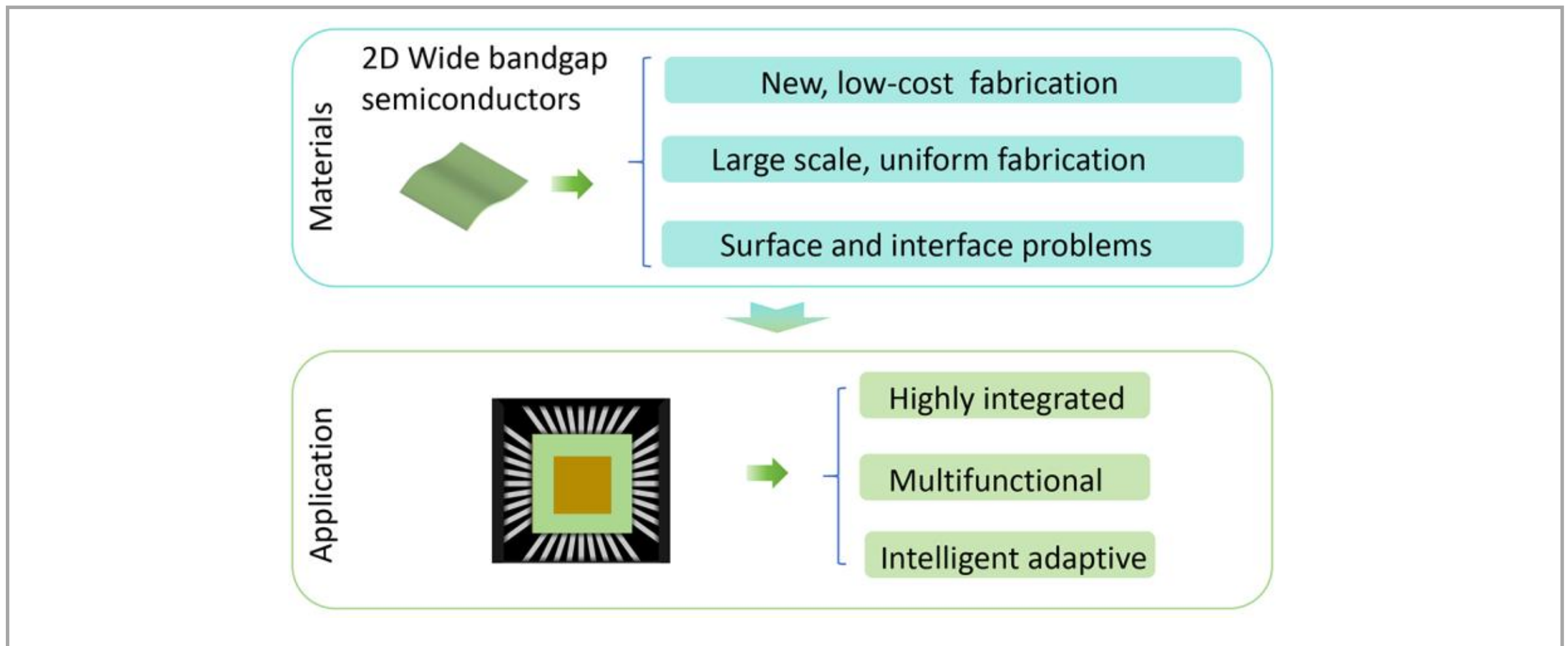

**Figure 2.** The future development trends of 2D wide bandgap semiconductors and its application in UV-C photodetection.

## Concluding remarks

2D ultrawide bandgap semiconductors have emerged as promising and attractive material platforms due to their high electron mobility and superior optoelectronic properties, enabling UV-C photodetector with high responsivity with fast response speed. Improvements in the fabrication of large-scale materials, defect passivation, and modification techniques of 2D ultrawide bandgap semiconductors will progress their integration into application. Further improvements in material properties and device design will enable us to fulfill future demands for multifunctional integration and artificial intelligence.

## Acknowledgements

This work is supported by the National Natural Science Foundation of China (No. 62374035, 52425308, 52402191, and 92263106), the National Natural Science Foundation of Jiangsu Province (No. BK20240633).

# 4.7. Metal Halide Perovskites

**Ruiheng Li[1], Guoqiang Peng[1] and Zhiwen Jin[1*]**

[1]School of Physical Science and Technology, Lanzhou Center for Theoretical Physics, Key Laboratory of Theoretical Physics of Gansu Province, Key Laboratory of Quantum Theory and Applications of MoE, and Gansu Provincial Research Center for Basic Disciplines of Quantum Physics, Lanzhou University, Lanzhou, Gansu 730000, China

E-mail: jinzw@lzu.edu.cn

### Status

Metal halide perovskites (MHPs), with the general formula $ABX_3$ (A = $Cs^+$, $MA^+$, $FA^+$; B = $Pb^{2+}$, $Sn^{2+}$; X = $Cl^-$, $Br^-$, $I^-$), combine tunable bandgaps, high carrier mobility, long carrier diffusion lengths, low exciton binding energies, and broad optical absorption, making them promising materials for next-generation optoelectronic devices.[1,2] Figure 1a illustrates the crystal structure and key advantages of $ABX_3$ perovskite materials.

However, their stability under strong UV-C irradiation remains a critical bottleneck, motivating the development of lower-dimensional variants. Lower-dimensional MHPs—such as quasi-2D Ruddlesden–Popper phases or zero-dimensional derivatives—offer enhanced operational stability by incorporating bulky organic spacer cations (e.g., $PEA^+$, $BA^+$) that suppress halide ion migration and interfacial degradation.[3,4] This structural confinement improves photo- and environmental durability but typically reduces in-plane carrier mobility due to quantum-well effects, highlighting a key trade-off between stability and transport efficiency in UV-C photodetectors.

Among MHPs, chloride-based compositions (e.g., $CsPbCl_3$) are particularly suited for UV-C photodetection because their wide bandgaps enable solar-blind operation and selective UV-C absorption. Building on this intrinsic advantage, recent advances in $CsPbCl_3$ nanocrystals have yielded high-performance devices with ~90% visible transparency, responsivity of ~1.9 A/W, and response times <50 ms.[5] These detectors illustrate the promise of chloride perovskites for next-generation wearable or disposable UV-C sensors.

Beyond nanocrystals, scalable thin-film deposition strategies—such as magnetron sputtering—have demonstrated stable pulsed UV detection with microsecond-scale responses. For instance, Bruzzi et al. reported $CsPbCl_3$ thin-film photodetectors operating reliably across a wide range of pulse frequencies (0.1–90 Hz), illumination intensities (250–500 $W/m^2$), and biases (10–30 V).[6]

By combining chloride composition, tunable dimensionality, and flexible low-temperature processing, MHPs open a pathway toward compact, cost-effective, and scalable UV-C photodetectors—poised to revolutionize public health, environmental diagnostics, and advanced sensing platforms.

### Current and future challenges

MHP UV-C photodetectors have demonstrated impressive responsivity and flexible integration potential, but overcoming several persistent challenges is essential to achieve long-term operational stability and practical deployment.

*UV-C-induced photochemical degradation*. UV-C photons can break chemical bonds within the perovskite lattice, particularly those involving halides and organic cations. In chloride-rich perovskites, continuous irradiation often leads to rapid halide segregation and phase transformation, causing crystalline degradation and performance loss within minutes to hours.[7]

*Ion migration, interfacial deterioration, and detection sensitivity limits.* The soft ionic nature of perovskites enables the migration of halide ions and A-site cations under illumination or applied bias, leading to electrode diffusion, interface reactions, and signal hysteresis.[8] UV-C exposure further accelerates ionic movement, amplifying interfacial degradation and long-term performance decay. In parallel, non-radiative recombination at grain boundaries and interfacial energy misalignment induce elevated dark current and increase the noise floor, significantly reducing detectivity—particularly under low-flux or pulsed UV-C conditions.[8] Together, these coupled mechanisms hinder reliable signal acquisition and pose critical barriers to achieving stable, high-sensitivity detection.

*Environmental stress coupling and toxic material concerns*. As illustrated in Figure 1b-c (schematic representations of perovskite crystal degradation under the effect of humidity (b), and UV light and oxygen exposure (c)), synergistic exposure to UV-C, moisture, and oxygen promotes degradation through photo-oxidation and hydrate formation. To mitigate such degradation, encapsulation strategies are often employed; however, conventional polymer encapsulants tend to yellow or crack under prolonged UV-C irradiation and humidity, reducing transparency and allowing further ingress of reactive species.[9] Meanwhile, lead-based perovskites like $CsPbCl_3$ offer outstanding UV-C sensitivity but pose environmental and regulatory concerns due to lead toxicity. Current lead-free alternatives still fall short: Sn-based compositions suffer rapid oxidation, while Bi- and Sb-based materials exhibit poor charge transport, limiting their practical viability.[10]

*Solubility limitations and challenges in large-area uniform deposition.* The low solubility of halide-rich perovskite precursors in commonly used solvents—often toxic, high-boiling-point solvents such as DMF and DMSO—hampers the formation of uniform, pinhole-free films over large areas, a problem particularly pronounced for wide-bandgap compositions suitable for UV-C detection.[11] Traditional spin-coating methods, while effective for small substrates, do not scale well, leading to inhomogeneous thickness, significant precursor waste, and morphological inconsistencies.[2] Scalable deposition techniques—including blade coating, slot-die, spray, and inkjet printing—require precise control over crystallization kinetics, solvent formulation, substrate wettability, and drying dynamics to achieve uniform coverage, reproducibility, and high device yield.

Resolving these challenges is critical for translating MHP-based UV-C photodetectors from lab prototypes into reliable tools for health, safety, and environmental applications.

### Advances in science and technology to meet challenges

Recent scientific breakthroughs and engineering innovations are laying the groundwork to overcome current obstacles in UV-C perovskite photodetectors.

*Dimensional and compositional engineering*. Transitioning from 3D to 2D or quasi-2D Ruddlesden–Popper perovskites has markedly improved stability under UV exposure. Embedding bulky organic spacer cations (e.g. $PEA^+$, $BA^+$) forms layered structures that confine charge carriers and reduce halide ion migration, enhancing resilience without compromising responsivity (Figure 1d).[3,12] Additionally, partial alloying of $Pb^{2+}$ with $Sn^{2+}$ or $Bi^{3+}$, and adjusting $Cl^-/Br^-$ ratios, enables fine tuning of the bandgap and defect states while reducing brittleness and toxicity.[13]

*Surface passivation and interface design.* Surface treatments such as self-assembled monolayers, fullerene derivatives, or small-molecule ligands (e.g. thiols, ammonium salts) effectively neutralize trap states at grain boundaries and interfaces. These strategies reduce non-radiative recombination and suppress dark current, significantly boosting detectivity and operational lifespan.[14] In parallel, ALD oxide coatings—such as ultra-thin $Al_2O_3$ or $TiO_2$—have emerged as conformal, pinhole-free barriers. These layers protect against moisture and ion migration while allowing UV-C transparency and charge transport.[15]

*Electrode and encapsulation innovations.* Replacing reactive noble-metal electrodes with carbon-based contacts or conductive oxides (e.g., ITO, AZO) enhances adhesion, moisture resistance, and thermal compatibility with flexible substrates—crucial for the long-term operation of UV-C photodetectors.[16]

Coupled with hybrid encapsulation, featuring organic polymer stacks reinforced with atomic-layer–deposited inorganics (e.g. ALD $Al_2O_3$), these innovations help prevent moisture and ion ingress while preserving device transparency and UV-C charge transport—significantly extending photodetector lifetimes under harsh conditions.[15]

*Scalable low-temperature fabrication*. Scalable deposition techniques such as blade coating, inkjet printing, slot-die coating, and vapor-assisted conversion are being tuned to produce uniform, large-area perovskite films with controlled thickness and grain orientation at temperatures below 150°C (Figure 1e).[17,18] Such methods are proving critical for manufacturing flexible and roll-to-roll platforms at industrial scale without compromising performance.

Together, these innovations mark a significant step toward UV-C perovskite detectors that balance high sensitivity, structural durability, and large-scale manufacturability—paving the way for their integration into public health, environmental monitoring, and compact sensing applications.

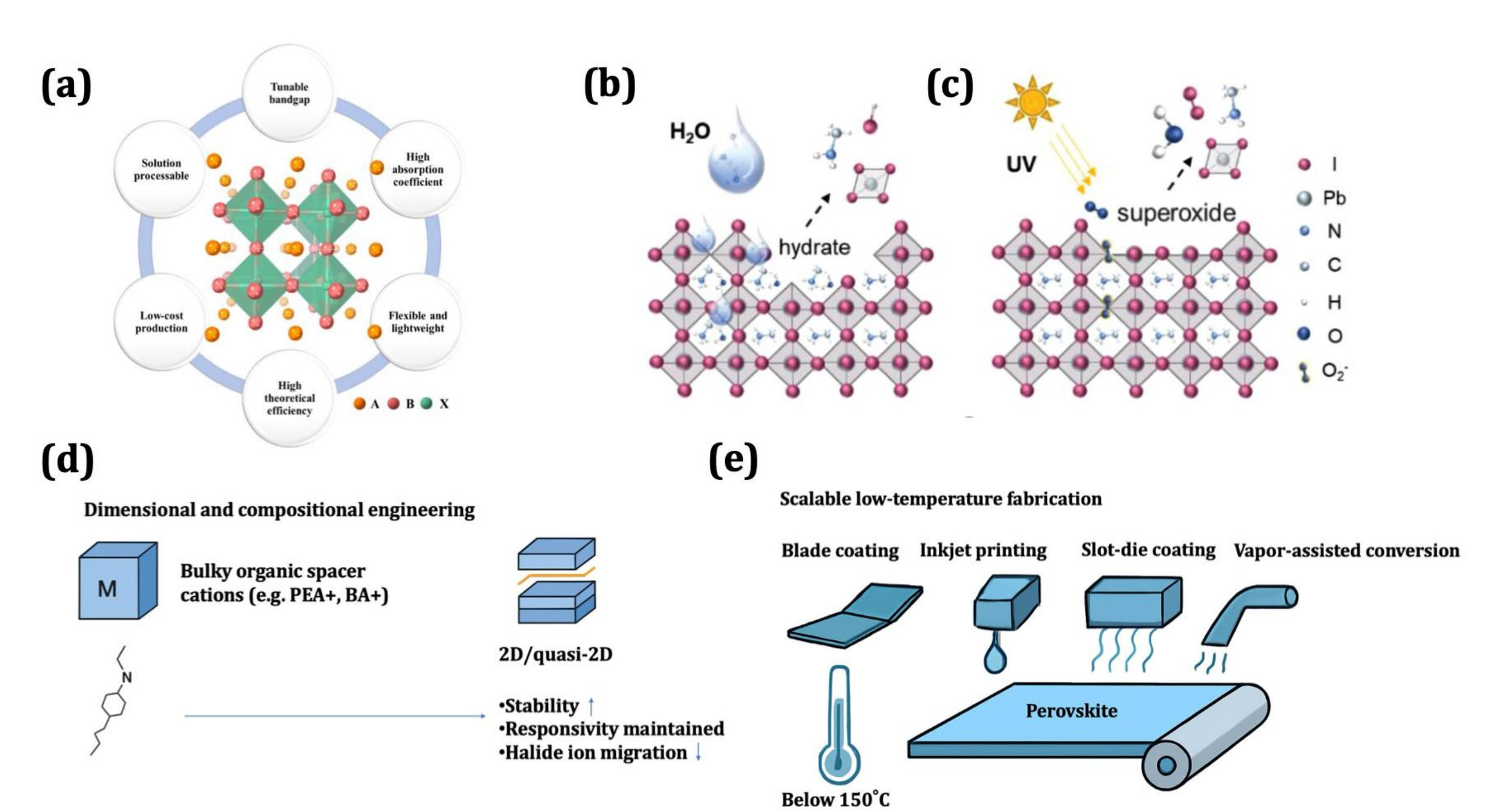

**Figure 1.** (a) Schematic illustration of the $ABX_3$ perovskite crystal structure, highlighting the corner-sharing $BX_6$ octahedra framework and A-site cations. The key optoelectronic advantages of this structure, including tunable bandgap, strong light absorption, and efficient charge transport, are also summarized.[19] Reproduced from Yitong Wang et al., Molecules, 2024, 29, 4976 (https://doi.org/10.3390/molecules29204976), licensed under CC BY 4.0 (https://creativecommons.org/licenses/by/4.0/). (b-c) Schematic illustration of the degradation of perovskite crystals under the influence of humidity (b) and UV light and oxygen exposure (c).[20] Copyright 2022 Wiley. (d) Schematic of the transition from 3D to 2D/quasi-2D perovskites, illustrating improved stability and suppressed halide ion migration. (e) Schematic illustration of scalable low-temperature fabrication techniques for metal halide perovskite films, including blade coating, inkjet printing, slot-die coating, and vapor-assisted conversion.

## Concluding remarks

MHPs offer unique opportunities for UV-C photodetection due to their tunable bandgaps, strong absorption, and compatibility with low-temperature fabrication. While notable progress has been made—particularly with chloride-based compositions such as $CsPbCl_3$, which enable solar-blind operation and transparent device architectures—key challenges including UV-induced degradation, ion migration, and large-area processing still limit their practical deployment. Ongoing advances in dimensional engineering, interface passivation, encapsulation, and scalable manufacturing are gradually overcoming these bottlenecks, paving the way for stable, sensitive, and cost-effective UV-C photodetectors for public health, environmental monitoring, and next-generation wearable sensing platforms.

## Acknowledgements

This work was funded by the National Natural Science Foundation of China (22279049, 12247101 and 12374393), the Fundamental Research Funds for the Central Universities (lzujbky-2023-eyt03 and lzujbky-2024-ey02), the Natural Science Foundation of Gansu Province (23JRRA1017), the State Key Laboratory of New Textile Materials and Advanced Processing Technologies (No. 03).

# 4.8. Microelectromechanical systems (MEMS)

**Lijie Li[1*], and Meiyong Liao[2]**

[1] Department of Electronic and Electrical Engineering, Swansea university, Swansea, SA1 8EN, United Kingdom
[2] Research Center for Electronics and Optical Materials, National Institute for Materials Science, Namiki 1-1, Tsukuba, Ibaraki 305-0044, Japan

*E-mail: L.Li@swansea.ac.uk

### Status

Microelectromechanical systems (MEMS) technology has commonly been used in development of free space optical elements, such as micro-optical scanner applied in display, biomedical imaging, and telecommunications [1] [2] . Recently there has been increasing development aiming at using MEMS in photodetectors to improve device performances. The fundamental mechanism is to apply mechanical strain/stress to the photodetector devices to realise performance tuneability or enhancement. There have been three directions along which MEMS help improve photodetector performances: 1) piezoelectric microstructures are used together with heterojunctions to create built-in potentials at the interface of heterojunctions to increase photocurrent using externally applied mechanical strains and to implement self-powered photodetectors. [3-7] 2) MEMS devices fabricated with wide bandgap materials are constructed to achieve multi-functional or dual-modality photodetectors that can convert the photo-signals with photo-electronic and photo-thermal-mechanical transduction mechanisms. [8, 9] 3) It is known that there is balance between photoresponsivity and photo-to-dark current ratio, recently MEMS technology has been used to overcome the limitation imposed by this balance [10].

### Current and future challenges

Research and development in photodetectors have been advancing significantly, especially with the rapid growth of activities in wide bandgap semiconductors. However, several challenges remain and are expected to persist in the future. Performance is paramount for photodetectors, and researchers are racing to develop new technologies to enhance it. It is well known that there can be a trade-off between photoresponsivity and dark current, as improving photoresponsivity may inadvertently increase the dark current, and vice versa [11]. In practical implementations, a large number of photodetector devices are often required, particularly for high spatial resolution imaging applications [12]. However, operating many devices simultaneously can lead to power consumption and loss issues. Furthermore, the response time of current photodetector devices based on wide bandgap semiconductors is typically slow [13], creating a demand for faster-response photodetectors. To address these challenges, MEMS technology—which has traditionally been used to miniaturise a wide range of sensors and actuators—can offer innovative solutions. The following section provides a summary of recent research efforts aimed at using MEMS technology to overcome the above challenges.

### Advances in science and technology to meet challenges

Three main research categories for MEMS being used for photodetector devices are summarised in terms of motivation:

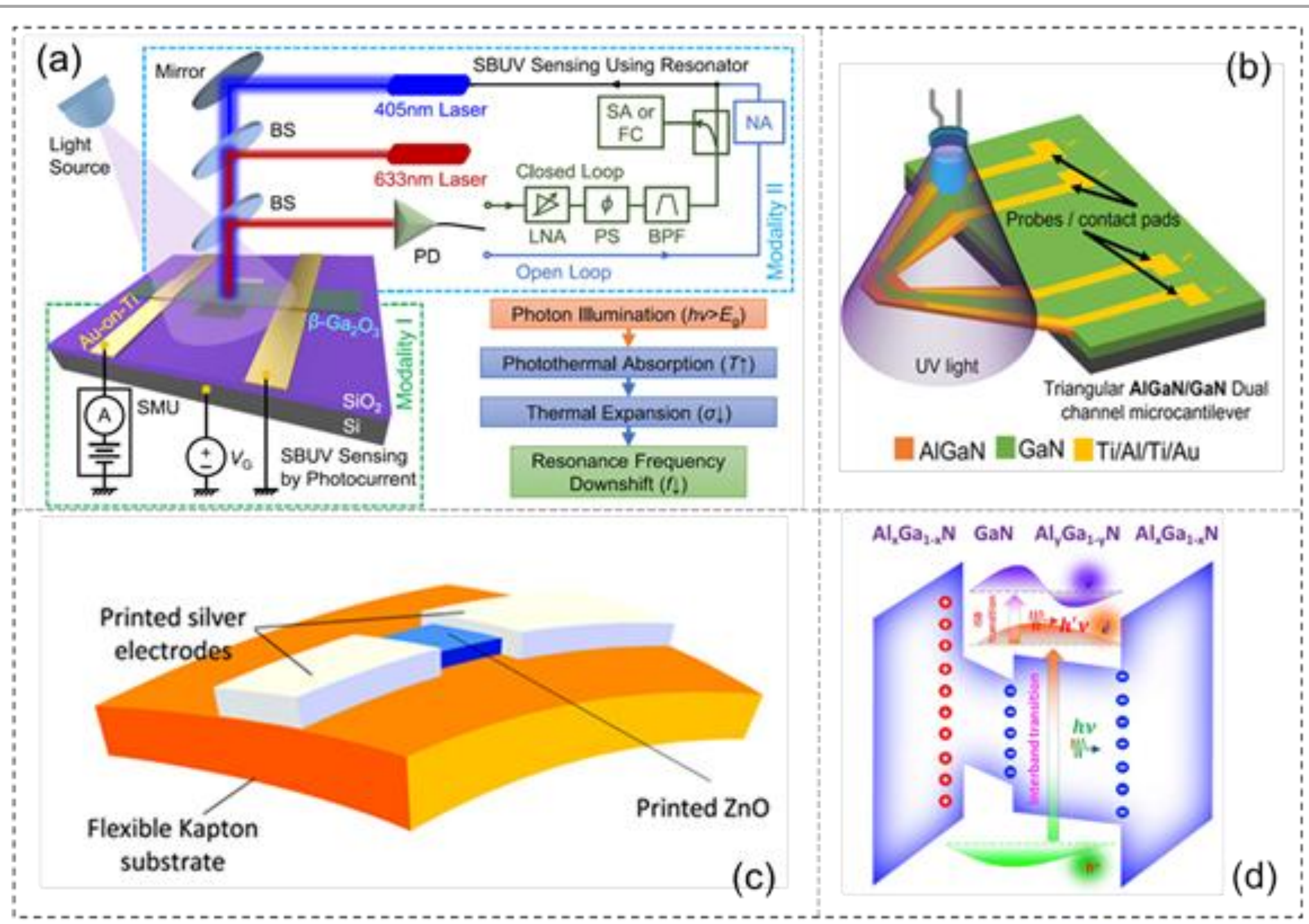

**Figure 1.** MEMS technologies promoting photodetector devices. (a) A dual-modality β-Ga2O3 UV photodetectors, Reprinted from [8], CC BY 4.0. (b) A triangular AlGaN/GaN photodetector, Reprinted from [10], Copyright (2022), with permission from American Chemical Society. (c) A printed flexible ZnO photodetector, Reprinted from [14], Copyright (2017), with permission from IOP Publishing (d) A quantum well piezotronic photodetector, Reprinted from [15], Copyright (2019), with permission from Elsevier

*Performance enhancement.* Interesting research conducted by Uppalapati et al. produced a triangle micro cantilever consisting of AlGaN/GaN two-dimensional gas channels with an intervening GaN layer [10]. The triangular suspended structure help in electric field shaping and concentration in the photoactive region, hence reducing the dark current and enhancing the photocurrent, resulting in high photoresponsivity (~$10^4$ A/W) and low dark current (~ pA). In [3], piezo-phototronic effect was used to enhance performances of Si/CdS heterostructure near IR photodetectors. It was reported that more than two orders of magnitude increase of photoresponsivity for the p-Si/n-CdS device, which has a maximum photoresponsivity of 14.5 A/W. This enhancement is theoretically elucidated as the band structure tuning in the junction by the piezoelectrically induced potential. Using a UV photodetector constructed by the ZnO/$Ga_2O_3$ heterojunction [4], it was unveiled that the responsivity can be increased 0.27 to 2.49 A/W with a tensile strain range of ±0.57% for the 254 nm illumination. However, the strain does not make significant impact on the 365 nm illumination. It was reported in [15] that mechanical strain can be used to enhance the performance of an AlGaN/GaN/AlGaN quantum well device, where quantum efficiency can be adjusted precisely by the mechanical strain. A MEMS photodetector features a lateral metal–polymer–metal configuration that responds across a broad optical spectrum (UV to mid-IR), in which the photoactive polyvinyl alcohol (PVA) layer is deposited on a suspended membrane to provide thermal isolation [16]. The device operates with integrated photoconductive and pyroelectric properties of the PVA, demonstrating responsivity of 0.53 A/W capable for broad-spectrum detections. Diamond material has sparked many research activities attributed to its ultra-wide bandgap, high thermal conductivity (22 W/mm K), high carriers' mobility, large breakdown electric field, and high thermal stability. Diamond is a good candidate for UV-C photodetection [17], but also a good candidate for mechanical resonators as it has large elastic modulus that leads to high resonant frequencies and quality factors. It is envisaged that combination of diamond's photo-electric and mechanical properties will set foundation for the next generation of high-performance photodetectors.

*Self-powering.* Self-powered UV photodetectors were presented with various combinations of heterojunctions where build-in electric potentials serve as the charges pump, separating photogenerated electron-hole pairs. As there is no external bias to this type of self-powered photodetectors (photovoltaic), responsivity is usually low comparing with the photoconductive devices integrated with an energy harvesting unit [5]. Self-powered devices are extremely desirable on the IoT where large arrays of photodetectors are required where power supply for large number of sensors becomes an issue. Piezotronic and piezo-phototronic effects use piezoelectrically generated charges to adjust the band diagram at the interface/junction, which subsequently modulates the optoelectronic carriers transportation, providing useful ways of improving the device performances. Wang, et al. [6] reported enhanced photoresponsivity for the CdTe nanowire photodetector operating at a broad optical frequency range of 325-808 nm. Piezo-phototronic effect was used in another example, where the photoresponsivity of a deep UV photodetector made of ZnO-$Ga_2O_3$ heterojunction has been increased three times with a mechanical strain of 0.042% [7]. Photoresponsivity of a hybrid organic-inorganic hybrid perovskite photodetector working at a broad visible wavelength range (~400 – 800 nm) has been improved significantly by piezo-phototronic effect [18]. ZnO is one of piezoelectric materials that have been used for micro sensors and actuators. When utilising the ZnO in photodetectors, its piezoelectricity is used for performance improvement. It was reported that the photocurrent of a n-ZnO/Si flexible membrane is increased by 22% under an external tensile strain [19].

*Multiple modalities.* A notable advancement for MEMS to improve the performances of photodetectors was reported by Zhang, et al. [8], in which a dual-modality solar blind UV photodetector was reported. The device is fabricated from a β-$Ga_2O_3$ micro-resonating structure, which can detector the UV light based on both the photo-electric and photo-thermal-mechanical effects. Boosted by a photo-field-effect transistor, the photoresponsivity can reach up to 63 A/W, and the mechanical photosensitivity was 250 Hz/nW. It was also concluded that the performances could be further improved by increasing the resonant frequency and mechanical quality factor. Currently the β-$Ga_2O_3$ MEMS devices are fabricated by transferring the β-$Ga_2O_3$ films to pre-patterned wafer. On-chip dry etching of the β-$Ga_2O_3$ microcantilevers was recently demonstrated in [20]. A middle UV photodetector has been realised with a resonating GaN/AlN double-end fixed bridge structure on a silicon substrate [9]. The mechanical resonant frequency is measured to be ~0.4 – 4 MHz, and mechanical photoresponsivity of -24 Hz/nW. This technique has potential applications in real-time UV photon sensing.

### Concluding remarks

Looking ahead, MEMS technology is expected to exert a growing influence on photodetector devices, particularly considering ongoing advancements in wide bandgap materials and the escalating demand in fields such as the IoT and artificial intelligence, where a high density of sensors is required. In this context, the low power consumption and multifunctionality offered by MEMS-based solutions are especially advantageous.

### Acknowledgements

LL would like to aknowledge acknowledge support by the UKRI project (UKRI595).

# 5. Applications of UV-C photodetectors and industry perspectives

# 5.1. Metrology

**Nasim Zarrabi[1*] and Sebastian Wood[1]**

[1] Electromagnetic and Electrochemical Technologies, National Physical Laboratory, Hampton Road, Teddington, TW11 0LR, United Kingdom

E-mail : nasim.zarrabi@npl.co.uk

### Status

Within optical metrology, the UV-C spectral range is less developed than the visible and near IR spectral ranges. UV-C is generally defined as the wavelength range 100 nm to 280 nm, however this overlaps with the vacuum ultraviolet (VUV) range from 10 nm to 200 nm [1] where strong atmospheric absorption further complicates measurements. Here we focus on the region 200 nm to 280 nm, though many of the same considerations extend to shorter wavelengths. The limitations of available sensing devices in the UV-C range is a key challenge since standard Si-based detectors exhibit poor quantum efficiency for wavelengths below 240 nm [2]. Additionally, the availability of UV-transmitting optical components is limited; materials such as fused silica and magnesium fluoride ($MgF_2$), which are required for UV-C transparency, are more expensive and challenging to fabricate [3]. These technological constraints hinder the development of traceable and accurate metrological systems in the UV-C domain, despite its growing importance in fields such as sterilization [4], semiconductor processing [5], environmental monitoring [6], and astronomy [7].

From a metrological point of view, there are two interrelated considerations: the requirements of metrology *for* UV-C photodetectors, and the applications of UV-C photodetectors *for* metrology.

*Metrology for UV-C photodetectors* involves the development of calibration methodologies, uncertainty quantification, and standardized measurement protocols to characterize photodetector performance. Key parameters include spectral responsivity, linearity, noise characteristics, frequency response, and long-term stability under operational conditions. A central requirement in this context is traceability to the International System of Units (SI), which is discussed below.

*UV-C photodetectors for metrology* have widespread uses as measurement devices for radiometric properties of UV-C sources such as optical power, irradiance, and spatial distribution. Accurate measurements of UV-C light sources are essential in many contexts, such as characterizing the output of UV lamps and lasers, monitoring flame emissions, and performing spectroscopic measurements. The accuracy and availability of such measurements are limited by the cost and performance of sensors in the UV-C range. There are many industrial and scientific uses of UV-C light across multiple sectors (as discussed in other chapters of this roadmap), each of which requires accurate measurements. Therefore, development of high-performance UV-C sensors has potential for widespread impact. UV-C photodetectors offering high quantum efficiency and low dark current would enable precise measurements even at low light levels. Furthermore, detectors with a fast temporal response and no dead-time would support accurate detection of rapidly varying or pulsed UV-C signals—an essential requirement in optical communications and time-resolved spectroscopy.

### Current and future challenges

Characterisation of a UV-C photodetector requires accurate measurement of several key performance parameters. These include the spectral responsivity (typically expressed in amperes per watt ($A{\cdot}W^{-1}$) or volts per watt ($V{\cdot}W^{-1}$) ), quantum efficiency (the percentage of incident photons that generate charge carriers in the electrical circuit), dark current (the current that flows in the absence of light), noise equivalent power (NEP) (the minimum detectable optical power), response time or bandwidth (how quickly the detector responds to changes in light intensity), linearity (how proportional the output is to the input light). Each of these parameters must be measured in a way that is traceable to the International System of Units (SI) to ensure accuracy and comparability across laboratories. Since a photodetector responds to light, most of these measurements require a light source with known spectral power, which is the optical power per unit wavelength ($W{\cdot}nm^{-1}$). This requires some kind of reference device, which could either be a standard light source or a calibrated photodetector.

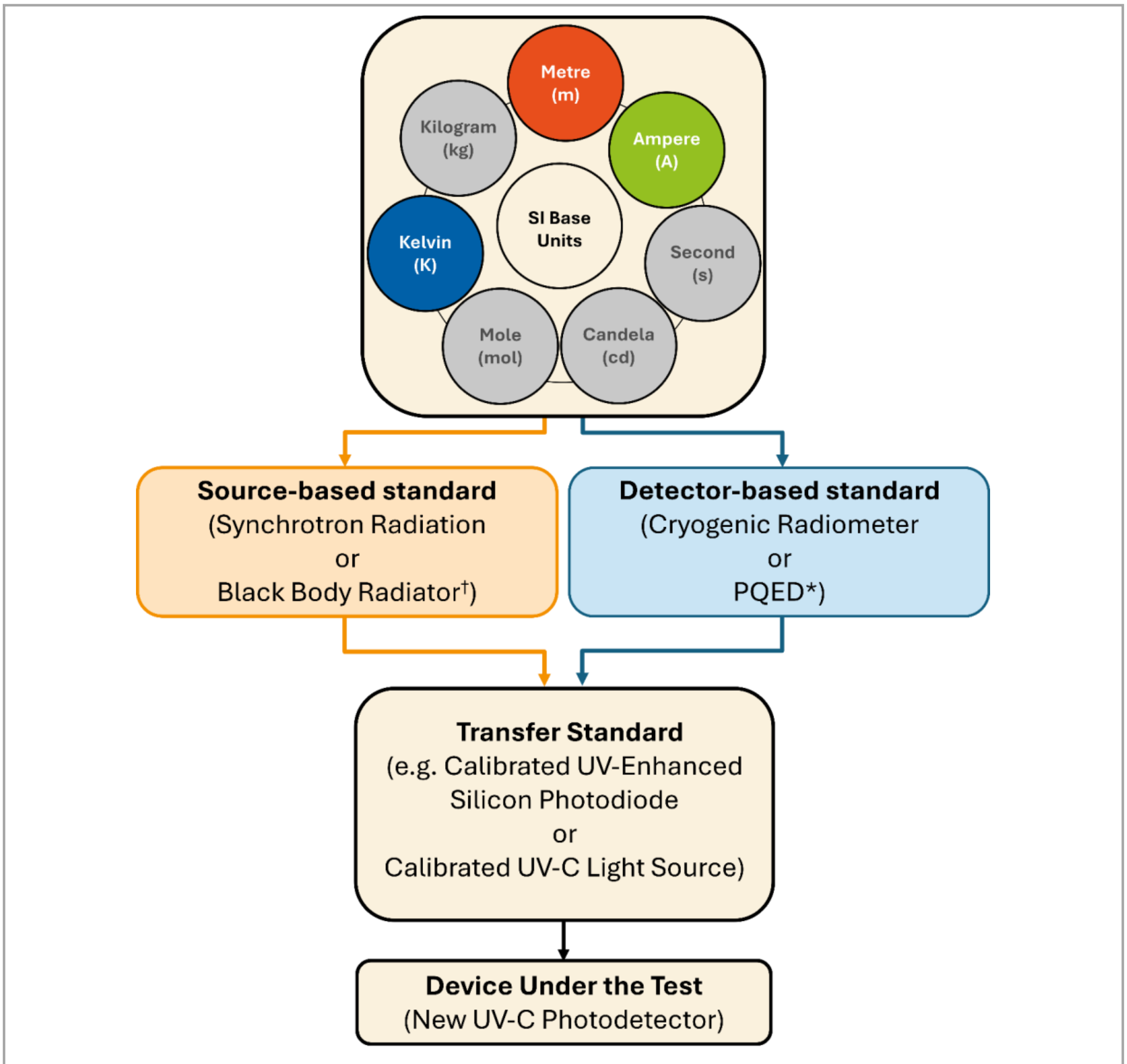

**Figure 1.** Traceability chain for optical power measurements in the UV-C region, showing links from the SI base units via primary realisation or radiometric units and transfer standards to a UV-C photodetector under test. †Black body radiators can be used for UV-C in principle but the high temperatures required prevent practical use for short wavelengths. *Predictable quantum efficiency photodetectors (PQEDs) have not yet been demonstrated as a suitable primary standard for the UV-C.

A substantial element of metrological research is to ensure traceability and minimise measurement uncertainties. This requires an unbroken chain of calibrations linking the laboratory measurement to fundamental physical constants. For optical radiometry, light sources and sensors are a complementary pair since either one can be used as a reference for the other. The traceability chain for photodetectors is illustrated in Figure 1.

Practical realisation of radiometric units comes from the SI kelvin, metre and ampere. We note that the SI Brochure describes this in the *mise en practique* for the candela although the candela itself is irrelevant to radiometric units. Both source-based and detector-based primary realisations are possible [8].

Source-based primary realisation can be achieved with synchrotron radiation, where the spectral radiant intensity of relativistic electrons in a storage ring can be calculated from first principles. This makes it an absolute source traceable to SI units through electrical and length measurements. Synchrotron radiation is particularly valuable because it covers a wide spectral range, including the UV-C region, and allows for precise control of photon flux over many orders of magnitude without altering the spectral shape. Alternatively, a Plankian (black body) radiator could be used, where the spectral radiance emitted by a cavity with high emissivity can be calculated using Planck's radiation law based on the thermodynamic temperature. This approach is traceable to the kelvin however producing high irradiance in the UV-C would require temperatures of the order of $10^4$ K , which are extremely challenging to achieve and accurately characterise experimentally [8].

Alternatively, detector based primary realisation can be achieved using cryogenic radiometers. These instruments measure optical power by comparing the heating effect of absorbed radiation to that of an equivalent electrical power. When operated at cryogenic temperatures (typically below 20 K), they achieve extremely low uncertainties. Another approach is the use of a predictable quantum efficient photodetector (PQED), which is a low-loss semiconductor device whose quantum efficiency can be accurately modelled based on the device structure [8]. However, Si PQEDs exhibit long-term instability for UV wavelengths[9]. These approaches to primary realisation of radiometric units are beyond the capabilties of most laboratories. Rather they are generally maintained by national metrology institutes, and disseminated using secondary standards. Again, this can be achieved either using a source or detector to transfer the calibration to a local measurement.

A common approach is to use a UV-enhanced Si photodiode or a deuterium lamp that has been calibrated against a primary (or secondary) standard. These calibrated devices serve as transfer standards: the photodiode can be used to calibrate local light sources whereas a calibrated light source can be used to characterize photodetectors. This approach ensures that measurements performed in the laboratory remain traceable to SI units, even in the absence of direct access to primary standard facilities.

A further gap in calibration is the absence of standardised procedures for UV-C spectroscopy. While best practices and reference materials are well established for vis-NIR regions, no equivalent standards currently exist for the UV. In particular, there are no standard reference materials for intensity calibration in UV photoluminescence or Raman spectroscopy. The development of traceable UV-C photodetectors could therefore play a key role in advancing quantitative UV-C spectroscopy, as detector-based calibration could provide improved intensity calibration, enhanced comparability between laboratories, and support the establishment of future measurement standards.

**Advances in science and technology to meet challenges**

At present the International Bureau of Weights and Measures (BIPM), which maintains the 'key comparison' database of measurement capabilities at global national metrology institutes includes UV-C detector calibration capabilities down to 200 nm achieving expanded uncertainties (95 % confidence interval) of at least 6 %. In contrast, uncertainties less than 1 % are readily achievable across the visible range[10]. Metrological progress in UV-C photodetectors encompasses both the development of new UV-C photodetectors and spectrally-matched light sources, due to their complementarity in connecting calibration chains. UV-sensitised Si photodetectors are readily available and extend the sensitivity of conventional Si to wavelengths of around 200 nm. However, the quantum efficiency remains low, and they are known to degrade with prolonged UV exposure. Wide bandgap semiconductors would be expected to offer a better match for UV photodetection in terms of both responsivity and operational stability.

A current European metrology research project (EMPIR Scale-Up) is seeking to extend the range of PQEDs into the UV range[9]. This approach requires high purity material and effective passivation of surface defects, but in principle provides primary calibration so can shorten calibration chains and give low uncertainty measurements despite poor quantum efficiency.

UV-C light sources are also valuable for the characterisation of UV-C photodetectors. Accurate characterisation requires a stable, well-defined, and broad-spectrum UV-C source. For decades, deuterium lamps have been the standard light source for UV applications. These lamps emit a continuous spectrum from approximately 160 to 400 nm, making them suitable for broadband UV measurements. Their relatively high intensity into the UV-C region and established use in spectroscopy made them a natural choice for photodetector testing. However, deuterium lamps come with significant limitations. First, their lifespan is limited to around 1,000–2,000 hours, requiring frequent replacement. Second, they require a warm-up period of several minutes to stabilize the output spectrum, which consume their lifetime, delays measurements and reduces throughput. Additionally, deuterium lamps are relatively expensive, both in terms of initial cost and ongoing maintenance, making them less economical for long-term or high-throughput applications.

Recent advances in UV-C LED technology offer promising alternatives. LEDs based on AlGaN or AlN substrates now provide longer lifetimes and compact designs. These features make them ideal for integration into modern, automated photodetector characterisation spectrometers. However, current UV-C LEDs have a key limitation: their narrow emission band. Most commercial devices emit at specific wavelengths with a bandwidth of around 20 nm, which restricts their usefulness in broadband spectral response measurements[11]. This makes them less suitable for applications requiring full-spectrum UV-C coverage, such as responsivity calibration across the entire UV-C range. The stability of the spectral output from UV LEDs is also a critical requirement for their use as reference light sources and their suitability in this respect is yet to be adequately demonstrated.

To address this, broadband UV-C LED sources are now emerging. For instance, phosphor-converted UV LEDs from companies like PhosphorTech combine UV-C chips with UVB/UVA phosphors to produce a broad, stable spectrum from 250 nm to over 850 nm. While these developments are promising, research and engineering efforts are still ongoing to fully match the spectral coverage and intensity of traditional deuterium lamps in all UV-C characterisation applications[12].

#### Industry perspectives

The metrology of UV-C photodetectors is essential because many industrial and scientific applications that rely on UV-C light sources and laser technologies require precise calibration and accurate measurement to ensure consistent and reliable performance. In the semiconductor industry, UV lasers are used for micromachining, wafer inspection, and defect detection—processes that demand tight control over laser intensity, beam uniformity, and exposure stability. Accurate photodetector metrology ensures these parameters are monitored and maintained, which is critical for producing high-performance microelectronic devices. UV-C photodetectors are also relevant to UV spectroscopy, where UV lasers enhance sensitivity and spatial resolution, enabling detailed chemical and structural analysis of semiconductor materials. Additionally, UV-C lasers are used in healthcare and environmental systems for disinfection, where reliable measurement ensures safe and effective operation. In all these cases, the performance and safety of the systems depend on the accurate and traceable measurement capabilities provided by well-calibrated UV-C photodetectors.

### Concluding remarks

Metrology of UV-C photodetectors is foundational to progress in the development and use of UV technology. Equally, such devices have a role to play in advancing the state of the art in UV optical radiometric metrology. This virtuous circle demands continued research and development to create high performance photodetectors that exceed the efficiency and stability of incumbent UV-enhanced Si photodiodes. The practical consideration of metrological traceability chains also highlights the importance of UV-C light sources as a complementary pair with corresponding photodetectors, and therefore attention must also be given to extending the spectral range of UV LEDs and demonstrating their stability as reference light sources.

### Acknowledgements

This work received funding from the UK Department for Science, Innovation and Technology (DSIT) through the National Measurement System Project. The authors acknowledge technical review by Martin Dury (NPL).

# 5.2. Astronomy

**Jesper Skottfelt[1*] and Susan E. S. Spesyvtseva[2,3]**

[1] Centre for Electronic Imaging, School of Physical Sciences, The Open University, Walton Hall, Milton Keynes, United Kingdom
[2] Department of Physics, University of Strathclyde, Glasgow, United Kingdom
[3] Institute of Photonics, SUPA, University of Strathclyde, Glasgow, United Kingdom

*E-mail: jesper.skottfelt@open.ac.uk

### Status

UV-C astronomy explores one of the most information-rich yet technically challenging regions of the electromagnetic spectrum. Because Earth's atmosphere is opaque to UV-C radiation, all observations must be conducted from space. Historically, UV astronomy has provided fundamental insights into the physics of hot stars, planetary atmospheres, the interstellar medium (ISM), and cosmic reionisation. Emission and absorption lines in this band trace highly ionised species such as O VI, C IV, Si IV, and He II, revealing information about hot plasmas, stellar winds, and intergalactic filaments at temperatures of $10^5$–$10^6$ K. These measurements are essential for modelling feedback processes in galaxies and understanding the thermal evolution of the circumgalactic medium [1].

UV-C observations also play a uniquely enabling role in exoplanet science. Observations in the UV-C band can probe oxygen and carbon isotopes and detect more complex organic molecules in exoplanetary atmospheres, offering clues to planetary habitability and potential biosignatures.

Over the past four decades, missions including IUE, HUT, FUSE, and GALEX have explored portions of the UV spectrum [1]. More recently, the Hubble Space Telescope, using UV instruments such as the Space Telescope Imaging Spectrograph [2] and the Cosmic Origins Spectrograph [3], has enabled transformational science across 90–300 nm. Hubble's UV observations have deepened our understanding of exoplanet atmospheres, intergalactic gas, and stellar composition. However, the Hubble Space Telescope is now operating well beyond its original intended lifetime, and so, continuity of UV-C observations is increasingly at risk.

In response, several new UV missions are under development, including UVEX [4] and CASTOR [5]. Most notable is NASA's upcoming Habitable Worlds Observatory, a flagship mission recommended by the 2020 Decadal Survey [6], which aims to search for biosignatures in Earth-like exoplanets and investigate galaxy and planet formation. Achieving its scientific objectives depends critically on extending UV sensitivity into the UV-C regime to allow characterisation of high-energy radiation environments and planetary atmospheres. This requirement places stringent demands on detector performance. In particular, the quantum efficiency must exceed 40%, while maintaining high radiation hardness and long-term stability for a mission lifetime of over 10 years. Further, the noise must be at the single-electron level and even achieve photon-counting for some science cases, such as exoplanet atmosphere characterisation.

In parallel, the emergence of smaller, modular spacecraft and CubeSat-class missions, is both broadening access to UV observations and enabling rapid technology validation [7]. This diversification of mission architectures increases demand for compact, efficient photodetectors capable of operating under varied power, thermal, and contamination constraints. Aligning technological progress with these evolving science drivers will ensure that UV-C astronomy remains a central tool in exploring the energetic and chemical complexity of the Universe.

**Current and future challenges**

Despite remarkable advances in space astronomy, many fundamental astrophysical questions remain unresolved because they require sensitive observations in the UV-C band. This region contains spectral fingerprints of some of the most energetic processes in the Universe, yet only a limited number of instruments have operated effectively at these wavelengths. Ensuring sustained and expanded access to UV-C observations is therefore a central challenge for future space astronomy.

*Exoplanet atmosphere characterisation.* One major frontier is the characterisation of exoplanet atmospheres. Many of the most diagnostic biosignature species, including ozone ($O_3$), molecular oxygen ($O_2$), and a range of carbon compounds, exhibit strong absorption features in the UV-C [8]. These measurements can also reveal isotopic ratios and complex organic signatures that help to discriminate between biological and abiotic processes. For hotter exoplanets the UV-C range is highly sensitive to the absorption of metal atoms and ions in the atmosphere [9] and Lyman-α absorption can be used as a probe of neutral hydrogen in the lower atmosphere of Earth-like exoplanets [10]. These signatures are often intrinsically weak and are typically measured against the continua of bright host stars. As a result, observations require detectors with exceptionally high quantum efficiency, ultra-low noise, and long-term radiometric stability[7]. Current detector technologies approach but do not yet fully meet these combined performance requirements.

Closer to home, Solar System bodies display UV-C spectral signatures from volatiles and surface chemistry, including water, carbon dioxide, and complex organics [12][14]. These observations help constrain planetary formation histories and assess astro-biological potential.

*Intergalactic and circumgalactic medium tracing.* Another major scientific challenge lies in tracing the intergalactic and circumgalactic medium, which is believed to contain a substantial fraction of the Universe's baryonic matter. Key absorption lines, including the H I Lyman series, O VI, and Si III fall within the UV-C and vacuum-UV regime, providing direct diagnostics of the temperature, density, and ionisation structure of diffuse gas [13][12][9]. Observing these features requires sensitive, wide-field spectrographs and detectors capable of maintaining uniform response across large focal planes.

*Understanding star formation and stellar feedback.* Within galaxies, UV-C observations are critical for understanding star formation and stellar feedback processes. The short-wavelength continuum of massive, young stars traces the ionising radiation that drives reionisation and regulates the escape of Lyman continuum photons [14][13]. Measuring these signals demands not only high sensitivity but also exceptionally precise calibration accuracy and control of instrumental systematics, as small uncertainties can lead to large measurement errors.

*Instrumentation challenges.* From an instrumentation standpoint, UV-C astronomy is uniquely challenged by material and contamination issues. Most optical materials strongly absorb UV-C photons, making optical throughput highly sensitive to surface quality, coatings, and even trace molecular residues. Telescope and instrument design and manufacture must therefore achieve extreme cleanliness and stability, often at significant cost.

Overall, the challenge for UV-C astronomy is to move beyond the limited and impermanent UV-C observational capability achieved to date and towards sustained, high sensitivity observational capability. Addressing the scientific questions outlined above requires detectors and instruments that surpass current performance limits in sensitivity, stability and scale. Without such advances, many of the most compelling questions in planetary science, astrophysics, and cosmology will remain unresolved.

**Advances in science and technology to meet challenges**

Addressing the scientific challenges of UV-C astronomy requires both incremental improvement and transformative innovations in UV-C detectors and instruments. Each key scientific objective, ranging from exoplanet atmospheric characterisation to tracing galactic evolution, places specific demands on sensitivity, stability, and scalability.

For exoplanet spectroscopy, the highest priority is achieving high quantum efficiency and low background over long integrations. Photon-counting detectors such as electron-multiplying charge-coupled devices (CCD) and advanced CMOS imagers are being developed to detect faint atmospheric signals in the presence of bright starlight. Techniques such as delta-doping and ALD passivation have enabled Si-based detectors to achieve quantum efficiencies exceeding 50 % at wavelengths below 200 nm, marking a substantial improvement over previous generation devices [14].

For studies of the intergalactic and circumgalactic medium, new wide-field spectrographs require large, uniform focal planes capable of ultra-low noise operation. Progress in techniques, such as CMOS tiling, and emerging technologies, such as wide-bandgap materials including AlGaN and diamond, enable scalable architectures with intrinsic solar-blindness. In addition, multichannel plate detectors with CsI or GaN photocathodes also offer excellent photon-counting speed and timing precision, as demonstrated by recent sounding rocket campaigns (e.g. CHESS, INFUSE.). The application of ALD coatings has improved MCP lifetime and stability, ensuring their continued relevance for high-sensitivity UV spectroscopy [15].

Studies of stellar evolution and reionisation rely on precise calibration and minimal systematic noise. Low-noise CMOS pixels, hybrid detector architectures, and radiation-tolerant designs are improving detector uniformity and resilience for long-duration missions [16][15]. Similarly, for Solar System targets, where radiation levels and contamination risks are high, progress in packaging, radiation-hardened readouts, and contamination-resistant materials is improving performance stability.

Looking further ahead, emerging super-lattice and superconducting detector technologies, such as microwave kinetic inductance detectors, transition-edge sensors, and superconducting nanowire single-photon detectors are showing promise for UV and optical astronomy [17]. These devices combine intrinsic photon counting and energy resolution with extremely low noise, though their cryogenic requirements currently limit space deployment.

At the system level, innovations such as tiled focal planes, micromirror arrays, and advanced reflective coatings ($MgF_2$, LiF, Al) are enhancing throughput and scalability [18]. Together, these advances are building a comprehensive and versatile detector ecosystem, that spans modernised MCP/CsI systems, high-quantum efficiency CMOS imagers, and emerging superconducting devices. By aligning these technological innovations with the demands of cutting-edge science, missions can achieve the sensitivity, precision, and reliability required to support the ambitious science goals of next-generation UV-C astronomy.

**Industry perspectives**

From an industrial perspective, UV-C astronomy represents a demanding but well-defined application domain that pushes photodetector and instrument technologies beyond the performance limits of standard commercial offerings. At the same time, these devices must also be manufacturable, reliable, and affordable. Scientific requirements for high quantum efficiency at short wavelengths, ultra-low noise, long term stability, and radiation tolerance necessitate highly specialised design, fabrication, and testing capabilities. As a result, industrial capability in this field is concentrated among a relatively small number of highly specialised suppliers with expertise in materials, vacuum-compatible processing, and space qualification, and who work closely with academic and government partners.

A central challenge for industry is translating laboratory-scale innovations into reproducible, manufacturable devices. Laboratory breakthroughs in passivation, coating, or radiation hardening must be translated into reproducible production processes delivering consistent detector performance across multiple units. Demonstrating reliability through environmental and radiation testing is essential for achieving the TRL 6 threshold required for major mission adoption. The long lead times of flagship missions, such as the Habitable Worlds Observatory, make early industrial engagement crucial.

Maintaining diversity in technology approaches is another important consideration. While advanced CMOS imagers are becoming the dominant architecture due to their scalability and integration with modern readout electronics, MCP-based photon-counting systems retain unique advantages in timing resolution and intrinsic solar blindness, due to the photocathode material used. Supporting both mature and emerging technologies allows industry to tailor solutions to a broad range of mission architectures, from small satellites to large observatories, while managing technical risk.

Robustness of necessary supply chains is a critical concern for industry, as UV-C detectors rely on specialised materials, coatings and fabrication processes that are not widely available in existing commercial markets. Components such as photocathodes, microchannel plates, and UV-compatible optical coatings often depend on a limited number of qualified suppliers. Continuity of production and timely access to test infrastructure is critical for sustaining capability across long mission development timescales.

Industry also plays a key role in system-level integration. Detector performance in the UV-C is strongly coupled to instrument design choices, including contamination control, thermal management, and optical configuration. Early collaboration between detector manufacturer and instrument teams can mitigate risks associated with contamination sensitivity and degradation, leading to improved performance.

At the global scale, competition is intensifying. UK, European and American expertise in scientific detector development remains strong, but the dominance of Asian manufacturers in the commercial CMOS sector is increasingly shaping the technical and economic landscape for space-based imaging. This increased competition can drive innovation and reduce costs, but it also emphasises the importance of sustained national investment in detector R&D to preserve domestic capability and ensure participation in future international collaborations.

Beyond astronomy, many of the technologies developed for UV-C space instrumentation have relevance in adjacent sectors. These overlaps provide opportunities for knowledge transfer and cost reduction, while applications in astronomy continue to drive development of the highest performance detectors.

Looking ahead, emerging disruptive detector concepts, such as superconducting or super-lattice devices, are at early stages of maturity but present both opportunities and uncertainties for industry. While these technologies are currently at low to moderate maturity levels, early engagement in the development of such technologies will position manufacturers to support future generations of UV observatories as these approaches evolve towards practical deployment. Forward-looking companies are beginning to invest strategically in these areas, balancing near-term commercial goals with the need to stay positioned for future breakthroughs.

In this context, UV-C detector development exemplifies how science priorities, industrial expertise, and policy support must align. Sustained cooperation between academia, industry, and space agencies will be essential to ensure that the next generation of missions can deliver on the scientific promise of UV astronomy.

## Concluding remarks

UV-C astronomy is entering a new era, driven by the need to extend humanity's UV view of the Universe beyond the lifetime of Hubble. Access to the 100–280 nm band provides diagnostics that are unavailable at longer wavelengths and is essential for addressing many of the most compelling questions in modern astrophysics, including probing exoplanet atmospheres, tracing the intergalactic medium, and understanding the evolution of galaxies and elements over cosmic time.

Realising the full potential of UV-C observations, requires continued advances in detector and instrument technology, from optimised CMOS and MCP systems to emerging superconducting architectures, as well as close coordination between science and engineering, and academia and industry.

As existing UV capabilities approach the end of their operational lifetimes, the development of new robust, space-qualified UV-C instrumentation becomes increasingly important. By aligning scientific ambition with technological innovation, the community can ensure that future missions deliver transformative science and preserve and extend the UV legacy that has defined modern astrophysics and shaped our understanding of the Universe.

**Acknowledgements**

The authors would like to thank Dr Jo Barstow and Dr Doug Jordan for input to this chapter.

# 5.3. Communications

**Jonathan McKendry[1,2], Christopher G. Leburn[1,2] and Daniel K. L. Oi[1*]**

[1] Department of Physics, University of Strathclyde, Glasgow, United Kingdom
[2] Institute of Photonics, University of Strathclyde, Glasgow, United Kingdom

*E-mail: daniel.oi@strath.ac.uk

### Status

Optical wireless communications (OWC) offers a high bandwidth, low latency resource that is robust to electromagnetic interference and with few regulatory constraints [1]. Growing demand for increased capacity and channel availability motivates the extension of OWC to the UV band as this region of electromagnetic spectrum has been comparatively little exploited. Most OWC implementations and research has concentrated on the visible, near IR, and short-wave IR wavelengths. The attractive characteristics of the UV-C region specifically are, firstly, the UV-C region is solar-blind with little to no natural or artificial background interference. Secondly, Rayleigh scattering of light is inversely proportional to the fourth power of wavelength, thus strongly scattered UV-C lends itself to non-line-of-sight (NLOS) communications, where scattered transmitted signals are used to overcome obstacles obscuring line-of-sight communications. Given the demand for high wireless data rates and, particularly for NLOS communications, high sensitivity, advances in solar-blind photoreceivers with high modulation bandwidths and high sensitivity, even single-photon sensitivity, are vital.

Beyond terrestrial applications, the UV-C band is also attractive for inter-satellite communications where there is no absorption and scattering of the atmosphere [2]. Satellite constellations are implementing optical crosslinks between satellites to achieve faster and more efficient data shuttling across the constellation without the requirement for scarce and expensive RF licences. Existing standards for satellite optical free-space communications, such as those established by the Consultative Committee for Space Data Systems (CCSDS) or Space Development Agency (SDA), are mainly targeted at C-band wavelengths, reflecting heritage in components and systems derived from optical fibre networks. Satellites are highly constrained in size, weight, and power (SWaP), thus reduction in the optical aperture required for a given channel link loss is highly desirable. Moving from 1550 nm to 250 nm wavelength would in principle allow a ~2.5 reduction in linear dimension, or equivalently a 15 times reduction in volume/mass of both the optical transmitter and receiver, keeping the link budget the same. The solar background, though not suffering atmospheric attenuation, is lower on a per photon basis at sufficiently short UV-C wavelengths compared with near and short-wave IR, also making it an attractive region to operate quantum communications. Source and receivers capable of operating at UV-C wavelengths in the conditions of space, mostly notably a high radiation environment, will be key to unlocking these avenues.

### Current and future challenges

The Shannon-Hartley theorem states that the channel capacity, $C$, the theoretical upper limit on the bit rate that can be transmitted under a set condition, is given by

$$C = B\log_2(1 + \frac{S}{N})$$

where $B$ is the bandwidth of the channel, $S$ is the average received signal power, and $N$ the average received noise power. A significant ongoing challenge for UV-C photodetection for OWC is that the rise and fall times of solar-blind photoreceivers are at the moment typically in the range of milliseconds to seconds, in contrast with nanoseconds from Si-based visible photoreceivers. This severely limits the photoreceiver's modulation bandwidth and thus, from Shannon-Hartley, the maximum data rates possible with these detectors.

Another challenge lies in the detection of weak UV-C signals that are typical of NLOS communications. Although scattering is considerably stronger in the UV-C (~0.1 to 0.5 $km^{-1}$ [3]) compared to visible or IR wavelengths, received signal strengths are nonetheless very weak over long distances. This necessitates the development of solar-blind single-photon sensitive photoreceivers. These photoreceivers can be fabricated from materials such as AlGaN, however, there remain significant challenges in epitaxial growth and p-type doping that must be overcome if these devices are to meet the required performance [4].

### Advances in science and technology to meet challenges

The range of UV-C wavelengths covered by LEDs has grown such that LEDs with peak emission wavelengths <230 nm have been demonstrated [5]. While external quantum efficiencies decline significantly at shorter wavelengths, as can be seen in Figure 1, UV-C LEDs with milliwatt output powers have been demonstrated, for example, at 226 nm [6] and <240 nm [7]. Furthermore, UV-C LEDs have modulation bandwidths on the order of several tens of MHz, or many hundreds of MHz if their light-emitting areas are micro-pixellated [8]. As such, UV-C LEDs are rapidly maturing as efficient and high-speed light sources for OWC, in a low size, weight, power and cost (SWaP-c) format. Future UV-C photodetectors having high modulation bandwidth and single-photon sensitivity would be well-positioned as complementary receivers for solar-blind OWC.

In addition to the emitters and detectors, the supply chain of supporting components will need to be developed to operate at such wavelengths. In particular, optical fibres and associated components will need to approach the same maturity as those developed for the telecom C-band. Greater availability of splitters, couplers, polarisers, dense wavelength division (de)multiplexing, and low-loss transmission material systems and fibres is desirable. Apart from employing more exotic materials, hollow-core silica fibres is a potential route for single-mode routing over short to medium distances, particularly within transmitter and receiver assemblies. To capitalise from the SWaP benefits afforded by the shorter wavelengths, the move to on-chip signal conditioning, modulation, and photonic integrated circuits compatible with UV-C is a priority.

### Industry perspectives

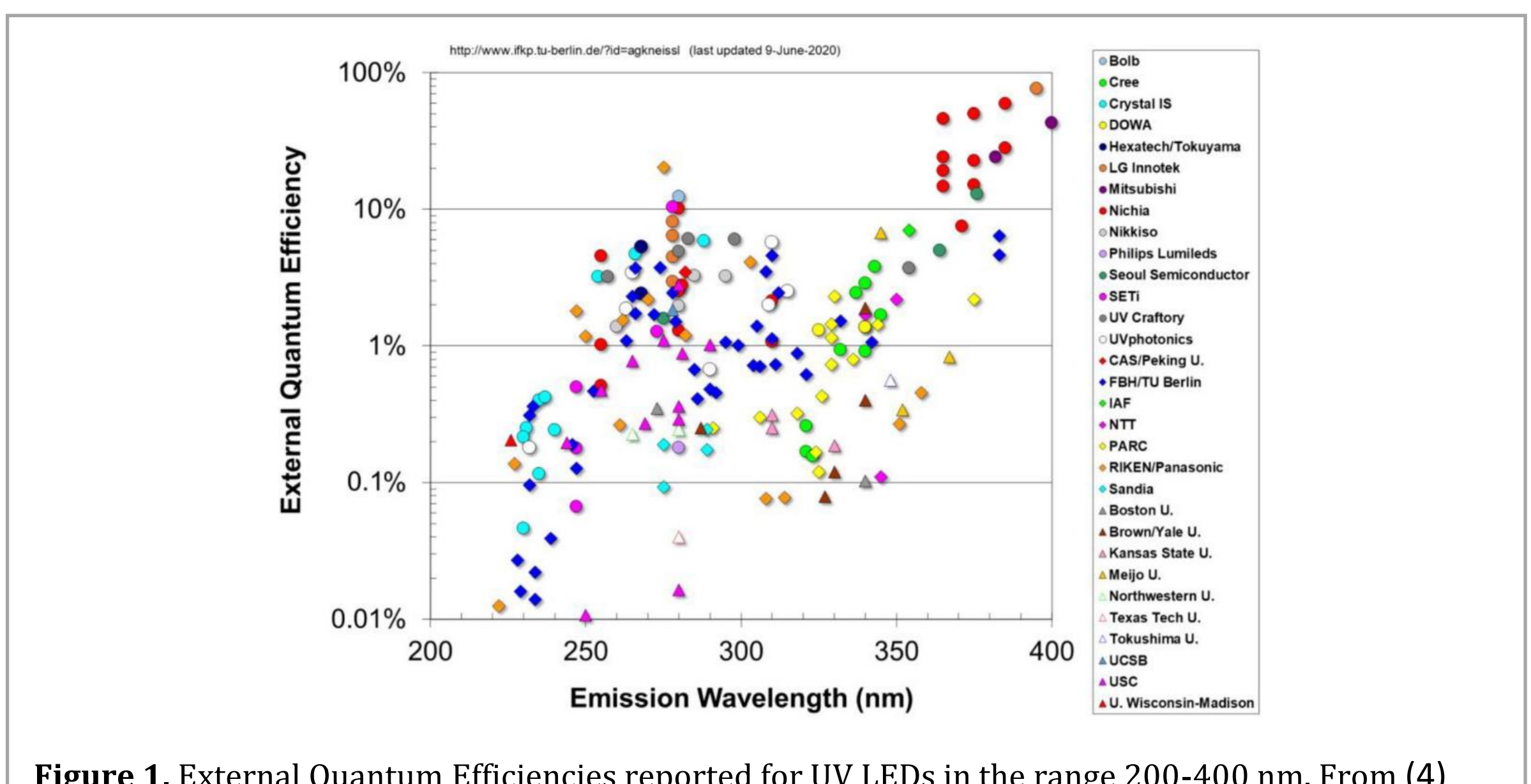

**Figure 1.** External Quantum Efficiencies reported for UV LEDs in the range 200-400 nm. From (4)

Industrial and commercial interest in OWC is strongest where radio struggles — capacity limits, congestion, interference/EMC constraints, licensing, security, or rapid deployment — and where optical links can deliver "fibre-like" performance without new physical infrastructure. Across OWC more broadly, industry is already investing in IR free space links for high-capacity backhaul/front-haul, temporary connectivity, campus links, and disaster recovery; and in light communications (e.g., LiFi) for RF-restricted or electromagnetic interference sensitive environments such as industrial sites, healthcare, aviation, and secure facilities, supported by emerging standards that improve procurement confidence.

Underwater optical wireless is also attracting pull (typically blue/green) for short-range, high-rate links to remotely operated vehicles, autonomous underwater vehicles, and subsea assets where acoustic links are bandwidth-limited. Within this landscape, UV-C OWC is a smaller but strategically distinctive niche, increasingly viewed less as "visible OWC at a different wavelength" and more as a mission-enabling mode where channel physics and operating constraints reward solar-blind operation and/or scattering-enabled NLOS.

In the UV-C band, much of the Sun's radiation is absorbed by the atmosphere, reducing background noise and enabling more robust daylight operation, while atmospheric scattering can support around-the-corner or less precisely aligned links (trading link budget for reduced pointing and line-of-sight constraints). These properties drive the most immediate UV-OWC interest in defence, security, and emergency response — ad-hoc tactical networks, RF-denied or jammed environments, and low-probability-of-intercept short-to-medium range links — as well as select industrial settings (refineries, chemical plants, tunnels/mines, ports, dense machinery) where RF multipath and interference are severe and where NLOS operation could simplify deployment for sensors, robotics, and local meshes.

In parallel, there is growing strategic interest in UV-C for space links — inter-satellite communications and quantum-enabled architectures — because there is no atmospheric absorption/scattering in space, existing optical communication standards and supply chains largely reflect C-band fibre heritage, and the solar background can be lower on a per-photon basis at sufficiently short UV-C wavelengths than in near or short-wave IR, strengthening the case for photon-efficient and quantum communications. However, industry's investment thesis is currently shaped by receiver limitations: the aforementioned response bandwidth constraint; and the dominant UV-C use cases (NLOS terrestrial links and long-range free-space links) are intrinsically photon-starved, demanding high sensitivity — ideally toward single-photon regimes. As a result, near-term commercialisation is likely to proceed through enabling components and modules (high-speed solar-blind detectors, UV-C emitters such as AlGaN LEDs/micro-LEDs, filters/optics, rugged packaging, and radiation-tolerant readout electronics) before scaling into integrated terminals and vertical solutions where UV-C's unique attributes justify complexity.

**Concluding remarks**

The pursuit of ever increasing communication bandwidth, the need for uncontested spectrum, and novel free-space applications provides strong motivations for operations in the UV-C region. The lack of background light, reduced diffraction, and potential NLOS operations are particular benefits that have yet to be fully exploited. Key to unlocking the potential of UV-C is the development of semiconductor UV-C sources and detectors with advantages in bulk, robustness, and mass production. However, there are several challenges in bandwidth, sensitivity, and device types to overcome in the course of technology platform maturation across the different wide-bandgap systems. There is also the need for a wider spread of UV-C compatible components to build full optical systems. Rapid progress in the characterisation and improvement of wide bandgap semiconductors is encouraging that these barriers may be overcome and UV-C communications emerges as a viable application solution.

**Acknowledgements**

J.J.D. McKendry acknowledges funding from the Future Telecoms Research Hub, Platform for Driving Ultimate Connectivity (TITAN), sponsored by the Department of Science Innovation and Technology (DSIT) and the Engineering and Physical Sciences Research Council (EPSRC) under Grants EP/X04047X/1 and EP/Y037243/1. D.K.L.O acknowledges funding from the UK Space Agency (ETP1-031, ETP4-052), the EPSRC Quantum Technology Hub in Quantum Communication (EP/T001011/1), the EPSRC Integrated Quantum Networks Research Hub (EP/Z533208/1), and the EPSRC International Network in Space Quantum Technologies (EP/W027011/1).

# 5.4. Environmental monitoring

**Ryan Pereira[1*], Graeme Moore[2], Tom Lendrem[3] and Christoph Wagner[4]**

[1] Lyell Centre, Heriot-Watt University, Edinburgh, United Kingdom
[2] Scottish Water, 6 Buchanan Gate, Glasgow G33 6FB, United Kingdom
[3] Badger Meter UK Ltd, Oldham, United Kingdom
[4] Badger Meter Austria GmbH, Vienna, Austria

*E-mail: r.pereira@hw.ac.uk

### Status

UV-C photodetectors, operating within the 200-280 nm wavelength range, have become increasingly relevant in environmental sensing, offering rapid, reagent-free analysis of water quality and organic matter dynamics. Its strength lies in the strong absorbance of aromatic and some inorganic compounds in this spectral range. This enables both direct and indirect quantification of key water quality parameters including suspended solids, nitrate, nitrite, bromide, thiosulfate, iron, manganese, chloramines, chlorates, chlorites, ozone, chemical and biological oxygen demand [e.g. 1, 2].

Modern UV absorbance spectroscopy relies on a range of light sources tailored to specific spectral needs. Deuterium lamps remain the standard for UV applications (160-400 nm), offering stable and continuous output ideal for dissolved organic carbon (DOC) and nitrate measurements. Xenon arc lamps provide broad-spectrum intensity across UV and visible ranges, suitable for high-sensitivity applications. Tungsten-halogen lamps, while limited to visible and near IR, are often paired with UV sources for full-spectrum analysis. Mercury lamps, with sharp emission lines, are used for calibration and targeted wavelength detection and disinfection applications. More recently, UV-C LEDs have emerged as compact, energy-efficient alternatives, offering narrow-band emission (e.g. 254 nm) for portable and in-line sensor systems. Each source presents trade-offs in intensity, stability, and spectral coverage, influencing instrument design and analytical performance in environmental monitoring.

In aquatic systems, UV-C photodetectors are widely used to directly assess DOC using absorbance at 254 nm, which correlates with aromatic carbon content [3] and indirectly to estimate biological oxygen demand and chemical oxygen demand using multiple wavelengths [4, 5]. When normalised to either higher absorbance wavelengths (e.g. 365 nm) or total DOC, established proxies provide insight into DOC concentration, molecular weight, aromaticity, potential source and reactivity [6-9]. Consequently, UV absorbance sensors are increasingly integrated into real-time monitoring platforms, enabling continuous assessment of water quality in fresh and marine waters [e.g. 10]. These systems support early warning for pollution events, track seasonal variability in organic matter, and inform real-time operations of water treatment and catchment-scale carbon flux models [1]. Beyond water, UV-C spectroscopy is emerging in air and soil monitoring. In atmospheric applications, UV-C photodetectors detect some volatile organic compounds, $NO_2$, $SO_2$, and $O_3$ via their strong absorbance in the UV-C range [11]. In soils, UV-C absorbance of waters in soil and groundwaters are being explored to characterise organic matter and nutrient profiles [e.g. 12].

Overall, UV-C photodetectors offer a versatile, scalable approach to environmental monitoring. Their ability to provide rapid, non-destructive measurements across diverse media positions them as a key technology in integrated environmental assessment frameworks.

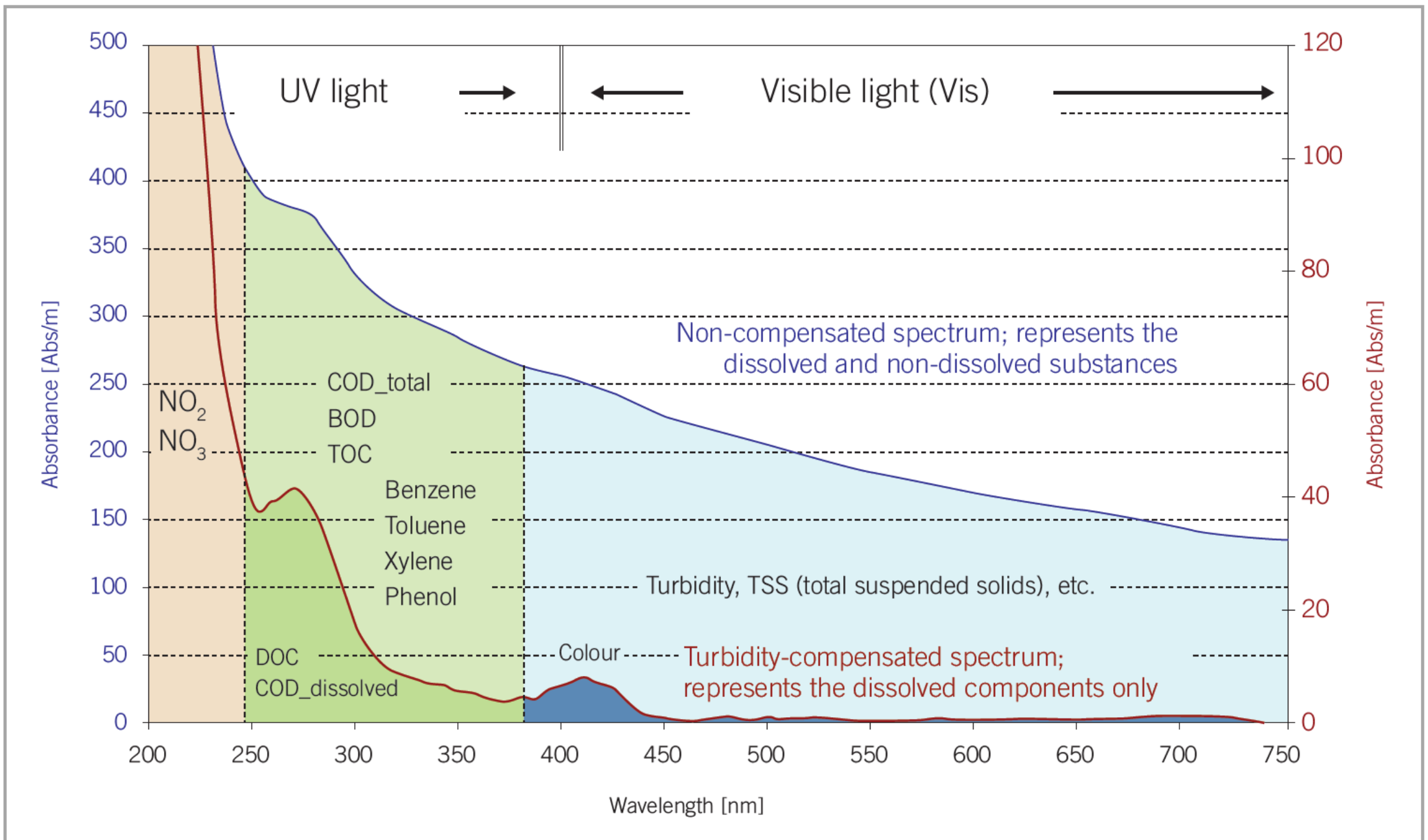

**Figure 1.** Typical UV/VIS spectrum (200-750 nm) of a water sample before (blue) and after (red) turbidity compensation highlighting spectral areas typically used for parameter evaluation (courtesy of Badger Meter).

### Advances in UV-C photodetectors for environmental monitoring

The deployment of UV-C photodetectors in environmental monitoring is constrained by a series of interrelated challenges, foremost among them being spectral complexity. In natural waters, the absorbance spectra of dissolved constituents often overlap, particularly in the UV-C range where aromatic organic compounds, nitrate, iron and other species exhibit strong but intersecting signals. This spectral overlap complicates the attribution of absorbance to specific analytes, reducing the reliability of single-wavelength proxies such as $UV_{254}$ (i.e. UV spectral absorbance at 254nm) and necessitating high-resolution optics and advanced signal processing to resolve individual contributions (Guan et al., 2024). Moreover, the relationship between UV-C absorbance and DOC is not always linear, particularly in systems influenced by photobleaching, flocculation, or microbial degradation [13]. Matrix variability further compounds these issues. Salinity, turbidity, and metal content can alter the optical behaviour of dissolved organic matter, shifting absorbance baselines and weakening the diagnostic ability of UV-C to determine DOC concentration and composition [14, 15]. Turbidity, in particular, introduces light scattering that distorts absorbance measurements (Figure 1), especially in high-flow or flashy systems that require correction approaches [16]. Standardisation across platforms is essential to ensure data comparability and interoperability. Calibration protocols must account for site-specific conditions, and inter-calibration between units is necessary to harmonise measurements across networks. Without consistent standards, the integration of UV-C photodetectors into distributed monitoring frameworks risks fragmentation and reduced data utility.

Scalability is increasingly important as UV-C photodetectors are integrated into IoT networks, unmanned aerial vehicles, and handheld platforms. These applications demand compact, low-power sensors with robust calibration and minimal maintenance. However, miniaturisation introduces trade-offs in optical path length, signal-to-noise ratio, and spectral resolution. The physical size of lamp sources and sensor housings can constrain deployment in mobile or embedded systems, particularly in remote or resource-limited environments. Operational barriers further limit photodetectors performance and scalability. Lamp lifetimes vary significantly across technologies, with mercury lamps offering stable output but requiring careful handling and disposal, while UV-C LEDs provide longer lifespans and rapid cycling but suffer from lower quantum efficiency and reduced intensity. Energy consumption remains a critical consideration, particularly for large-scale deployments or continuous monitoring systems. Mercury lamps typically operate at higher power levels, whereas LED-based systems offer energy savings but may require denser arrays to achieve comparable irradiance. Fouling presents an additional operational constraint impacting optical surfaces, attenuating UV transmission, adding species with intersecting signals, and degrading signal quality requiring rigorous maintenance protocols and fouling minimisation mechanisms.

Ultimately, the success of UV-C based photodetectors for environmental applications depends on their ability to disentangle overlapping signals and compensate for matrix-induced variability. Continued refinement of optical design, signal processing, and calibration protocols will be essential to expand their utility in dynamic and heterogeneous environmental systems both now and in the future.

**Advances in science and technology to meet challenges**

Recent developments in UV-C photodetectors technology have focused on improving analytical sensitivity, miniaturisation, and integration with real-time monitoring platforms. The shift from mercury lamps to UV-C LEDs has enabled compact, energy-efficient designs suitable for distributed networks, unmanned aerial vehicles, and handheld systems. LEDs now dominate commercial probes for nitrate, DOC, and dissolved oxygen detection, while broadband lamps (e.g. Xenon, Deuterium) remain relevant for multi-variable sensing. Photodiodes are the most common detectors, though mini-spectrometers using CCD/CMOS sensors are emerging, offering full-spectrum detection in fingertip-sized formats [17].

Sensor design has evolved to accommodate environmental constraints. Absorbance probes use variable path lengths (0.3–50 mm) to target specific concentration ranges, with longer paths suited to low-concentration monitoring and shorter paths for high-load environments. Cleaning mechanisms including wipers, compressed air, ultrasonic waves are now standard to mitigate fouling. Flow-through designs and nanoparticle coatings further enhance stability and reduce noise [18]. While UV–Vis spectroscopy offers a scalable alternative to costly lab-based sampling, signal processing innovations have significantly improved spectral interpretation. Machine learning models may offer significant advancement in addition to traditional regression approaches (e.g. Partial Least Squares Regression, Lasso, Stepwise) to predict analyte concentrations from absorbance spectra, especially when signal to noise ratios are at their limit. Signal processing techniques such as wavelet transforms and derivative analysis further enhance feature isolation. The integration of these models into sensor firmware enables autonomous, adaptive monitoring, advancing low-cost, continuous environmental assessment [19].

Importantly, UV-C photodetectors outputs are designed for seamless integration into big data frameworks. High-frequency data streams from fixed, mobile, and remote platforms are ingested into cloud-based systems, enabling centralised analysis. This facilitates predictive modelling, early warning systems, and real-time management across spatial and temporal scales. When UV-C data is combined with hydrological, meteorological, and remote sensing inputs, it enhances our understanding of ecosystem dynamics and treatment performance. This integration is particularly powerful when paired with advanced molecular analyses such as Liquid Chromatography – Organic Carbon Detection – Organic Nitrogen Detection (LC-OCD-OND), Fourier Transform Ion Cyclotron Resonance Mass Spectrometry (FT-ICR-MS), and Nuclear Magnetic Resonance (NMR) spectroscopy [20-22]. Together, these approaches offer the potential to resolve temporal and spatial variations in target analytes, calibrated against broader environmental constituents, thereby supporting investigations into carbon cycling and pollution dynamics

**Industry perspectives**

Industry adoption of UV-C photodetectors reflects a balance between innovation and operational constraints. In water monitoring, UV-C photodetectors are well-established for organics (DOC) and nitrate monitoring and compositional analysis, particularly in regulated treatment contexts. However, broader deployment is shaped by concerns around calibration and fouling potential impacting the ability to interpret and react to real-time data. Furthermore, UV-C LED systems are gaining traction in decentralised and remote applications, but their limited spectral range, power output and cost currently limit wider uptake. UV transmittance, expressed as a percentage, has been widely adopted to measure how much light passes through water. Low UV transmittance values suggest high turbidity or organic load, which can impair water treatment processes such as UV-C based disinfection. UV transmittance has further been adopted by the water industry for use in dose control/monitoring of organic matter removal [23]. Photocatalysis (involving the activation of semiconductor materials such as $TiO_2$ by UV-C) is an emerging technology of interest, but not yet practical technology for municipal treatment. However, this approach may offer an indirect approach to monitor organic pollutants such as pharmaceuticals, endocrine disruptors, and other emerging contaminants in real-time via an indirect measurement.

## Concluding remarks

UV-C photodetectors represent a powerful tool for environmental monitoring, offering real-time insights into organic carbon dynamics, disinfection efficacy, and water quality trends. Their application spans fresh and marine waters, engineered systems, and process waters with growing importance in climate-sensitive and regulatory-driven contexts. Current deployments rely heavily on $UV_{254}$ absorbance as a proxy for DOC and aromaticity, complimented by additional data sources. Advances in sensor design, algorithm development, and data analytics improve reliability and scalability. However, challenges remain. Interferences, sensor fouling, and calibration variability limit the accuracy of UV-C measurements in complex environments requiring considered monitoring design and implementation coupled with robust and experienced interpretation. The transition to UV-C LED technologies introduces new considerations around energy efficiency, system design, and sensor placement. Industry feedback underscores the need for robust monitoring strategies and regulatory alignment to support adoption. UV-C technologies are becoming increasing common in the water industry, with well-established applications for microbial disinfection and water quality monitoring. The transition from mercury-based lamps to UV-C LEDs reflects broader trends toward sustainability and supports a move away from other lamp technologies. Meanwhile, UV based photodetector instrumentation continues to enhance process control treatment, serving as a surrogate for organic carbon measurement and a tool for optimising coagulation and disinfection.

Future research should focus on developing standardised correction algorithms for matrix effects, validating sensor performance across diverse water types, and exploring the integration of UV-C sensors into autonomous and distributed monitoring networks. The refinement of LED technologies, coupled with advances in materials and data science, will be pivotal in realising the full potential of UV-C sensors in environmental applications. As the water industry evolves toward more sustainable and data-driven practices, UV-C photodetectors will play a central role in bridging analytical precision with operational efficiency. Their continued development and deployment will enhance our capacity to monitor, manage, and protect aquatic environments in a changing world.

## Acknowledgements

RP acknowledges support from the European Research Council (ERC) under the European Union's Horizon 2020 research and innovation programme (Grant Agreement No. 949495) and would like to thank Dr J Bischoff for discussions of this chapter.

# 5.5. Fire detection

**David Maestre[1], Emilio Nogales[1], David J Rogers[2], Eric Sandana[2], Michel Chamberlin[2], Ferechteh H. Teherani[2] and Bianchi Méndez[1, *]**

[1] Departamento de Física de Materiales, Universidad Complutense de Madrid, 28040 Madrid, Spain
[2] Nanovation, 8 route de Chevreuse, 78117 Châteaufort, France

*E-mail: bianchi@ucm.es

### Status

UV radiation plays a critical role in safety applications, including flame and fire detection, and partial-discharge monitoring [1]. Vacuum UV and the UV-C wavelengths coming from the Sun are blocked in the stratosphere by ozone, but many common flames and electrical discharges generate UV-C photons in the troposphere and thus they are detectable at long range because there is little to no UV-C background signal. Some examples are the UV-C signatures in the 200-260 nm range of methanol and hydrogen flames or the 185 – 260 nm emission from gasoline fire. Hence UV-C signals can play a role in remote optical fire detection/localisation in forests, large-scale facilities, cultural assets or transport vehicles, to name a few. Indeed, state-of-the-art remote optical fire sensors already combine an IR sensor with a high sensitivity UV-C PMT. The latter eliminates false positives from the IR sensors but requires high-voltage driving, has a limited lifetime and is fragile. For these reasons, there is a drive to replace PMTs with solid-state UV-C photodetectors which offer compactness, robustness, potential for integration with electronics and autonomous low voltage (or even self-powered) operation [2,3].

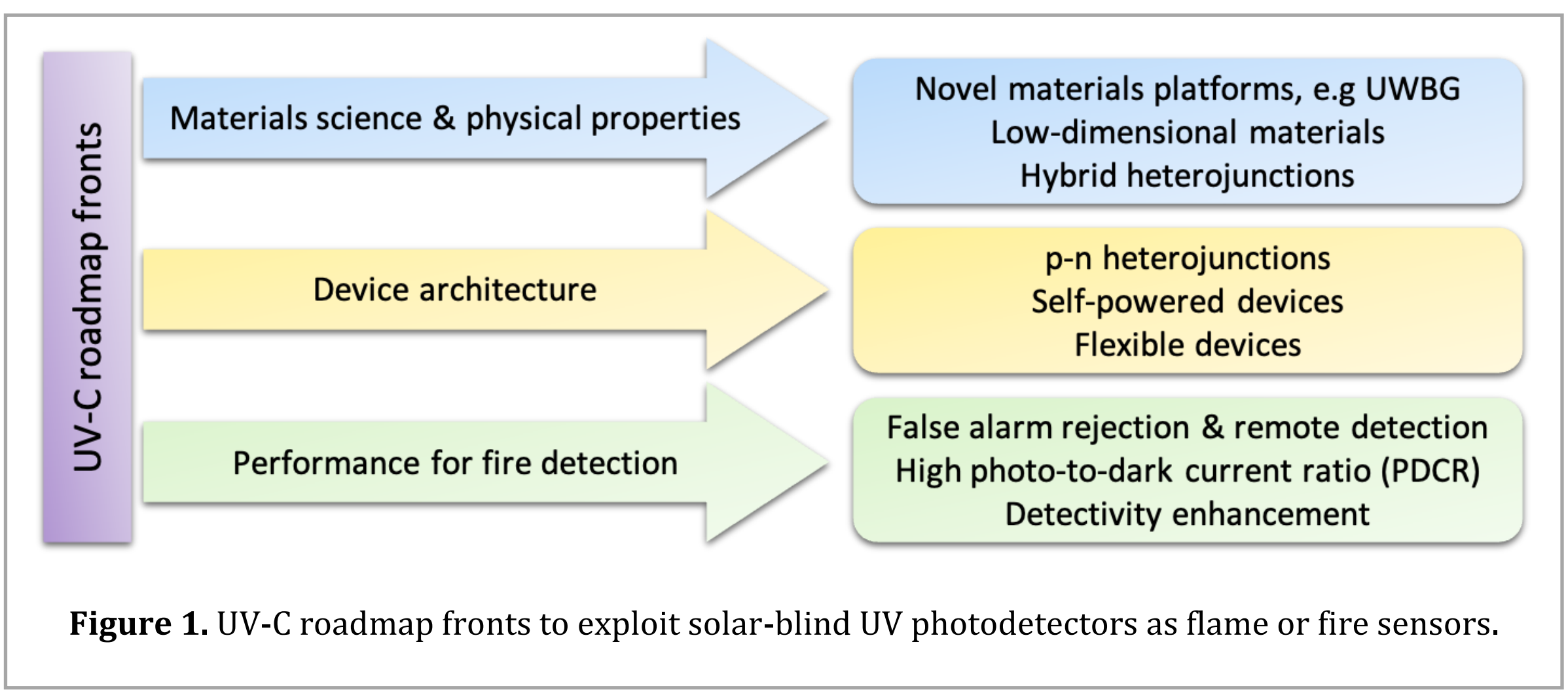

**Figure 1.** UV-C roadmap fronts to exploit solar-blind UV photodetectors as flame or fire sensors.

The solid-state UV-C detection roadmap going forward encompasses several advancing fronts in which novel materials and smart device co-designs are key to achieving the necessary performance for flame and remote fire sensing (Figure 1). A whole range of ultrawide bandgap materials are currently being considered for use as active components in such sensors. Among them, GaN (alloyed with Al), and $Ga_2O_3$, are the most suitable materials due to their solar blindness, high material quality and wide area manufacturability. On the other hand, the potential of low-dimensional and hybrid materials as efficient photo-absorbers is also being developed. The prospects of novel materials families, as well as innovative p-n heterojunctions capable of self-powered operation are also under development. Another approach is the exploration of flexible and lightweight devices. In terms of performance, the goals are to achieve high UV-C to solar rejection ratio to achieve sufficient noise reduction to get true detectivities above $10^{10}$ Jones.

**Challenges and advances in science and technology**

Each front in the UV-C photodetection roadmap has its challenges. Generally, photon detectors can be designed based on optimising internal or external photoelectric phenomena. Usually, a p-n junction or a Schottky barrier diode is at the core of the photodetector [1]. Regarding materials, an in-depth comprehension and control of the electrical, optical, crystallographic, morphological and defect structure properties of ultrawide bandgap materials is clearly needed. Low carrier mobility caused by intrinsic defects and the lack of p-type conductivity (in the case of $Ga_2O_3$) are some of the main issues restricting the performance of detectors based on ultrawide bandgap materials.

New trends look at low-dimensional materials, such as 2D materials [4], nanowires [5] or nanoparticles [6] to enhance light detection. In this regard, both materials science and fundamental physics are required to propose suitable device architectures for the photodetectors. Now, the development of methods to build heterojunctions based on thin films, nanomaterials and/or hybrid designs represents an active area of research. Excellent solar rejection, tuneable spectral responsivity, high sensitivity, high signal to noise ratio, low dark signal, and relatively fast response time are some of the performance criteria that position $Ga_2O_3$ as one of the most promising materials platforms for UV-C photodetection. Moreover, innovative features, such as the potential for autonomous photovoltaic operation, flexible devices or multifunctionality are potentially valuable assets for future solid-state UV-C sensors [2,3].

Advances in both experimental and atomistic modelling are contributing greatly to the comprehension of the microscopic mechanisms underlying UV-C photodetection. In recent years, defect/doping engineering has become a critical approach to control the optical and electronic properties in ultrawide bandgap semiconductors. In the case of UV-C detectors, although n-type doping and native defects can give high responsivity [7], they are not desired because they enhance the dark current/noise and induce persistent photoconductivity which are application killers for low level UV-C light detection in fire sensing. Therefore, controlling the synthesis of ultrawide bandgap materials is crucial to achieve the desired electronic band structure, defect chemistry and doping. In this sense, solar-blind UV-C detectors exhibiting ultrahigh detectivity have been designed recently by developing defect-poor $Ga_2O_3$ layers [8] or by applying pressure to modify the defect structure [9]. In another approach, the exploitation of less common $Ga_2O_3$ polymorphs, such as $\kappa$-$Ga_2O_3$ have also been tested successfully in fire detection [10].

Heterojunctions and Schottky diodes incorporating 2D materials, wide-bandgap nitrides, or conductive polymers can enhance charge separation and support self-powered operation. Layered materials such as TMDs or h-BN offer bandgaps spanning the near-UV to deep-UV [11], although only a small subset is suitable for solar-blind fire detection [12]. Another materials family worth mentioning is Metal–organic frameworks (MOFs) since they provide tuneable bandgaps and simple processing routes, with recent demonstrations of MOF-based UV-C photodetectors detecting alcohol flames under laboratory conditions [13].

The latest trends in science-based technologies employ deep-tech concepts to generate disruptive solutions in the fabrication of ultrawide bandgap semiconductors heterojunctions by innovative methods, such as machine learning. This work involves modelling and simulations capable of covering large length scales, in order to assess/optimise the performance for UV-C photodetection and to provide insight into the electrical transport of the heterojunctions and their dependence on various structural and external parameters. As a recent example, machine learning models using linear regression and neural network approaches have been reported to simulate the external quantum efficiency of a UV photodetector based on MgZnO thin film transistor [14].

In general, rapid advances in solid state UV-C photodetectors are driving towards application in flame and/or fire detection. The critical parameter combination requiring further improvement are sensitivity, detection limit, wavelength selectivity and time response, to be a practical solution for the early warning of fires. In some cases, sensitivity and detection limit parameters are overestimated due to the difficulties in the proper assessment of noise. Nevertheless, detectivities above $10^{10}$ Jones have been very recently reported in self-powered devices based on GaN/Sn:$Ga_2O_3$ and Li:NiO/$Ga_2O_3$ heterojunctions at 254 nm and 222 nm, respectively [5,15], with response times in the range of tens of milliseconds. This high sensitivity is possible due to a high photo-to-dark current ratio. It should also be mentioned that depending on the material platform, not all of these parameters are easily optimized simultaneously, and a trade-off, for example, between sensitivity and time response needs to be reached. In this sense, further work on novel materials and designs is needed in order to develop the protocols for an accurate determination of the detectivity.

### Industry perspectives

The global UV sensor market is projected to reach nearly USD 1 billion by 2035, driven by applications in environmental monitoring, safety systems, and industrial automation. A significant share relates to UV-C flame detectors and UV-C/IR multispectral devices used in oil & gas, chemical plants and power generation facilities [16]. Traditional UV-tube and UV-C/IR systems currently dominate the market due to their long-range sensitivity, resistance to false alarms and existing certifications (e.g., EN 54-10, SIL, ATEX). Industrial adoption of solid-state UV-C detectors remains at an early stage, however. In spite of the extensive studies in the last years on the materials side, there are few reports on operational environment demonstration or detector qualification so far. To go further in the TRL, scientists and engineers should work in a synergistic way to unravel fundamental and technological issues and attain solid-state UV-C photodetectors with appropriate performance and qualification [17].

To transition from laboratory prototypes to commercial flame-detection products, solid-state photodetectors must demonstrate low false-alarm rates, stable detection of standard test fires, fast recovery, and robust operation under temperature cycling, contamination and vibration.

Encouragingly, the increasing availability of ultrawide bandgap materials from emerging foundries, combined with advances in packaging, system diagnostics, and smart-sensor integration, is narrowing the gap between prototype and product. Moreover, features such as self-powered operation, flexible substrates and multifunctionality (UV-C + temperature or IR sensing) may provide unique advantages in next-generation fire-safety systems.

### Concluding remarks

Advances in ultrawide bandgap materials—especially high-quality $Ga_2O_3$ films grown on appropriate substrates—have strengthened the prospects for compact, robust solid-state UV-C photodetectors for flame and fire detection. Key performance parameters such as sensitivity, selectivity and response time continue to improve through defect engineering, heterostructure design and low-dimensional materials. To translate these advances into industrial adoption, future work must focus on stability, reproducibility and certification readiness.

Solid-state UV-C detectors offer compelling benefits in terms of miniaturisation, integration and low-voltage operation. As materials quality improves and system-level reliability is demonstrated, these devices may complement or eventually extend beyond the capabilities of traditional UV-tube and UV/IR fire-detection technologies.

### Acknowledgements

The authors acknowledge funding from Agencia Estatal de Investigación of Spain (MCIN/AEI/10.13039/501100011033) through Grants PCI2023-143388, PID2021-122562-NBI00 and PID2024-155966NB-I00.

# 5.6. Missile warning

**Yana Suchikova[1], Marina Konuhova[2] and Anatoli I. Popov[2*]**

[1] Scientific Department, Berdyansk State Pedagogical University, 69061 Zaporizhzhia, Ukraine;
[2] Institute of Solid State Physics, University of Latvia, Kengaraga 8, Street, LV-1063 Riga, Latvia

*E-mail: popov@latnet.lv

### Status

Missile warning systems constitute a critical component of modern defence and aerospace platforms, providing early detection and tracking of missile launches to enable timely countermeasures and threat mitigation [1]. Contemporary missile warning system architectures are inherently multisensor, typically combining IR, radar, and UV channels to achieve robust detection across diverse operational scenarios [2]. Within this multispectral framework, UV photodetection in the solar-blind UV-C range occupies a distinct and strategically important role [3].

The primary advantage of UV-C photodetection lies in its intrinsic insensitivity to solar background radiation at ground and near-ground levels, owing to strong atmospheric absorption below ~280 nm [4]. This "solar-blind" characteristic enables high signal-to-noise ratios during daylight operation, where conventional IR-based systems often face elevated false alarm rates due to solar reflections, clouds, or hot surfaces. In missile warning applications, UV-C sensors are particularly effective at detecting early-stage UV emissions associated with missile exhaust plumes, enabling rapid cueing of complementary sensing modalities and enhancing overall system reliability [5, 6]. These UV signatures typically originate from broadband combustion-related emissions within the solar-blind UV-C spectral region below ~280 nm. A conceptual representation of the role of UV-C photodetection within a multisensor missile warning architecture is shown in Figure 1.

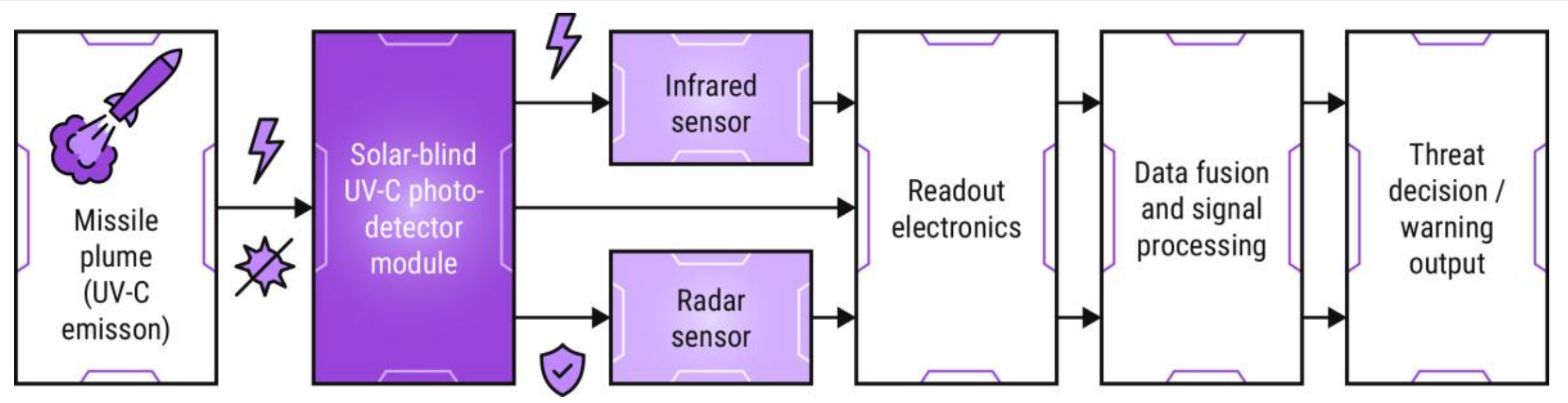

**Figure 1.** Conceptual architecture illustrating the role of solar-blind UV-C photodetection within a multisensor missile warning system. UV-C photodetectors complement IR and radar channels by enabling early-stage plume detection with reduced sensitivity to solar background radiation, thereby enhancing overall detection reliability and reducing false alarm rates.

Historically, UV detection in missile warning systems relied on vacuum-based technologies, including PMTs and photocathode devices. While these technologies offered high sensitivity, they were limited by size, fragility, power consumption, and long-term reliability. Over the past two decades, there has been a clear transition toward solid-state UV-C photodetectors, driven by advances in wide bandgap semiconductor materials and microfabrication technologies [7]. This shift has enabled more compact, rugged, and energy-efficient UV sensors compatible with modern airborne and spaceborne platforms [8].

Several wide-bandgap material platforms have emerged as leading candidates for solar-blind UV-C photodetection, including $Ga_2O_3$, AlGaN alloys, diamond, and, to a lesser extent, BN-based and oxide-derived systems. These materials offer intrinsic solar-blind bandgaps and are compatible with planar device architectures. However, their levels of technological maturity and integration readiness vary significantly. While AlGaN-based devices have demonstrated relatively advanced device engineering and array-level integration, emerging oxide-based platforms such as $Ga_2O_3$ are attracting increasing attention due to their ultra-wide bandgap, high breakdown fields, and potential for low-cost scalable processing [9, 10].

Currently, UV-C photodetectors are most commonly deployed as complementary elements within multisensor missile warning system architectures rather than as standalone detection solutions. Their primary function is to enhance early warning capability, reduce false positives, and improve confidence in threat identification [1-3]. Despite substantial progress at the material and device levels, broader adoption of UV-C photodetection in operational missile warning systems remains constrained by challenges related to long-term stability, device uniformity, environmental robustness, and system-level integration. Addressing these issues requires a coordinated perspective that links materials science, device engineering, and application-driven system requirements—an approach that underpins the roadmap considerations discussed in the following sections [10].

**Current and future challenges**

Despite the demonstrated relevance of solar-blind UV-C photodetection for missile warning systems, several persistent challenges continue to limit large-scale deployment and long-term operational reliability. These challenges span multiple levels, from system-level requirements to device physics and industrial qualification, and must be addressed in an integrated manner to enable broader adoption.

*System-level detection requirements*. At the system level, missile warning applications impose exceptionally stringent requirements on detection reliability. Ultra-low false alarm rates are essential, as spurious detections can lead to unnecessary countermeasures, mission disruption, or safety risks. Although solar-blind UV-C detection intrinsically suppresses solar background interference, false alarms may still arise from non-missile UV sources, sensor noise, or transient environmental effects. Achieving reliable discrimination under varying atmospheric conditions, viewing geometries, and operational altitudes remains a key challenge. In addition, missile warning systems demand rapid response times, often in the microsecond-to-millisecond range, to enable timely threat assessment and cueing of complementary sensors [11].

*Device-level performance trade-offs*. At the device level, trade-offs between sensitivity, dark current, and temporal response continue to constrain performance. Wide-bandgap UV-C photodetectors often exhibit low dark currents, which are advantageous for noise suppression, but may simultaneously suffer from reduced responsivity or slower carrier dynamics due to defect-related trapping and surface states [12]. Long-term stability is another critical concern: prolonged exposure to UV radiation, elevated temperatures, and radiation-rich environments can lead to parameter drift, responsivity degradation, or increased noise levels over time. Ensuring reproducible device characteristics across large-area arrays remains particularly challenging, especially for emerging material platforms with limited industrial processing heritage [13].

*Array integration and qualification.* Integration-related challenges further complicate the translation of laboratory-scale devices into operational missile warning systems. UV-C photodetectors must be compatible with focal-plane-array architectures, readout electronics, and system-level calibration protocols. Variations in pixel-to-pixel performance, temperature-dependent behaviour, and ageing effects complicate signal processing and increase the burden on downstream data fusion algorithms. Moreover, the lack of universally accepted standards for UV-C detector qualification in missile warning contexts hampers direct comparison between technologies and slows technology readiness advancement [13].

*Manufacturability and supply chain.* From an industrial and operational perspective, reliability, manufacturability, and supply chain robustness often outweigh peak laboratory performance metrics. Defence and aerospace stakeholders prioritise technologies that demonstrate stable operation over extended lifetimes, tolerance to harsh environmental conditions, and predictable failure modes. Emerging UV-C detector platforms must therefore address not only scientific and technological challenges, but also issues of scalability, reproducibility, and certification. These constraints define the critical bottlenecks that future advances in materials science and device engineering must overcome to meet the demanding requirements of missile warning systems [4]. These challenges can be broadly grouped into system-level requirements and corresponding technological bottlenecks, as summarised in Table 1.

**Table 1.** Key system-level requirements for UV-C photodetectors in missile warning systems and associated technological challenges.

| System-level requirement | Relevance for missile warning systems | Key technological challenges |
|---|---|---|
| Solar-blind selectivity | Reliable daytime operation with minimal background interference | Defect-induced leakage, surface states |
| Fast temporal response | Early detection and rapid cueing of complementary sensors | Carrier trapping, contact optimisation |
| Low false alarm rate | Operational reliability and mission safety | Noise suppression, signal discrimination |
| Long-term stability | Extended deployment with minimal recalibration | UV-induced degradation, thermal drift |
| Array uniformity | Integration into focal plane arrays | Process variability, yield control |

## Advances in science and technology to meet challenges

Addressing the challenges associated with UV-C photodetection in missile warning systems requires coordinated advances across materials science, device architecture, and system integration. Recent progress in these areas indicates several promising directions that can bridge the gap between laboratory demonstrations and operational deployment [4, 10].

At the materials level, significant effort has been devoted to improving intrinsic solar-blind selectivity and suppressing defect-mediated recombination. Wide-bandgap semiconductors with bandgaps exceeding the UV-C photon energy threshold inherently reject longer-wavelength radiation, reducing reliance on external optical filters [7, 8]. Advances in crystal growth, doping control, and defect engineering have reduced background conductivity and improved carrier transport, thereby directly enhancing signal-to-noise performance under low-photon-flux conditions typical of early-stage missile plume detection [4, 12]. Surface and interface passivation strategies have also emerged as critical enablers for stabilising device characteristics and mitigating long-term degradation under continuous UV exposure [4, 14].

Device-level innovations increasingly emphasise architectures that balance sensitivity, speed, and robustness. Optimised metal–semiconductor–metal, Schottky barrier, and p–n junction photodiode configurations have demonstrated improved temporal response and reduced noise through tailored contact engineering and electric field management. In parallel, there is growing interest in device designs that minimise carrier trapping and persistence effects, which can otherwise compromise fast transient detection. Importantly, advances are not limited to individual detector elements; uniformity and reproducibility across multi-pixel arrays are now recognised as equally critical performance metrics for missile warning applications [13].

Beyond individual devices, progress in process integration and packaging has played an essential role in enhancing operational readiness. The development of fabrication routes compatible with large-area substrates and standard microelectronic processing has improved yield and scalability, addressing long-standing manufacturability concerns. Packaging strategies designed to withstand thermal cycling, vibration, and radiation exposure have further increased the suitability of UV-C photodetectors for airborne and spaceborne environments [4].

At the system level, advances in readout electronics and signal processing are enabling more effective utilisation of UV-C sensor data. Improved readout integrated circuits enable low-noise operation and high-speed signal acquisition, while emerging data-fusion approaches allow UV-C channels to be combined with IR and radar inputs in a complementary manner. Rather than relying solely on absolute detector performance, system-level optimisation increasingly focuses on the coherent integration of multispectral information to enhance detection confidence and reduce false alarms [1, 2].

Collectively, these advances indicate a gradual shift from material-centric optimisation toward application-driven design philosophies. Continued progress will depend on close alignment between materials research, device engineering, and system-level requirements, ensuring that technological developments directly address the operational constraints of missile warning systems.

### Industry perspectives

From an industrial standpoint, the adoption of UV-C photodetectors in missile warning systems is less driven by laboratory performance records and more by demonstrated reliability, manufacturability, and integration readiness. Defence and aerospace stakeholders operate under conservative risk frameworks, where incremental performance gains are often secondary to predictable behaviour, long-term stability, and compliance with qualification standards. As a result, the transition of emerging UV-C photodetector technologies into operational systems is typically gradual and highly selective [4, 15].

One of the primary considerations for industry is TRL. While several UV-C detector platforms exhibit promising sensitivity and selectivity at the device level, only a subset demonstrate the reproducibility and yield required for array-level implementation. Uniformity across detector elements, minimal parameter drift over time, and compatibility with established readout and packaging technologies are essential prerequisites for system integration. Devices that fail to meet these criteria may remain confined to experimental or niche applications, regardless of their peak performance metrics.

Reliability under harsh operational conditions represents another decisive factor. Missile warning systems must operate across wide temperature ranges, withstand mechanical stress, and operate in radiation-rich environments, often for extended service lifetimes with limited maintenance opportunities. Industrial stakeholders therefore prioritise detector platforms with well-understood failure mechanisms and robust degradation pathways. In this context, materials and device architectures with simpler interfaces and fewer process-sensitive steps are often favoured over more complex but less predictable alternatives.

Supply chain resilience and scalability further influence technology selection. Defence-oriented production demands secure access to raw materials, stable fabrication routes, and the ability to scale manufacturing without introducing unacceptable variability. Technologies that rely on specialised or geographically constrained supply chains may face additional barriers to adoption, particularly in dual-use or export-controlled contexts. Consequently, compatibility with existing semiconductor manufacturing infrastructure is increasingly viewed as a strategic advantage [4].

Looking ahead, industry perspectives on UV-C photodetection increasingly emphasise system-level optimisation rather than standalone sensor performance. The value of UV-C detectors is maximised when they are seamlessly integrated into multisensor architectures, where data fusion and intelligent processing enhance overall detection confidence. As missile warning systems evolve toward greater autonomy and software-defined functionality, UV-C photodetectors are expected to play a complementary but indispensable role, provided that ongoing advances continue to align scientific innovation with the practical constraints of industrial deployment.

## Concluding remarks

Solar-blind UV-C photodetection has established itself as a strategically important component of modern missile warning systems, providing capabilities that are difficult to achieve with IR or radar sensing alone. Its intrinsic immunity to solar background radiation enables reliable early-stage detection of missile exhaust plumes, thereby improving situational awareness and reducing false alarm rates within multisensor architectures. As such, UV-C photodetectors are best understood not as standalone solutions, but as enabling elements that enhance the overall robustness and confidence of missile warning systems.

The current landscape reveals a clear shift in emphasis from material-specific optimisation toward application-driven performance metrics. While advances in wide-bandgap semiconductor platforms continue to expand the design space for solar-blind detection, the decisive factors for deployment increasingly lie at the interfaces between materials, devices, and systems. Long-term stability, reproducibility, integration compatibility, and qualification readiness now define the primary benchmarks for technological success.

Looking forward, further progress will depend on close alignment between fundamental research and system-level requirements. Coordinated development across materials engineering, device architecture, and industrial integration is essential to translate laboratory advances into operational capabilities. Within this framework, UV-C photodetection is poised to remain a critical contributor to next-generation missile warning systems, supporting the evolution toward more reliable, adaptive, and resilient defence sensing platforms.

## Acknowledgements

This research was funded by the Ministry of Education and Science of Ukraine under the project 0125U000156 for Y.S, while M.K. and A.I.P were supported by Latvian research project lzp-2023/1-0453, “Prediction of long-term stability of functional materials under extreme radiation conditions”.

# 5.7. Gas detection

**Luís F. da Silva[1] and Eduard Llobet[2,3,4*]**

[1] LM2N, Federal University of São Carlos, São Carlos, Brazil
[2] Universitat Rovira i Virgili, MINOS, School of Engineering, Tarragona, Spain.
[3] IU-RESCAT, Research Institute in Sustainability, Climatic Change and Energy Transition, URV, Vila-seca, Spain.
[4] TecnATox - Centre for Environmental, Food and Toxicological Technology, URV, Tarragona, Spain.

*E-mail: eduard.llobet@urv.cat

### Status

Many atmospheric pollutant gases such as ozone ($O_3$), nitrogen oxides ($NO_x$), sulphur dioxide ($SO_2$), hydrogen sulphide ($H_2S$) or vapours such as benzene, toluene, ethylbenzene, and xylenes (BTEX) exhibit strong, characteristic absorption bands in the UV-C range. This has been exploited via developing UV-C spectrophotometry as a technique for the highly selective detection of these pollutants [1]. The evolution of UV-C LEDs and photodetectors in the last years is raising the attractiveness of the technique, enabling the lowering of costs and miniaturization. The reported limits of detection lie generally between a few hundred part-per-billion (ppb) and units of part-per-million (ppm) with a linear detection range up to a hundred ppm [1]. Such performance is promising for developing widespread air quality control and environmental monitoring systems. However, because ambient moisture shows significant UV absorption for wavelengths under 200 nm, nitrogen is often used as a carrier. If ambient air is used instead as carrier, a de-humidifying filter needs to be included at the inlet of the measurement chamber. Although effective, both approaches are cumbersome to implement.

A different approach has consisted of using UV-light to boost the chemo-resistive performance of a wide range of gas responsive semiconductor nanomaterials such as metal oxides, carbon-based nanomaterials, mono and dichalcogenides and their hybrids [2]. All these nanomaterials generally require being thermally activated at temperatures well above room temperature for them to display significant and reversible gas responses. Upon UV excitation of the semiconductor nanomaterial, electron-hole pairs are photogenerated and separated under the effect of an electric field. The photogenerated charge carriers favour the desorption of surface species including strongly adsorbed oxygen species or the formation of surface vacancies, and the adsorption of weakly adsorbed oxygen species. These new weakly bound surface oxygen species can easily take part in redox reactions with the target gases, even if the gas sensitive films are kept at room temperature, generating measurable changes in the electrical resistance of the films [2-4]. This approach has been reported to activate the room temperature gas response of different semiconductor nanomaterials such as ZnO, $SnO_2$, $In_2O_3$, $TiO_2$, $WO_3$, $MoS_2$, $WS_2$, ZnS, $SnS_2$, $PtSe_2$, InSe and their hybrids, achieving remarkable response and fair selectivity to nitrogen dioxide, ozone, sulphur dioxide, hydrogen or formaldehyde, toluene, acetone and ammonia vapours with limits of detection ranging from a few ten ppb to few ppm [2-9]. Studies report rarely the use of UV-C excitation [10,11], being the UV-A the most used by far, as the latter better match the bandgap of the semiconductors studied as gas sensitive films. Implementing UV-C excitation is worth considering as it enables studying high-bandgap, monolayer metal oxides, which offer high surface area and carrier mobility for achieving superior sensing performance [12].

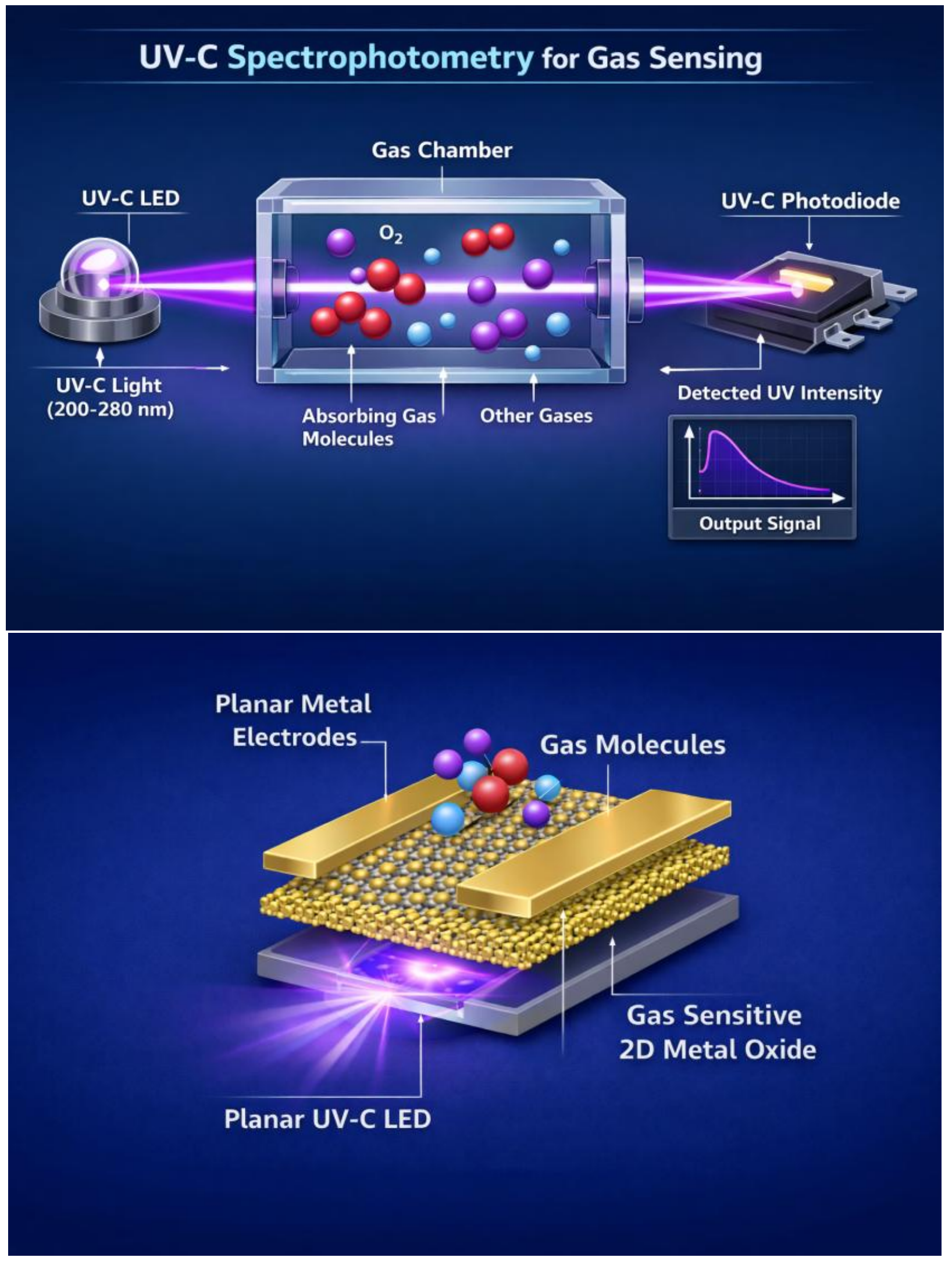

**Figure 1.** Cartoon of affordable UV-C spectrophotometry using an UV-C LED and photodiode for gas sensing (top). Artistic view of an integrated UV-C light activated chemo-resistive gas sensor employing a two-dimensional metal oxide as gas sensitive material (bottom). Generated by the authors using Open AI resources.

### Current and future challenges

In UV-C spectrophotometry the future research aims at developing portable devices that challenge the analytical performance of lab-based spectrophotometers (see Figure 1, upper panel).

*Ameliorating the performance of UV-C light emission and detection components*. UV-C LEDs face the challenges of needing highly stable power supply sources, they still experience susceptibility to thermal fluctuations and suffer from parasitic emission and thermally induced noise. UV-C photodetectors need improved spectral selectivity, higher rejection to the visible range, reduced dark current and enhanced stability upon systematic exposure to UV-C radiation.

*Miniaturization and optimization of optofluidic components.* The miniaturization and optimization of the optofluidic components (e.g., the gas cell) are current challenges for achieving an efficient UV transmission, an easy alignment and longer optical paths, but also for enabling a simpler component integration and overall fabrication process.

UV-light activated resistive gas sensing operating at room temperature, as an alternative to the traditional thermal activation of semiconductor materials has proved its feasibility and advantages. Yet, gaining a deeper understanding on the sensing mechanisms associated to the UV light excitation of the very wide range of gas sensitive semiconductor nanomaterials, would help establishing the technology and facilitate industry uptake.

*Gaining deeper knowledge on UV-C activated gas sensing.* Further understanding of mechanisms is needed, particularly, for new materials such as $Ga_2O_3$, g-$C_3N_4$, AlGaN, MgZnO, 2D semiconductors, metal halide perovskites, etc., being researched now and those considered in the future.

*Long term stability and selectivity.* Other very relevant challenges that need to be addressed are the long-term stability of the gas sensitive nanomaterials operated under highly energetic UV-C light and the currently experienced cross sensitivity to environmental factors (e.g., ambient moisture and temperature) and to other gaseous species, which compromise sensor performance.

*Achieving integration for mass production at reduced cost.* As simplicity and affordability are key features of chemo-resistive sensors, another challenge is achieving the integration of the UV-C light source, the gas sensitive nanomaterial and the interdigitated transduction elements within a single device (see Figure 1, lower panel). This would enable, on the one hand to ensure that a homogeneous UV excitation of the whole gas sensitive film is achieved and, on the other hand to ease device miniaturization and compactness. To this end, challenges relative to their manufacturing technology for achieving a reliable mass production need addressing.

### Advances in science and technology to meet challenges

In LED devices, nitride semiconductors research would help achieving low-power UV-C emitters and very low parasitic emission. Efforts for achieving constant power supply and improved thermal management would help minimizing the experienced thermal noise, thus resulting in enhanced stability. Research in innovative UV-C absorbing materials for narrow beam photodiodes that can be directly coupled to the optical system avoiding the use of filters is much needed. Sustained improvement in SiC photodiodes shows promise for achieving high-performance, narrow beam ($\leq$ 10 nm) deep-UV devices. Similarly, amorphous $Ga_2O_3$ is well suited, given its wide bandgap and low manufacturing costs, for adsorbing UV-C without interferences from the visible and IR ranges. However, improving film quality via controlling growth defects is needed to realise its promise [13]. A route worth exploring is the doping of metal oxides such as ZnO with heteroatoms (e.g., F or/and Ag) [14] or $SnO_2$ with rare earths [15] in order to suppress native visible defects in metal oxides for achieving UV-C absorption without parasitics. Further efforts are needed in the design of miniaturised gas cells and adapting hollow core waveguides for UV-C applications shows good prospects for achieving improved sensitivity and portability [1].

Considering UV-C activated chemo-resistive sensing, advances are needed towards the use of wide gap, two-dimensional semiconductor nanomaterials. While transition metal chalcogenides show susceptibility to oxidation under standard environmental conditions, monolayer or few layer metal oxides such as $ZrO_2$, $GeO_2$ or $SnO_2$ show bandgaps that range between 4 and 6 eV, superior oxidation resistance and high carrier mobilities peaking above 3,000 $cm^2\ V^{-1}\ s^{-1}$ [12]. Thus, efforts are needed towards the scalable production of such 2D metal oxides and their integration in highly sensitive and stable devices, able to operate under standard environmental conditions [16]. Mastering interfacial properties (e.g., defects) and uniformity in hybrid materials such as metal oxides/graphene will help achieving high sensitivity, selectivity, stability and short response and recovery times (i.e., few seconds) [17]. In that sense, using UV-C light treatments should be developed further as a way to desorb surface species including oxygen ions and hydroxyls and help reduce surface defects [18]. Further research for gaining a deeper understanding on the chemo-resistive sensing mechanisms under UV-C light excitation should be directed towards the implementation of *in operando* spectroscopies techniques (e.g. XPS, XAS, FTIR, impedance, and photoluminescence). In the case of UV-C-light activated gas sensors, photoluminescence spectroscopy studies conducted while the sensor resistance changes due to cyclic exposures and recoveries to gases are recorded could be particularly helpful at understanding photon-gas-matter interactions.

### Industry perspectives

Commercial UV gas analysers comprise UV-light sources and spectrometers. Configured as desktop, rack-mounted or portable gas analysers, they can selectively and accurately detect different gases such as $SO_2$, $NO_x$, and $O_3$. These are high-end products, widely used in research, environmental monitoring, industrial process control, and emission monitoring. The global UV Gas analyser market is projected to grow from US$ 203 million in 2024 to US$ 290 million by 2031, at a Compound Annual Growth Rate of 5.2% (2025-2031) [19]. This growth is driven by increasingly stringent environmental regulations worldwide. In the next years the evolution of UV-C LEDs and photodetectors coupled to the miniaturization of the associated optofluidics can be a game changer, enabling simpler, more affordable, hand-held or even wearable gas analysers to become available. This would extend their current applications to the food industry, the monitoring of occupational exposure to harmful gases or to the pharmaceutical and medical industry.

Besides, there is also a strong demand for gas sensors and detectors with high sensitivity, selectivity, stability, low-power operation and, in some cases, intrinsically safe to operate in potentially flammable/explosive environments. In this sense, the UV excitation of room-temperature operated chemo-resistive gas sensors and sensor arrays paves the way for the development of ultralow-power devices with many applications such as IoT enabled, portable/wearable hydrogen leak detectors, inexpensive and spatially distributed air quality monitoring nodes ($NO_x$, $SO_2$, $O_3$, BTEX), industrial safety and personal safety devices or point of care non-invasive diagnostic devices (e.g., breathalysers), only to cite a few. Market acceptability and the realisation of the potential impact of this technology in the different application fields will heavily depend on the success of achieving monolithically integrated UV-light activated gas sensors. The structure of a micro-LED photoactivated gas sensor should comprise a stack of contact electrodes to drive a UV micro-LED platform, electrically insulated electrodes to measure the gas sensing signals and a gas-sensitive film [20]. The use of standard MEMS fabrication processes to achieve such devices will be key for scaling them up to mass production with high device-to-device reproducibility at very low cost.

### Concluding remarks

Many harmful compounds, either naturally produced or from anthropogenic origin possess distinctive absorption signatures in the UV-C range. UV-C spectrophotometry for gas detection offers high accuracy and selectivity in high-end applications of environmental monitoring. The evolution of UV-C LEDs and UV-C photodetectors will enable the emergence of miniaturised gas detectors to challenge the accuracy of lab-based spectrophotometers. To this end, new materials, dopants and their manufacturing need to be researched for enhanced stability, supressing parasitic emission, minimising thermal noise, narrowing absorption bands, showing better immunity to changes in environmental conditions and reaching low power operation. In contrast, middle-end and low-end gas detection applications are often addressed using affordable sensors such as chemo-resistors. In the last years, the room-temperature operation of UV-C LED activated chemo-resistors has shown its potential. The successful monolithic integration of UV-C LEDs with two-dimensional, high bandgap, gas sensitive semiconductor materials will pave the way for setting a new generation of affordable gas sensors with superior performance. Besides the efforts needed for improving UV-C LEDs, sustained research in new gas sensing semiconductor materials amenable to standard clean room fabrication processes and efforts to gain a deeper understanding in the UV-activated gas sensing are still needed. It can be expected that these advances will result in impactful changes in the global market of gas sensing analysers and detectors.

### Acknowledgements

E.L. is supported by the Catalan Institution for Research and Advanced Studies via the 2023 Edition of the ICREA Academia Award. L.F.S. is supported by the São Paulo Research Foundation (under grant No. 2022/02927–3) and Brazilian National Council for Scientific and Technological Development (under grants No. 300153/2025–2 and 403501/2024–5).

# 5.8. Medical diagnostics

**Sangjin Yoon[1], Dohyung Kim[1], Sangwoo Hong[1] and Seung Hwan Ko[1*]**

[1] Department of Mechanical Engineering, Seoul National University, Seoul, Republic of Korea

*E-mail: maxko@snu.ac.kr

### Status

In medical diagnostics, UV-C photodetection should be viewed as more than a comparison of detector materials. It supports spectral readout, imaging, and quantitative reporting within diagnostic workflows. Such approaches have been explored for blood-cell analysis, cytology-based assessment of premalignant and malignant oral lesions, and classification of abnormal tissue specimens such as colorectal adenocarcinoma using intrinsic UV-derived contrast. Historically, UV-C use in diagnostic studies often remained at the laboratory scale because UV-C sources, optical components, and detectors were not widely available or straightforward to integrate into compact systems. As UV-C photonic components have improved and become more accessible, assembling label-free biomolecular imaging and analysis platforms has become more feasible. These advances include broader availability of UV-C emitters such as UV-C LEDs, improved planar imaging detectors/detector arrays, and better spectral filtering optics that help isolate the wavelength region of interest [1].

Current discussions of UV-C photodetection in medical diagnostics typically cover several related directions. One research direction emphasizes label-free UV measurements, where intrinsic absorption, Raman-related signatures, and autofluorescence contrast provide information while reducing reliance on dyes or labels [2]. Another direction examines whether UV-derived spectral features from clinical specimens such as plasma or tissue can serve as indicators for sample-level classification or monitoring [3, 4]. In parallel, UV-C adoption in healthcare environments has increased, which creates a practical need to verify delivered dose and spatial distribution under real operating conditions. This need has driven growing interest in monitoring and verification methods from an industry perspective [5].

The topic remains important for clinical translation. Label-free UV approaches can reduce sample preparation and consumable use, and they may shorten turnaround time and cost. Moreover, as UV-C adoption expands in hospitals, outcomes can depend on operational and environmental variables such as wavelength selection, delivered dose, humidity, and temperature. This dependence strengthens the role of UV-C photodetection as enabling infrastructure for measurement and monitoring [5]. Further progress is expected from improvements at the workflow level. Examples include more reliable signal extraction at lower irradiation, integrated analysis that combines spectral and imaging information for automated interpretation, and clinical procedures that are defined more consistently across sites. As these elements mature, UV-based diagnostics may move from qualitative visualization toward reproducible and reportable quantitative outputs. Sensor-enabled quality assurance can also support reliable deployment and operational optimization of UV-C technologies in clinical environments.

### Current and future challenges

*Quantitative reproducibility.* A major challenge in applying UV-C photodetection to medical diagnostics is achieving reproducible quantification despite the complexity of clinical specimens and the variability of real operating environments. In the UV-C band, measurements can be affected by strong absorption/scattering and limited penetration depth, making signals sensitive to sample geometry and surface conditions; consequently, results may vary across specimens and sites [2, 6].

*Signal quality versus photodamage.* Another key limitation is the fundamental trade-off between signal quality and photodamage. UV-C irradiation can induce ionization, radical formation, bond cleavage, and secondary reactions with surrounding molecules. Consequently, a common strategy in visible imaging—reducing irradiance while compensating with longer acquisition—may not sufficiently suppress UV-C-induced molecular degradation [1]. For this reason, protective approaches have been investigated, including strategies that suppress damage while still enabling strong Raman-type readout. Overall, these efforts highlight the need for system-level optimization that balances photodamage control with reliable quantitative measurement [7]. From a workflow perspective, the practical target is to define acquisition conditions that provide adequate diagnostic contrast while keeping photochemical alteration below a threshold that would bias downstream interpretation.

*Biological variability and protocol design.* Biological uncertainty and protocol design are also important in UV-C–related diagnostic or verification contexts. Microorganisms can exhibit photoreactivation and dark repair after UV-C exposure, meaning that inactivation can be partially reversed; therefore, standard operating procedures (SOPs) and validation studies should explicitly specify readout timing and post-exposure conditions [8]. This requirement is particularly relevant when UV-C measurements are used for verification or comparative studies, where inconsistent post-exposure conditions can lead to apparent discrepancies that are methodological rather than technological.

*Safety and operational integration.* As deployment expands beyond controlled laboratory settings, safety and operational integration become increasingly significant barriers. In healthcare environments, UV-C effectiveness can depend on wavelength, dose, relative humidity, and temperature, and practical factors such as shadowing and environmental variability can reduce performance unless robust operating protocols are used [5]. In addition, ozone generation can become a concern when emission spectra include wavelengths below ~240 nm, and human exposure risks require engineering controls and placement guidelines [9, 10]. Thus, future progress will depend not only on incremental detector improvements but also on system-level standardization, including clinical endpoints, multi-site validation, safety interlocks, and operational SOPs. Establishing such a framework will be essential for translating UV-C photodetection from promising demonstrations into reliable, routine use across diverse clinical and healthcare environments.

**Advances in science and technology to meet challenges**

Several lines of development are helping to address the challenges outlined above. On the detector side, progress in ultrawide-bandgap semiconductor materials—including β-$Ga_2O_3$ and Al-rich AlGaN—has improved the feasibility of solar-blind photodetection in clinical settings [11, 12]. Because these materials are inherently insensitive to visible and near-UV ambient light, they can provide improved signal-to-noise ratios without bulky external optical filters, thereby simplifying the design of point-of-care UV-C measurement systems.

At the platform level, integration of UV-C sources and detectors onto common substrates is advancing through on-chip approaches. In a recent demonstration, a cascaded n–p–n AlGaN/GaN photodiode array combined with a deep-neural-network-based spectral reconstruction achieved sub-nanometre spectral resolution and sub-10 ns response time within a miniaturized chip format, enabling spatially resolved single-shot imaging of organic analytes in the UV range [13]. While full clinical validation of such integrated devices remains ongoing, the trend toward on-chip UV spectroscopy offers a pathway to bring laboratory-grade analysis closer to the patient.

On the computational side, deep learning methods are being explored to extract diagnostic information from low-signal UV data more effectively. Approaches such as virtual staining—where label-free autofluorescence images of unlabelled tissue are computationally mapped to histology-equivalent representations using generative adversarial networks—may reduce the required irradiation dose while still providing useful diagnostic contrast [14, 15]. This is directly relevant to the photodamage challenge: if sufficient contrast can be obtained at lower UV-C exposure levels, the trade-off between signal quality and specimen integrity becomes more manageable. Automated interpretation pipelines may also help reduce operator-dependent variability, although their reliability in diverse clinical populations requires further evaluation.

In terms of clinical workflow, microscopy with UV-C surface excitation has demonstrated the ability to provide optical sectioning of fresh tissue surfaces using approximately 280 nm light, without the need for conventional fixation and physical sectioning steps [16]. This capability is of particular interest for intraoperative settings, where rapid tissue assessment can inform surgical decisions.

More broadly, progress is also needed in defining standardized acquisition and readout protocols. Since biological responses such as photoreactivation can influence measured outcomes depending on post-exposure conditions, advances in measurement technology will need to be accompanied by corresponding improvements in procedural standardization and multi-site validation frameworks to support reproducible clinical use.

**Industry perspectives**

The UV-C industry is transitioning from mercury-vapor-based sources to solid-state alternatives, driven in part by the Minamata Convention mandate to phase out mercury. While the largest commercial segments remain in water treatment and surface disinfection, the availability of UV-C LEDs with improved output power and operational lifetime has reached a level where integration into diagnostic instruments is becoming practical [17]. In particular, the maturation of AlGaN-based emitters and detectors operating in the 260–280 nm range is supporting the development of compact, mercury-free measurement platforms relevant to nucleic acid analysis and label-free spectroscopic sensing.

This hardware maturation has also supported the emergence of companies developing integrated diagnostic solutions, including systems based on UV-excited tissue imaging and computational analysis. The commercial interest in reducing diagnostic turnaround time and reducing dependence on consumables appears to be a common driver across these efforts. However, the transition from laboratory prototypes to routine clinical deployment requires not only device-level performance but also system-level reliability and workflow compatibility.

From a regulatory standpoint, the 2022 revision of the ACGIH threshold limit values for UV radiation has influenced the broader UV-C landscape [9, 18]. The revised guidelines, which separated eye and skin exposure limits and substantially increased the permissible daily dose at 222 nm, are most directly relevant to germicidal applications. Nevertheless, they also carry implications for medical environments in which UV-C photodetection systems may operate alongside or within disinfection infrastructure. In such settings, ensuring that diagnostic and germicidal UV-C deployments are compatible and that cumulative exposure remains within established limits becomes a practical design and operational consideration.

A notable gap in the current industry landscape is the absence of standardized, purpose-built UV-C dosimetry solutions for medical applications. While photodetectors and radiometers are available for general UV measurement, few are specifically designed, calibrated, and certified for the dose verification needs of clinical diagnostic or therapeutic workflows [19]. Addressing this gap will likely require collaboration among detector manufacturers, instrument developers, and regulatory bodies to define performance standards and calibration protocols appropriate for medical use. While miniaturization and unit cost are highly important as platforms move toward point-of-care and integrated formats, we consider solar-blind rejection ratio, low-irradiance sensitivity, calibration standards, and reliability to be the most critical attributes for medical diagnostic use. This is because they jointly determine whether adequate signal to noise ratio and quantitative reproducibility can be achieved while keeping photodamage within diagnostic tolerance.

Looking forward, commercial progress in this space will depend on system-level integration and workflow compatibility. Diagnostic modules that can be incorporated into existing clinical procedures with minimal additional training or infrastructure are more likely to see adoption. As hardware, software, and regulatory frameworks continue to mature in parallel, UV-C photodetection may become a more routine component of rapid, label-free diagnostic workflows in healthcare settings.

### Concluding remarks

In medical diagnostics, UV-C photodetection is advancing from laboratory-scale studies toward routine, label-free quantitative analysis. While transitioning to real-world clinical use, a fundamental challenge remains the trade-off between signal quality and UV-induced specimen photodamage. Fortunately, parallel technological breakthroughs—such as ultrawide bandgap semiconductors, on-chip sensor integration, and deep learning-based computational methods like virtual staining—are effectively mitigating these limitations by enabling high-contrast imaging at lower irradiation doses. However, overcoming device-level hurdles is only part of the solution. As the industry shifts toward compact, solid-state emitters, successful clinical deployment will increasingly depend on system-level standardization. Moving forward, developing purpose-built medical dosimetry, establishing robust SOPs, and ensuring seamless workflow compatibility will be critical to establishing UV-C technologies as reliable diagnostic infrastructure in healthcare environments.

### Acknowledgements

This work was supported by the National Research Foundation of Korea (NRF) grant funded by the Korea government(MSIT) (No. RS-2025-11092968).

# 6. Concluding remarks

**Fabien Massabuau**[1]

[1] Department of Physics, SUPA, University of Strathclyde, Glasgow, United Kingdom

E-mail: f.massabuau@strath.ac.uk

UV-C photodetectors are entering an important phase of development, driven by rapid advances in wide bandgap semiconductors, combined with improved performance and availability of UV-C LEDs and a growing demand across wide ranging applications. This roadmap highlighted a diverse technology landscape, spanning mature incumbent material solutions such as SiC as well as emerging and exploratory platforms including $Ga_2O_3$, AlGaN, BN, diamond, MgZnO, 2D materials, metal halide perovskites and MEMS-based devices. Applications of UV-C photodetectors in metrology, astronomy, communications, environmental monitoring, fire detection, missile warning, gas sensing, and medical diagnostics were presented together with perspectives as to what industry is seeking to adopt a technology.

Perhaps a key takeaway is that while proof-of-concept demonstrations understandably focus on figures-of-merit of a specific material platform, it will be important to consider more industry-focused prioritisations. Across the materials chapters, the common priorities rotate around the idea of improving figure-of-merit in a scalable, system-integrated context and at lower cost. However, the industry perspectives reveal that suitable technology will not necessarily be selected based on classical figures-of-merit alone, but other factors such as solar rejection ratio, device reliability and supply chain must be seriously accounted for.

Considering these constraints now can make the difference between a technology that will be deployed industrially and a technology that remains confined to research and niche applications. With coordinated effort between materials research, device engineering, metrology, industry and end-users, UV-C photodetectors can move beyond proof-of-concept demonstrations toward robust, deployable technologies with substantial scientific and societal impact.